\documentclass[a4paper,fleqn,usenatbib,useAMS]{mnras}

\usepackage[T1]{fontenc}
\usepackage{ae,aecompl}
\usepackage{newtxtext,newtxmath}
\usepackage{multirow}

\usepackage{subfigure}

\usepackage{graphicx}	
\usepackage{amsmath}            
\usepackage{amssymb}            
\usepackage{natbib}

\title[The central regions of MW galaxies]{The central spheroids of  Milky Way mass-sized galaxies}

\author[Tissera et al.]{ 
        Patricia B. Tissera$^{1,2}$\thanks{patricia.tissera@unab.cl}, 
Rubens E. G. Machado$^{1,3,4}$,
Daniela Carollo$^{5,6}$,
Dante Minniti$^{1,2,7}$,
\newauthor
 Timothy C. Beers$^{8}$
Manuela Zoccali$^{2,9}$,
Andres Meza$^{1,10}$\\
$^{1}$Departamento de Ciencias Fisicas, Universidad Andres Bello,  Av. Republica 220, Santiago, Chile.\\
$^{2}$Millennium Institute of Astrophysics, Av. Republica 220, Santiago, Chile.\\
$^{3}$Universidade Federal de Ouro Preto, Departamento de Fisica, Campus Universitario Morro do Cruzeiro, 35400-000, Ouro Preto, Brazil.\\
$^{4}$Universidade Tecnologica Federal do Parana, Rua Sete de Setembro 3165, 80230-901 Curitiba, Brazil.\\
$^{5}$INAF-Osservatorio Astronomico di Torino, 10025 Pino Torinese,
Italy.\\
$^{6}$Research School of Astronomy and Astrophysics, The Australian National University, Canberra, ACT 2611, Australia.\\
$^{7}$Vatican Observatory, V00120 Vatican City State, Italy.\\
$^{8}$ Department of Physics and JINA Center for the Evolution of
the Elements, University of Notre Dame, Notre Dame, IN 46556, USA.\\
$^{9}$Institute of Astronomy, Av. Vicu\~na Mackena, Pontificia
Universidad Cat\'olica de Chile.\\
$^{10} $Facultad de Ingenier\'ia, Universidad Aut\'onoma de Chile, Pedro de Valdivia 425, Santiago, Chile.
}

\date{Accepted 2017 September 18. Received 2017 September 18; in original form 2017 February 02}

\pubyear{2016} 

\begin{document}

\label{firstpage}
\pagerange{\pageref{firstpage}--\pageref{lastpage}}

\maketitle

\begin{abstract}

We study the properties of the central spheroids located within 10 kpc
of the centre of mass of Milky Way mass-sized galaxies simulated in
a cosmological context. The simulated central regions are dominated by
stars older than 10 Gyr, mostly formed in situ, with a contribution of
$\sim 30$ per cent  from accreted stars. These stars formed in well-defined
starbursts, although accreted stars exhibit sharper and earlier ones. The
fraction of accreted stars increases with galactocentric
  distance, so that at  a
radius of $\sim 8-10$ kpc a  fraction of $\sim 40 $ per
cent, on average, are detected. Accreted stars are slightly younger,
lower metallicity, and more $\alpha$-enhanced than in situ stars. A
significant fraction of old stars in the central regions  come from
a few ($2-3$) massive satellites ($\sim 10^{10}M_\odot$). The bulge
components receive larger contributions of accreted stars formed in
dwarfs smaller than $\sim 10^{9.5}M_\odot$. The difference between the
distributions of ages and metallicities of old stars  is thus linked to
the accretion histories -- those central regions with a larger fraction of
accreted stars are those  with contributions from more
massive satellites. The kinematical properties of in situ and accreted
stars are consistent  with the latter being supported by their
velocity dispersions, while the former exhibit clear signatures of
rotational support. Our simulations demonstrate a range of
characteristics, with some systems  exhibiting a co-existing bar and
spheroid in their central regions, resembling in some respect the
central region of the Milky Way.  

\end{abstract}


\section{Introduction}

 The formation of galaxies in a hierarchical universe involves the
continuous accretion of lower mass systems \citep{wr78}. As a
consequence, the so-called history of galaxy assembly is a complex
process,  with large variations at a given mass that also encode the
effects of their formation environments.
The Milky Way (MW) offers a unique perspective to test models of galaxy formation
because stars can be individually observed and their properties, such
as kinematics, dynamics, and chemical composition, can be more
accurately derived in the four main stellar components: bulge, thin and
thick discs and stellar haloes \citep[see][for a detailed discussion of
the MW components]{bhg2016}. The co-existence of these components in
the central region is now being explored in more detail as the result
of larger and more precise observational data in the bulge
\citep[e.g.,][]{gran2016,zoccali2016}, stellar halo
\citep[e.g.,][]{santucci2015,carollo2016,helmi2016} and the disc 
\citep[e.g.,][]{minniti2017}.

Bulges of galaxies, once thought to be simple systems supported by
velocity dispersion, have been reported to be more complex structures,
with bars, pseudo-bulges, classical bulges, etc., or even a combination
of them \citep[e.g.,][]{macarthur2009,sanchezb2011,morelli2012}. The
bulge of the MW is a clear example of this complex scenario
\citep{gonzalez2015}.
 Observational results suggest the co-existence of two stellar
populations: low-metallicity, $\alpha$-enhanced stars associated with an
spheroidal component and metal-rich, solar-level [$\alpha$/Fe] stars
related to a bar structure \citep[e.g.][]{babusiaux2010,gonzalez2013,
johnson2012,ness2012,zoccali2008,zoccali2016}. Recent observations have
reported the existence of very old stars traced by RR Lyrae populations
\citep{minniti2016}, as well as young stars  
\citep{dekany2015}, within the bulge region.


Contemporary studies show that the MW halo  also comprises
diverse stellar populations \citep{carollo2007,carollo2010}, with a
significant number of structures and overdensities related to satellite
accretion \citep[][]{hw1999,gilbert2013,ibata2013,deason2015,
refiorentin2015,grillmair2016, merritt2016}. 

\citet{carollo2007,carollo2010} and \citet{beers2012}  report that stars
in the diffuse inner and outer haloes exhibit different spatial
distributions, kinematics, dynamics, and chemical composition,
suggesting different formation mechanisms. According to these works, the
inner halo  population dominates the region within $15-20$ kpc, and
is  composed of stars with metallicities in the range $-2.0 <$
[Fe/H] $< -1.0$; it is -- on-average -- non-rotating with higher binding
energies. The outer halo  population is mostly present at r $>
15-20$ kpc,  with metallicities [Fe/H] $< -2.0$; it  also
exhibits a high net retrograde rotation ($\sim -80$ km/s) and
lower binding energies.  These properties have been confirmed by a
number of observational studies \citep{dejong2010,nissen2010,an2013,
hat2013,kaf2013,an2015,das2016}. Signatures of retrograde motion and
less-bound orbits for outer-halo stars have been recently detected in 
studies using data from the first release of the Gaia mission
\citep{helmi2016}, confirming previous claims. 

Still in the Local Group, large surveys conducted of the stellar halo of
M31 have revealed its properties with unprecedented accuracy.
\citet{gilbert2013} reports a significant decrease of metallicity as the
projected galactocentric distance increases, going from [Fe/H]$\sim
-$0.4 in the inner-most fields, at r $<$ 20 kpc, to [Fe/H] $\sim -$1.4
in the outer-most fields, at r $>$ 90 kpc. Outside the Local Group,
observation of six nearby massive disc galaxies with the Hubble Space
Telescope have revealed their stellar haloes with  sufficient
precision to estimate that, in half of them, there  exists a clear
negative color gradient, reflecting declining metallicity profiles
\citep{monachesi2016,harmsen2017}.

Different numerical approaches have been used to study the formation of
bulges. Cosmological simulations predict the action of different
channels of bulge formation via major collapse, gas inflows, mergers, and
secular evolution \citep[e.g.,][]{tissera1998, governato2009,obreja2013,
okamoto2013}. These simulations describe the formation of structure
in a cosmological context, including the physics of baryons by applying
subgrid modelling. 
Simulations of individual isolated galaxies are
valuable numerical tools. These models consider alternative processes to
explain the formation of the MW bulge from the thin/thick disc, for
example \citep[e.g.,][]{dimatteo2015,debattista2016}.

Numerical simulations of MW-like galaxies in a cosmological context show
that their stellar haloes formed with  significant contributions from
accreted stars. In particular, the outer regions of the haloes are
dominated by this process \citep[e.g.,][]{bullock2005,helmi2011,
tissera2013,cooper2015}.  However, the inner regions are found to be
assembled from a variety of stellar populations formed by different
processes, such as in situ star formation from gas accreted into the
central regions by the merging of more massive, gas-rich satellite
galaxies, or in situ star formation from in-falling gas in the main
galaxies \citep[e.g.,][]{zolotov2009,tissera2012,mcc2012,obreja2013,
tissera2014}.  While in global terms there is agreement between
numerical results coming out of the simulations, there are differences
between them when they are examined in detail and compared to observations
\citep{harmsen2017}. Therefore, progress remains to be made before
a robust description of the stellar haloes is achieved, since
simulations still resort to subgrid physics and approximations in order to
model the relevant physics \citep[e.g.,][]{tissera2013,cooper2015,
pillepich2015}.

In this paper, we focus  our analysis on the central regions of a
set of MW mass-sized haloes from the Aquarius project \citep{scan09}.
Special attention is  given to old stars which are thought to be
formed in the first galaxies. This is of particular interest, given the
 recent detections of old RR Lyrae stars in the bulge-halo
\citep{minniti2016} and the disc-halo \citep{minniti2017} regions by the
VVV survey. Furthermore, \citet{santucci2015} and \citet{carollo2016}
identified the so-called ``ancient chronographic sphere (ACS)'' in the
inner region of the MW's stellar halo, which is  claimed to be 
populated by stars older than 10 Gyr. The ACS  is distinguished from the
slightly younger  outer-halo region, which includes individual structures as
young as 9-9.5 Gyr \citep{carollo2016}.

To carry out our study, we employ a subset of four simulated haloes
 that have central regions with different characteristics.  We
stress that these simulations are not expected to reproduce the MW in
detail. They are used to learn about the process of  galaxy formation and the
imprints that they might have left on the properties of galaxies at
$z=0$. Our main focus is to understand to what  extent observations can
be reproduced, and make predictions that might be tested with on-going
or future surveys.  We note that the simulated haloes of the
Aquarius runs have approximately a factor of two more stellar mass than
that expected for the MW \citep{bhg2016}. Nevertheless, there are
several properties that can be reproduced by these haloes. From these
similarities, and principally, from the differences, we expect to
set contraints on the subgrid physics, and to learn about the assembly of
stellar haloes \citep{tissera2012,pillepich2015,cooper2015}.

This paper is organised as follows. In Section 2, the main
characteristics of the analysed simulations, the code used to run them,
and the analysed haloes are described. In Section 3 we  analyse the
properties of the in situ and accreted stars in the central 10 kpc
region, and focus on the properties of the old stellar populations and
their formation mechanisms. In Section 4, the velocity distributions of
the stars in the central regions are  considered.  Finally, in
Section 5, we summarise our results.

\begin{figure*}
\hspace*{0.1cm}\resizebox{4.cm}{!}{\includegraphics{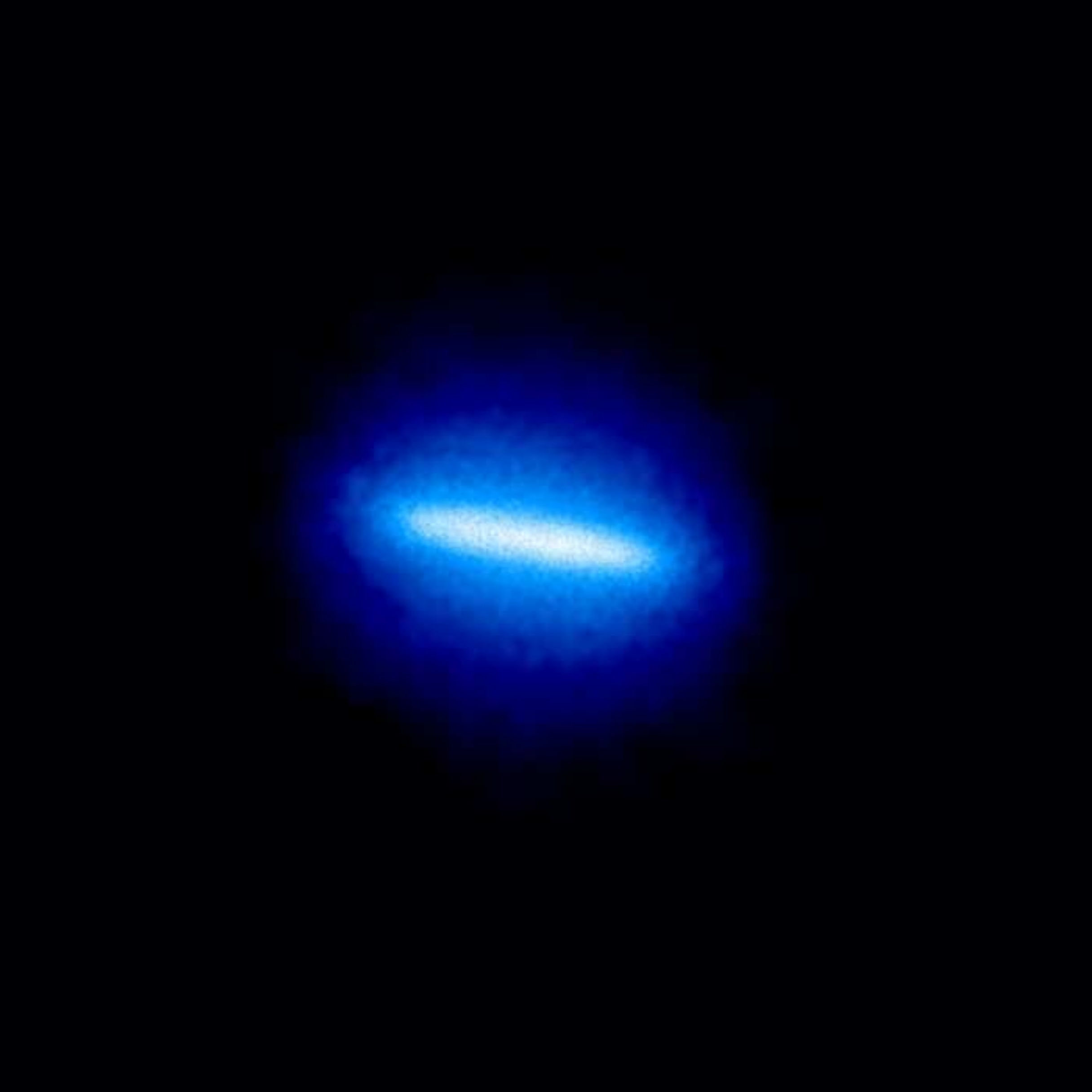}}
\hspace*{0.1cm}\resizebox{4.cm}{!}{\includegraphics{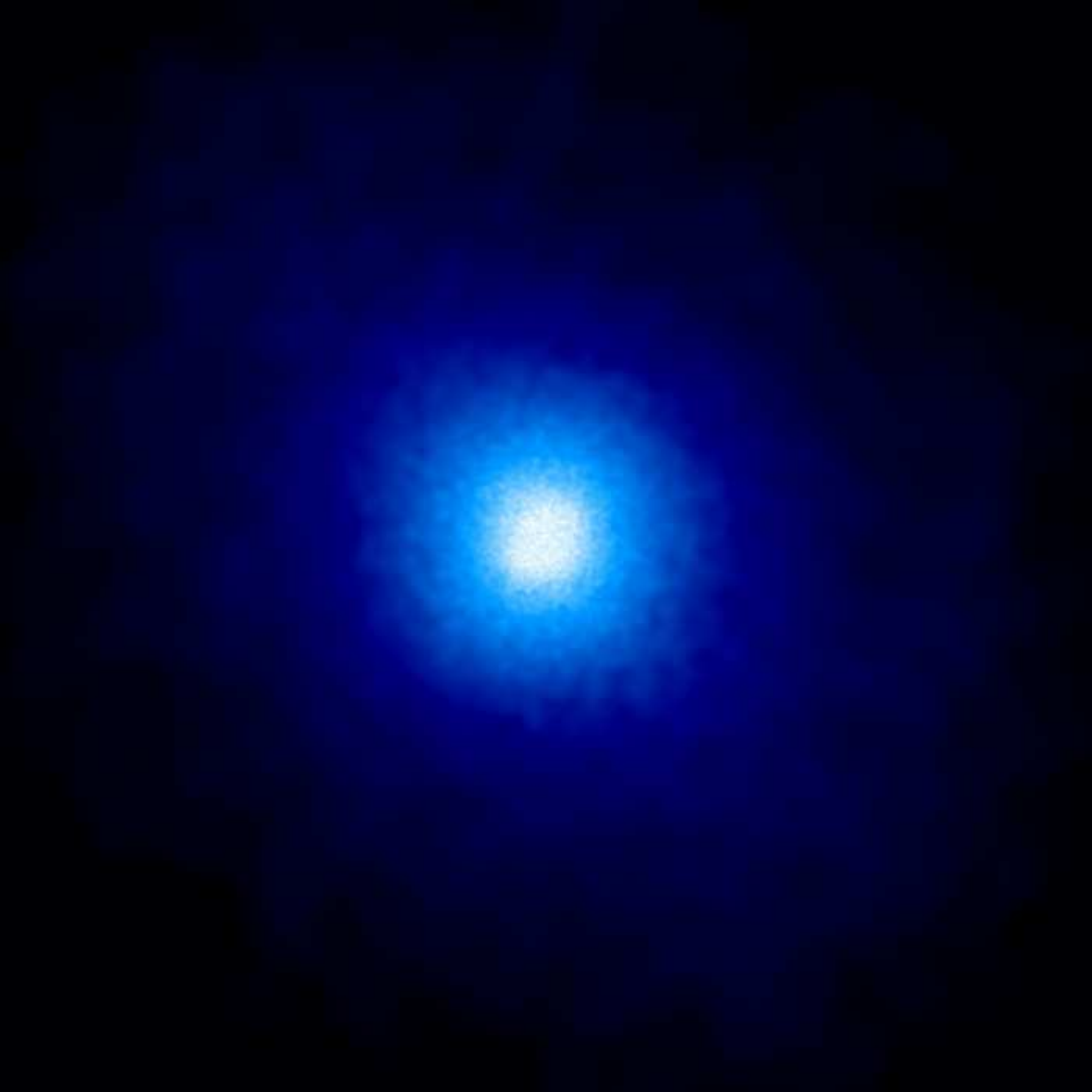}}
\hspace*{0.1cm}\resizebox{4.cm}{!}{\includegraphics{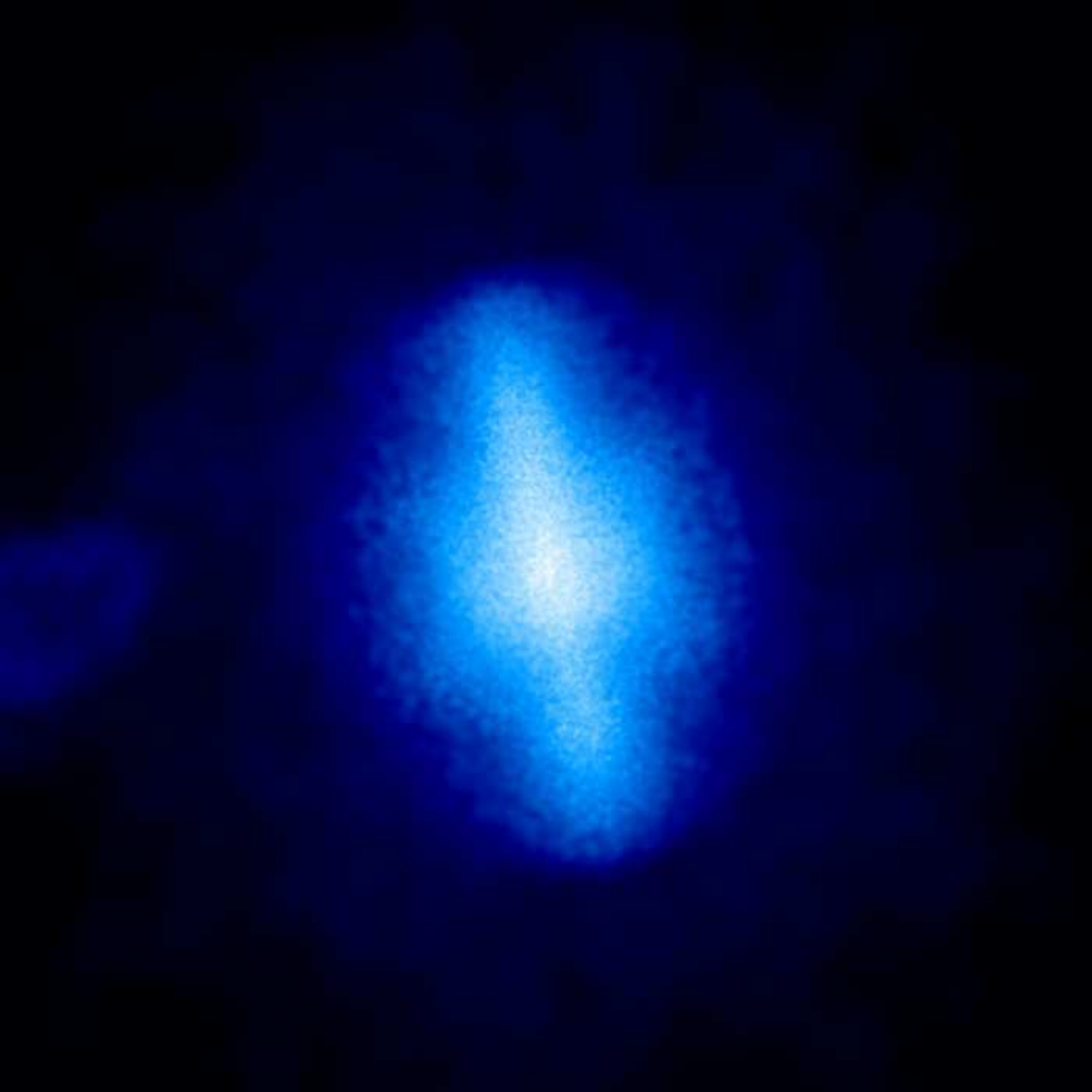}}
\hspace*{0.1cm}\resizebox{4.cm}{!}{\includegraphics{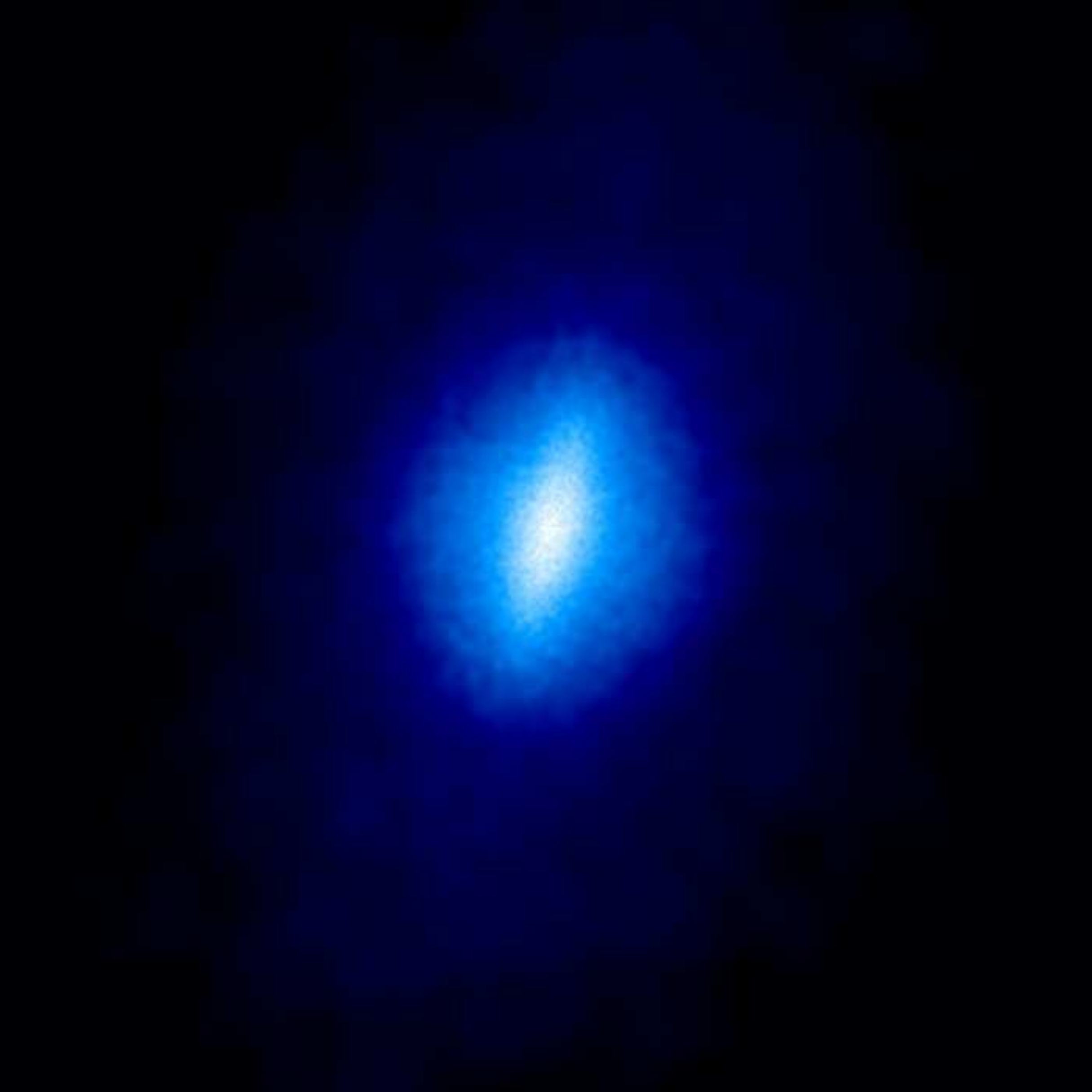}}\\
\hspace*{0.1cm}\resizebox{4.cm}{!}{\includegraphics{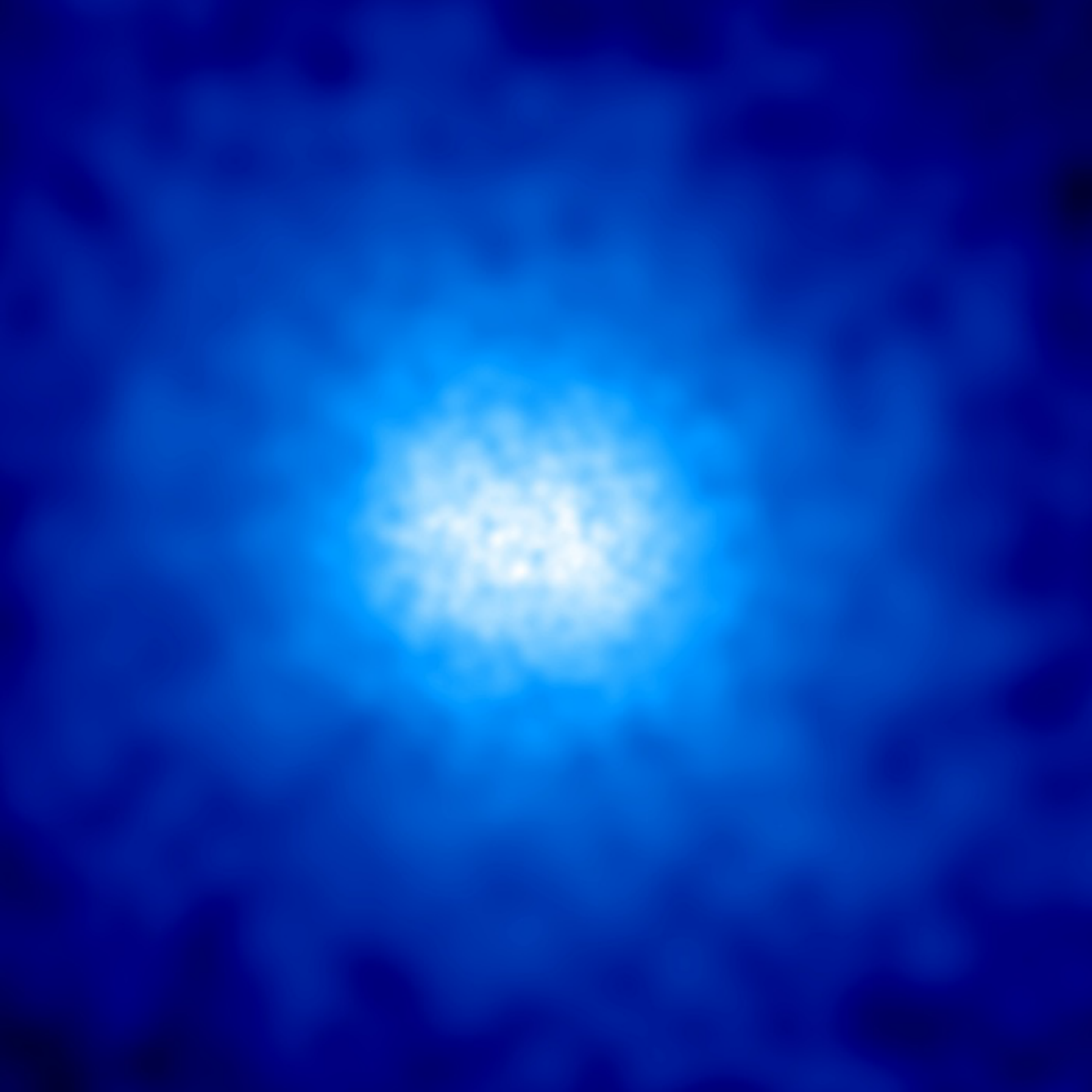}}
\hspace*{0.1cm}\resizebox{4.cm}{!}{\includegraphics{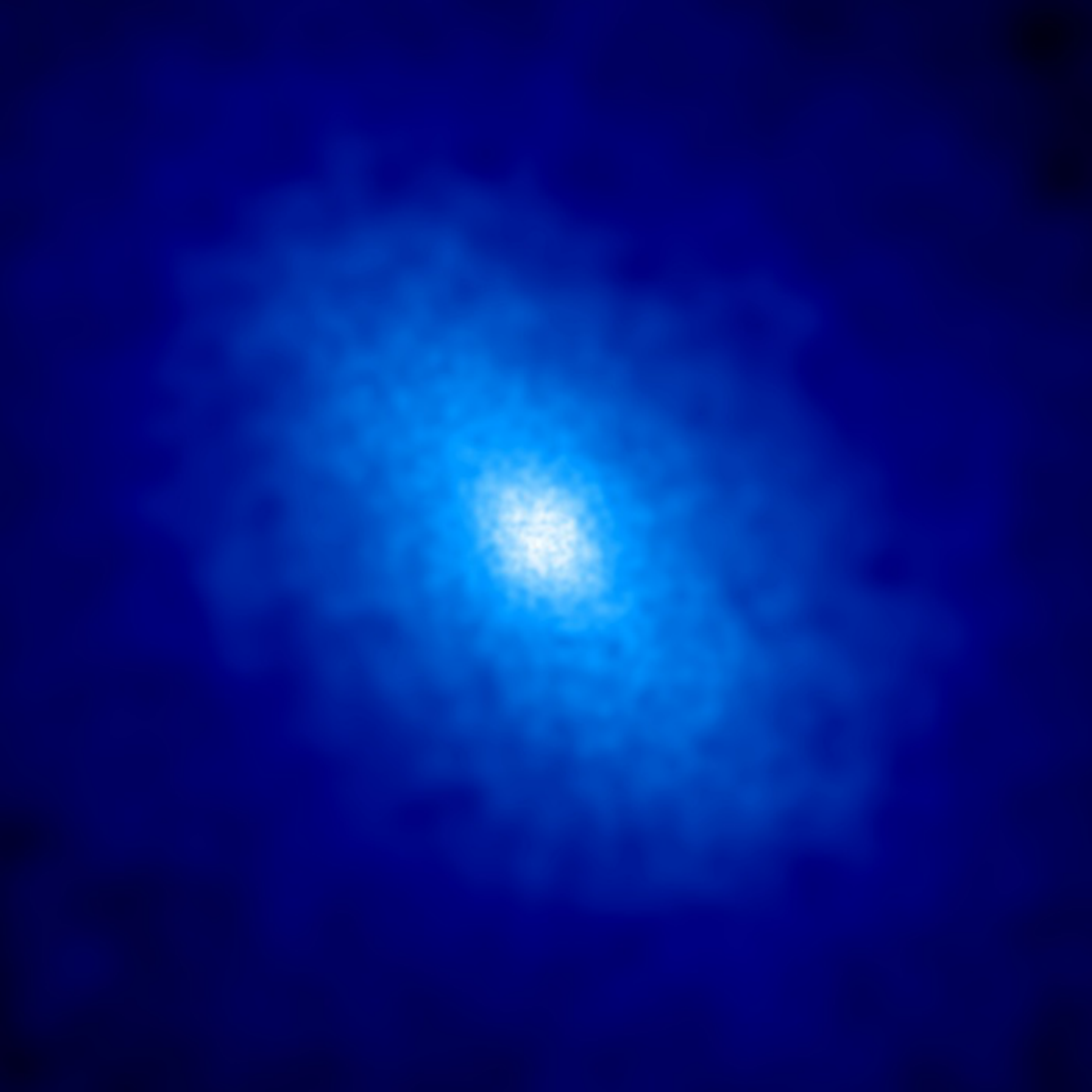}}
\hspace*{0.1cm}\resizebox{4.cm}{!}{\includegraphics{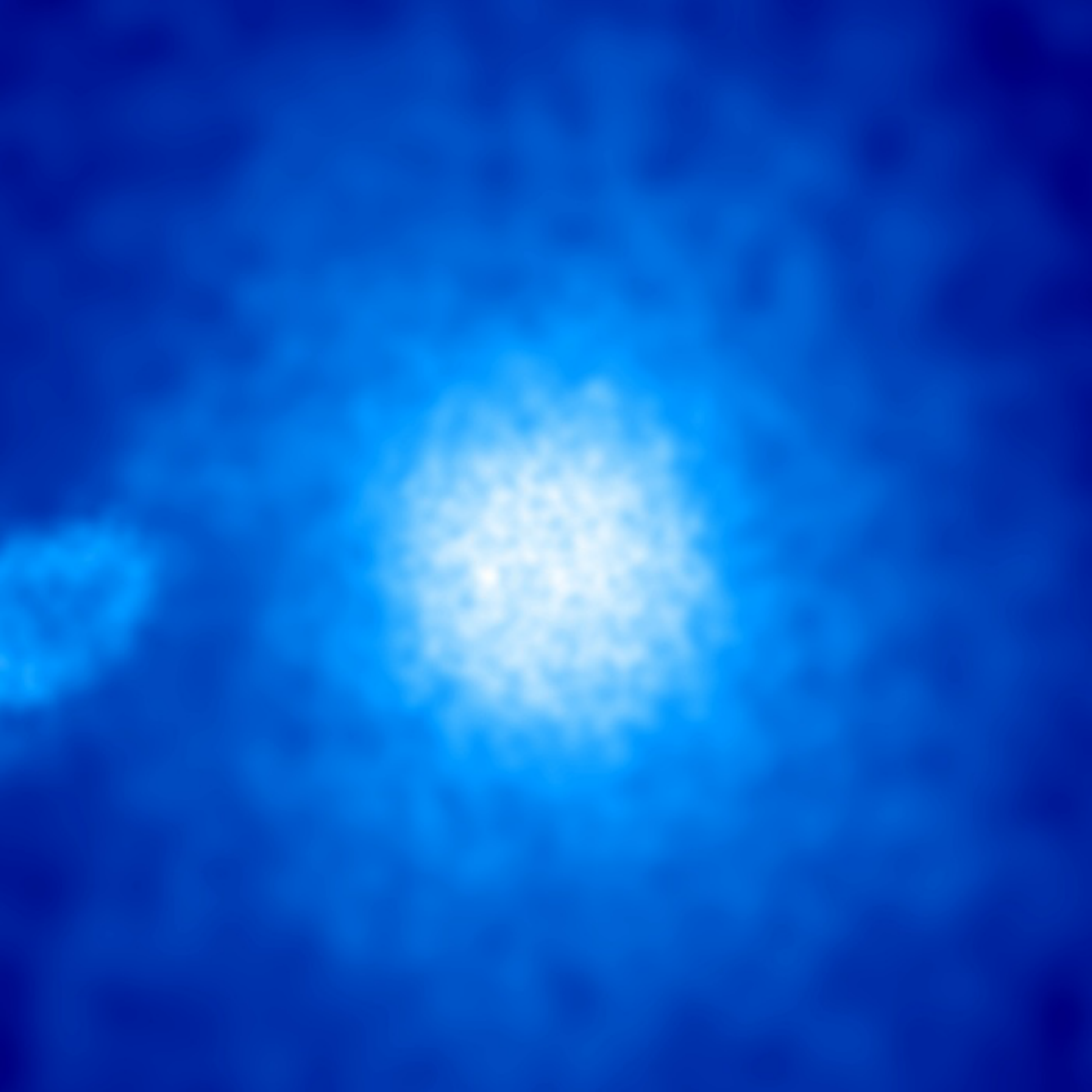}}
\hspace*{0.1cm}\resizebox{4.cm}{!}{\includegraphics{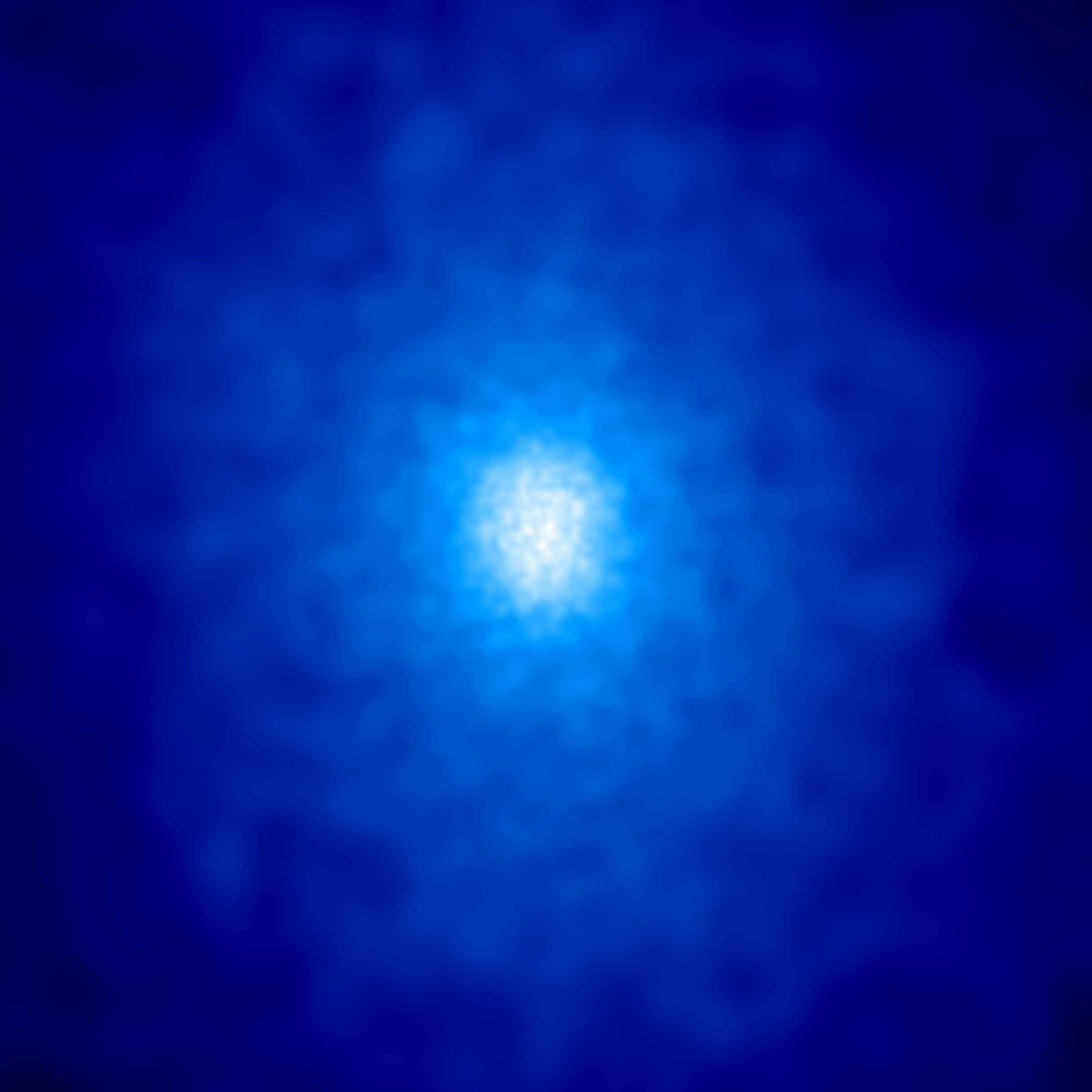}}
\caption{Projected density  maps of stars of the inner 10 kpc
    regions formed   in situ (upper panels) and in satellites which
  were later  accreted (lower panels) in the central regions of the
  four analysed galaxies: Aq-A, Aq-B, Aq-C and Aq-D (from left to
  right panels)  at $z=0$. 
}
\label{mapas}
\end{figure*}

\section{Numerical Experiments}

In this paper, we analyse a galaxy subset from the Aquarius Project
performed with a version of P-Gadget 3 \citep{scan09}. This code
includes chemical evolution and supernova (SN) feedback. The Aquarius
haloes have  virial masses around $10^{12} {\rm M_{\odot}}$, and were
selected from the Millenium-II Simulation. The initial conditions are
consistent with $\Lambda$CDM cosmogony with the following adopted
cosmological parameters: $\Omega_{\rm m}=0.25,\Omega_{\Lambda}=0.75,
\Omega_{\rm b}=0.04,
\sigma_{8}=0.9, n_{s}=1$ and $H_0 = 100 \ h \ { \rm km~s^{-1}
  Mpc^{-1}}$ with $h =0.73$. The analysed haloes correspond  to
level 5 resolution. The dark matter particle masses are of the order of
$10^{6}{\rm M_\odot }h^{-1}$, and the initial gas particle masses 
are about $2 \times 10^{5}{\rm M_\odot} h^{-1}$. The maximum physical
gravitational softenings are in the ragen $0.5-1$ kpc$~h^{-1}$. The
 selected sample has a variety of central regions (e.g., two galaxies with
central bars are included), and we analysed the properties of their old
stellar populations in detail, including their  spatial
distributions.  The
Aq-C-5 simulation we include in our analysis is often taken as a reference halo
\citep{aumer2013}.

As mentioned above, our version of {\small GADGET-3}
\citep{springel2005} considers updated treatments for chemical
enrichment, stochastic star formation (SF), and metal-dependent
radiative cooling. It also includes a multi-phase model for the
interstellar medium (ISM), and the SN feedback prescription from
\citet{scan05,scan06}.  Such SN feedback successfully triggers
galactic mass-loaded winds without the need to introduce mass-scale
parameters, so that it self-regulates according to the potential well of
the systems. The modeling of galactic winds contributes to the alleviation
of the angular momentum loss problem that produces discs that are too
small in the numerical simulations \citep{navarro1991, navarro2000}. As
SN feedback models have become more sophisticated, the resulting simulated galaxies
are more comparable to observations \citep[e.g.][]{scan08, dalla2008,
brook2012}.

Here we also summarise the main aspects. The reader is refereed to
\citet{scan08} for a detailed explanation of the SN feedback model. The
adopted ISM multiphase model permits the  co-existence of dense and
diffuse gas phases \citep{scan06}. Each gas particle defines its own hot
and cold phases, defined by applying local entropy criteria. This allows
gas particles to hydro-dynamically decouple from the low entropy ones
(in the case that they are not in a shock front). The energy for each
cold gas particle  is stored in a reservoir until it fulfils the
required conditions to be promoted  into the hot phase. The chemical
model adopts the nucleosynthesis productions of \citet{WW95} and
\citet{thie1993} to describe the enrichment by Type II (SNII) and Type
Ia (SNIa), respectively. The stars with masses larger than 8 M$_{\odot}$
are assumed to end as SNII, and a value of 0.0015  SNIa per year 
 is
adopted for the rate of SNIa, as discussed in
 \citet{mosconi2001} and \citet{jimenez2015}.We assume a Salpeter Initial Mass
Function in order to compute the number of stars of a given mass. The
chemical model assumes that the ejection of enriched material and energy
by SNIa has a time delay randomly selected in the range $[0.1,1]$ Gyr
\citep[see ][for a discussion on this point]{jimenez2015}, and that the 
SNII explode according to the lifetimes of \citet{rait1996}. Each SN
event releases $0.7 \times 10^{51 } $erg s$^{-1}$,  which is distributed
into the hot and cold gas surrounding each stellar particle. The energy
released by each SN event is injected into the cold gaseous phase as a
fraction $\epsilon_{c} =0.5$, with the rest being pumped into the hot
medium. Chemical elements are distributed into both the cold and hot gas
phases using the same $\epsilon_{c}$.

\begin{figure*}
\resizebox{4.3cm}{!}{\includegraphics{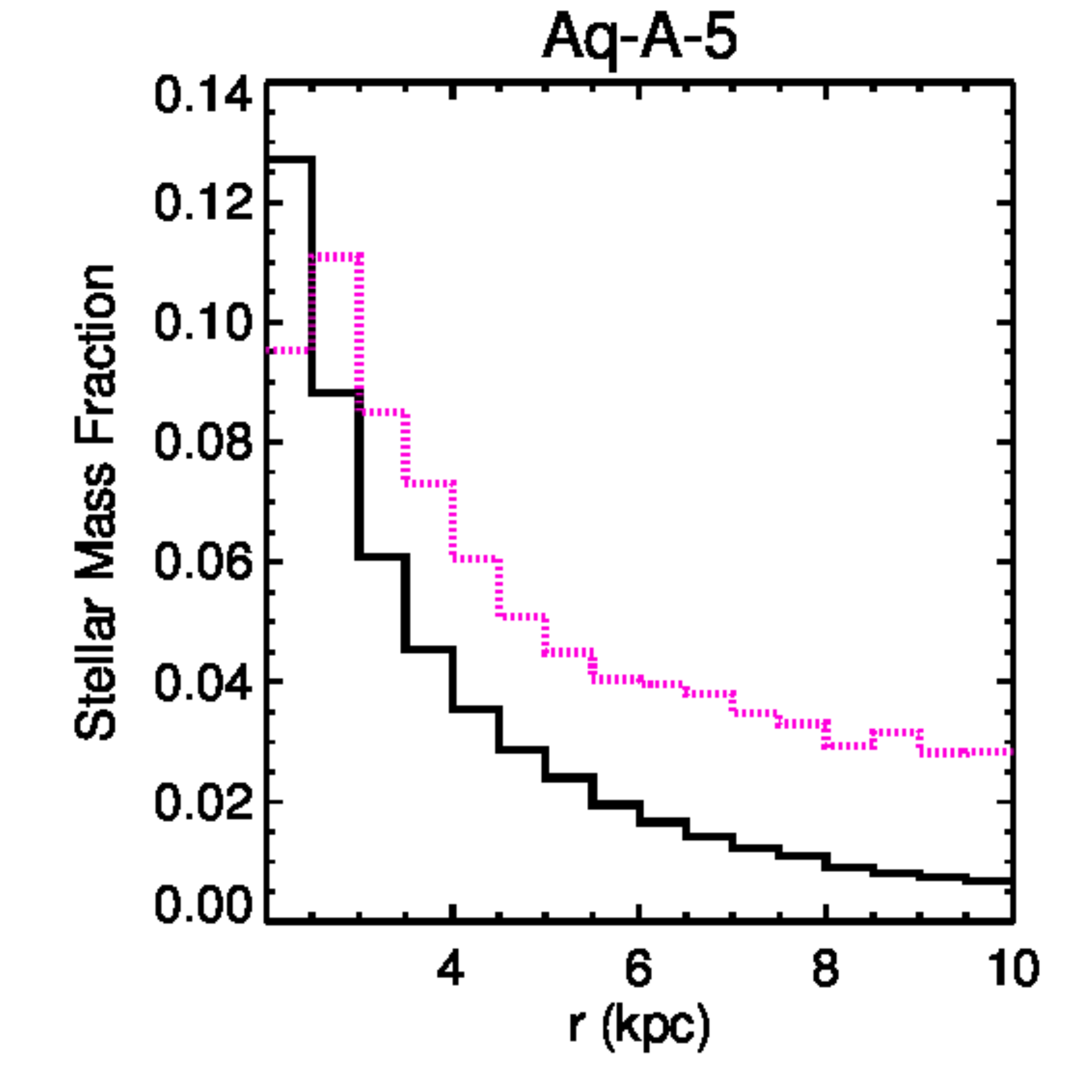}}
\resizebox{4.3cm}{!}{\includegraphics{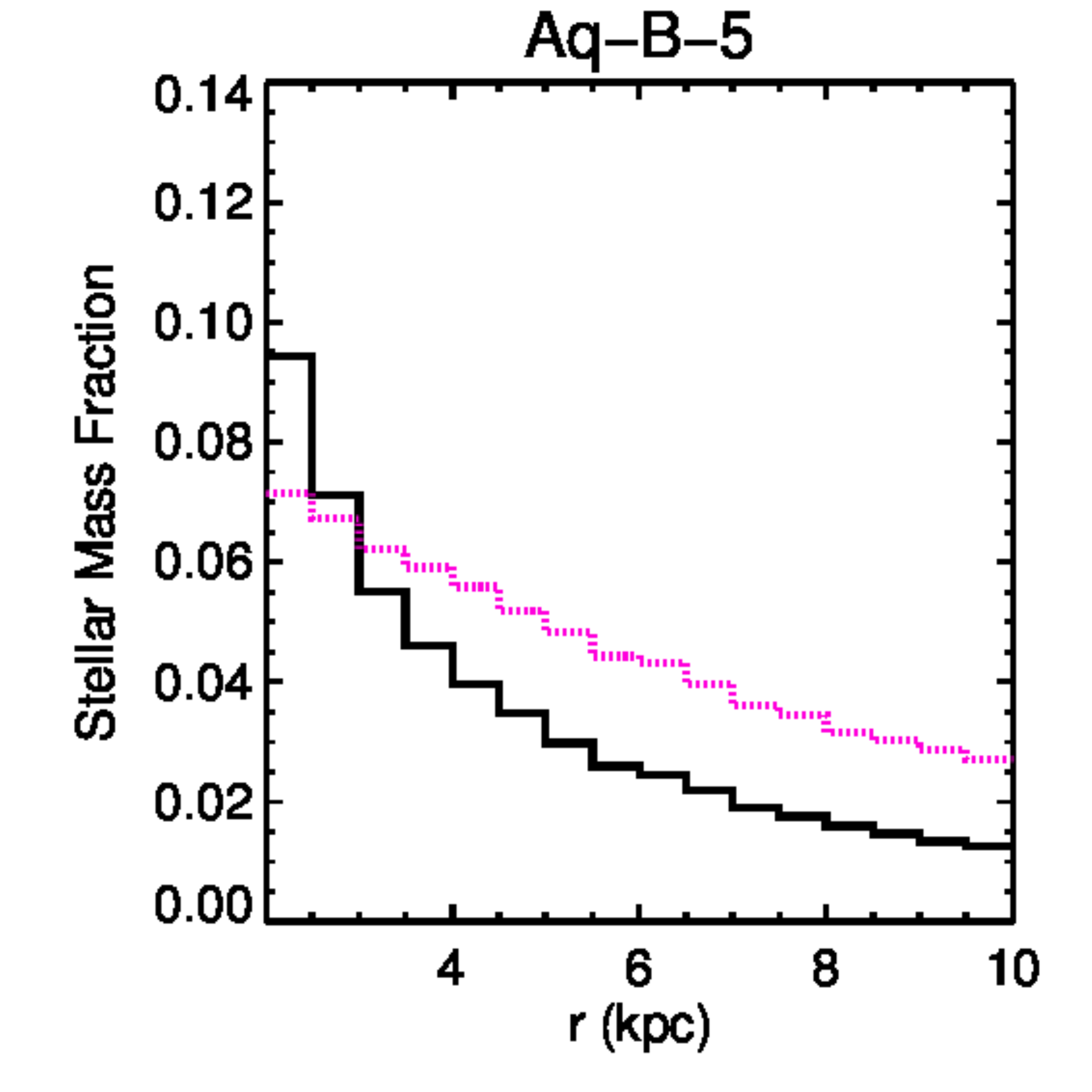}}
\resizebox{4.3cm}{!}{\includegraphics{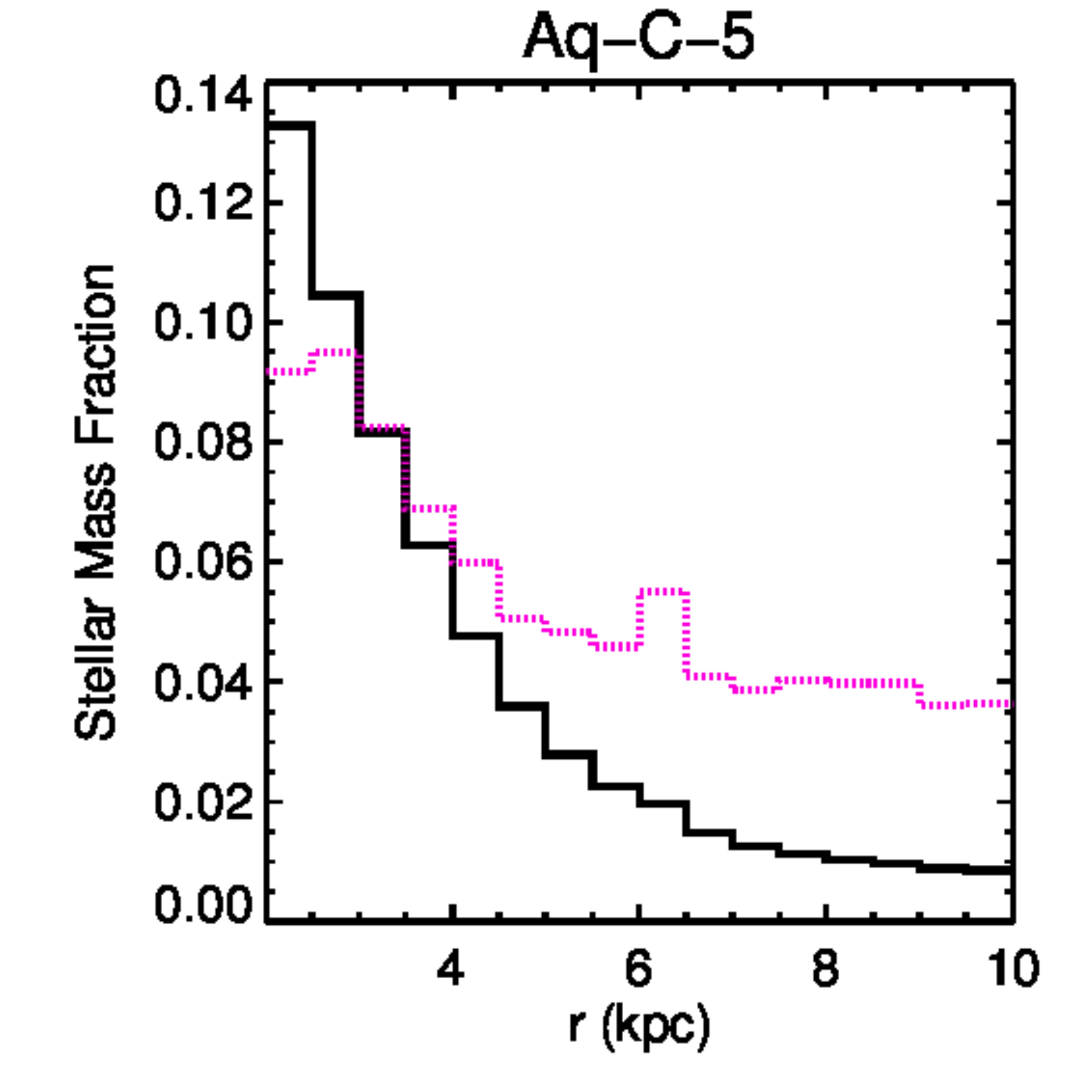}}
\resizebox{4.3cm}{!}{\includegraphics{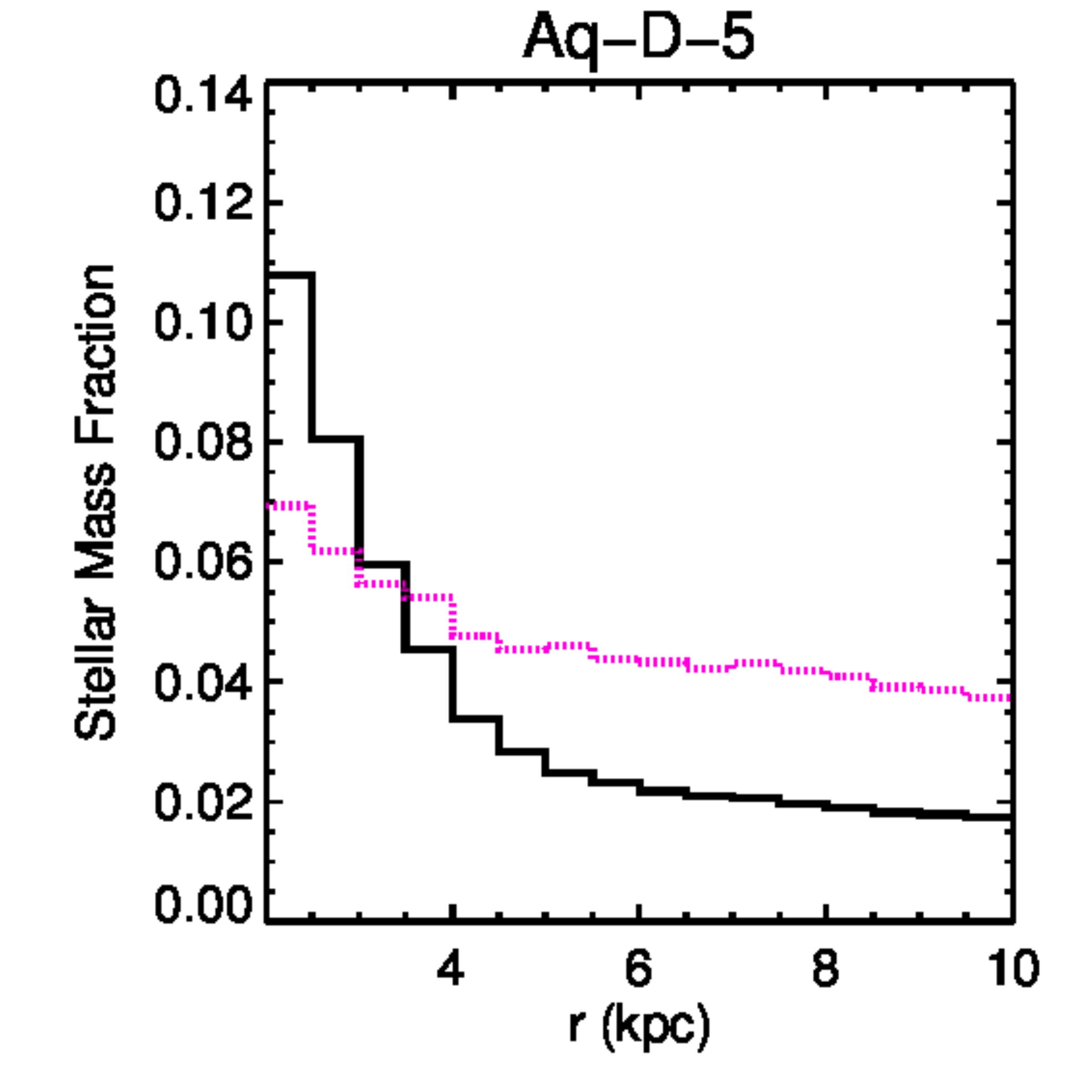}}\\
\resizebox{4.3cm}{!}{\includegraphics{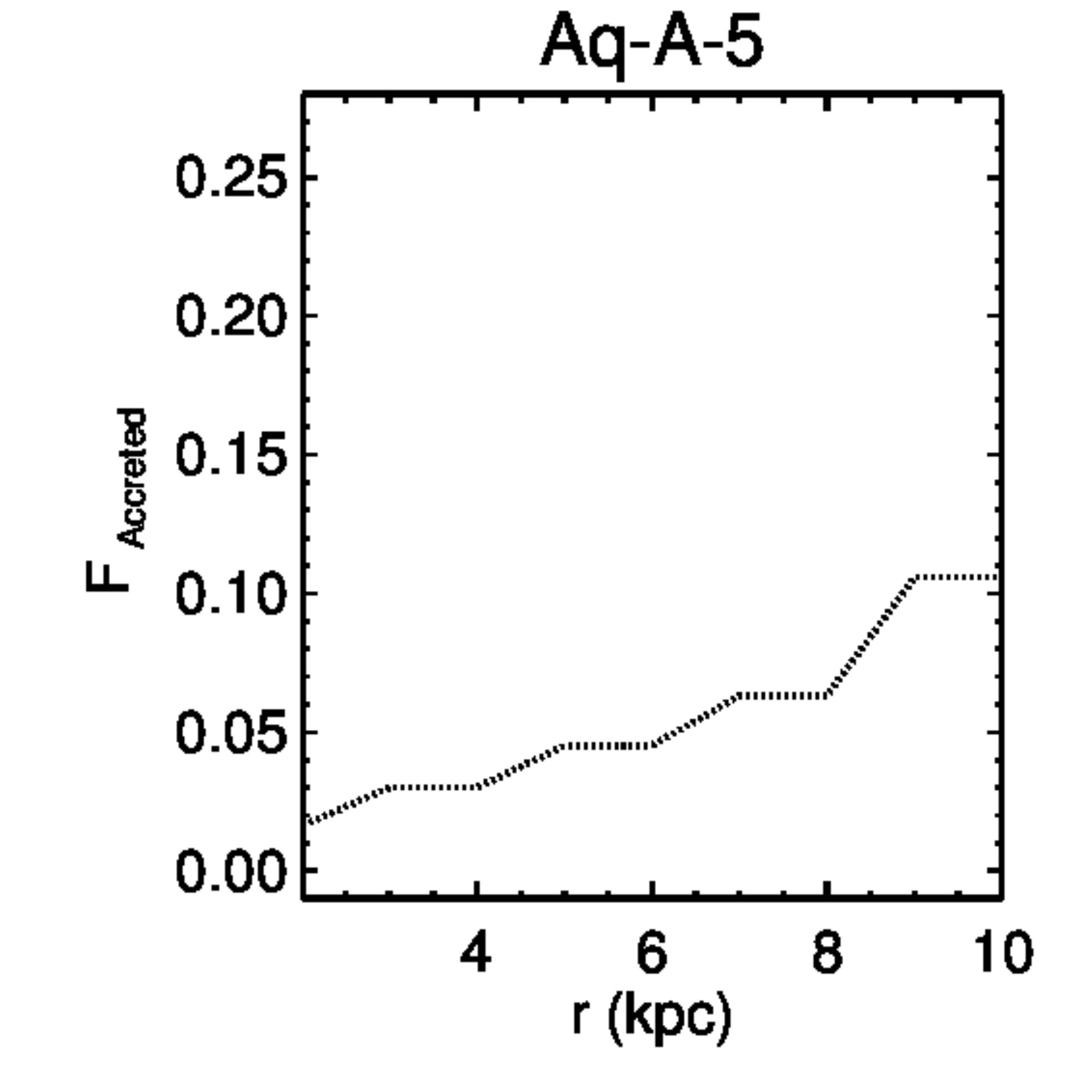}}
\resizebox{4.3cm}{!}{\includegraphics{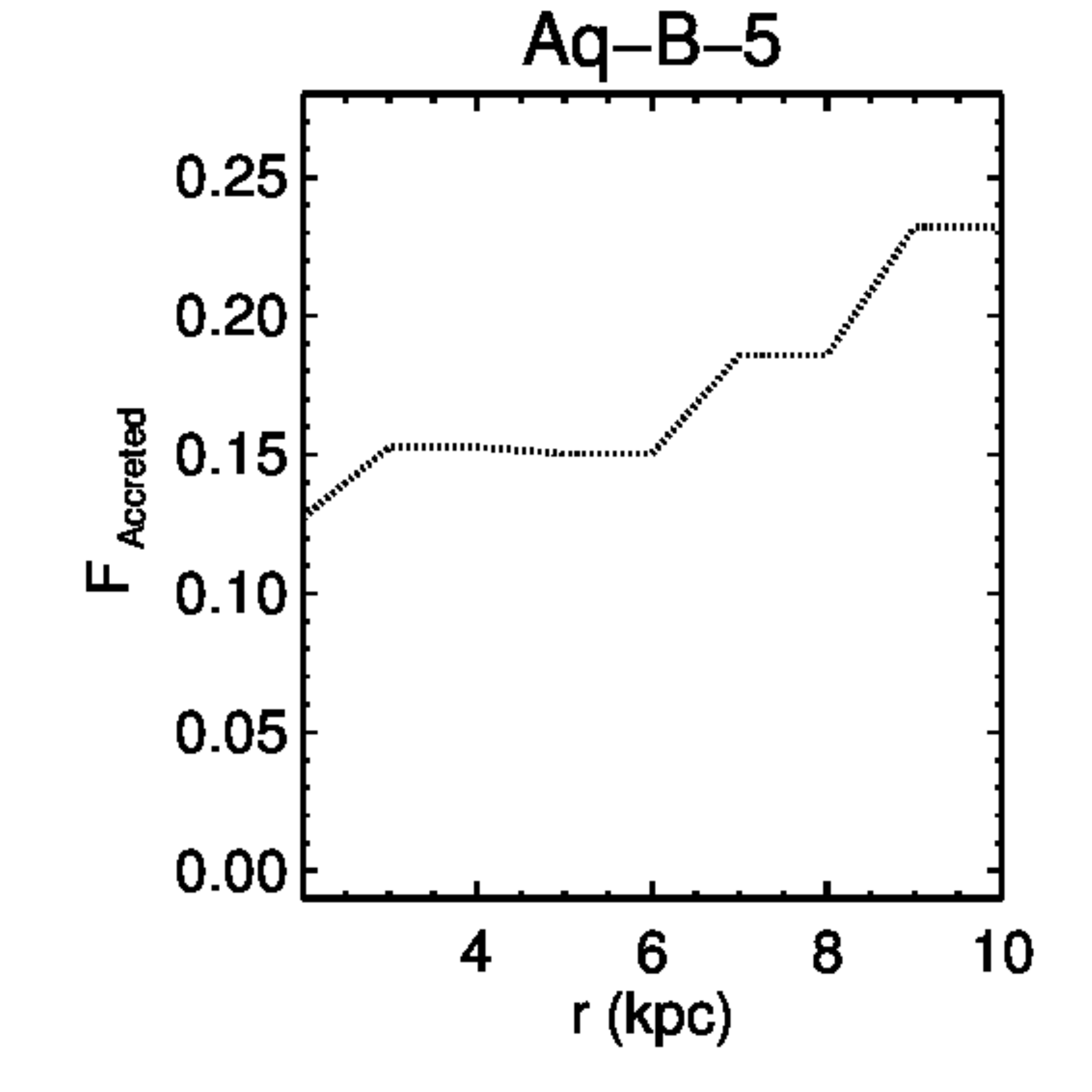}}
\resizebox{4.3cm}{!}{\includegraphics{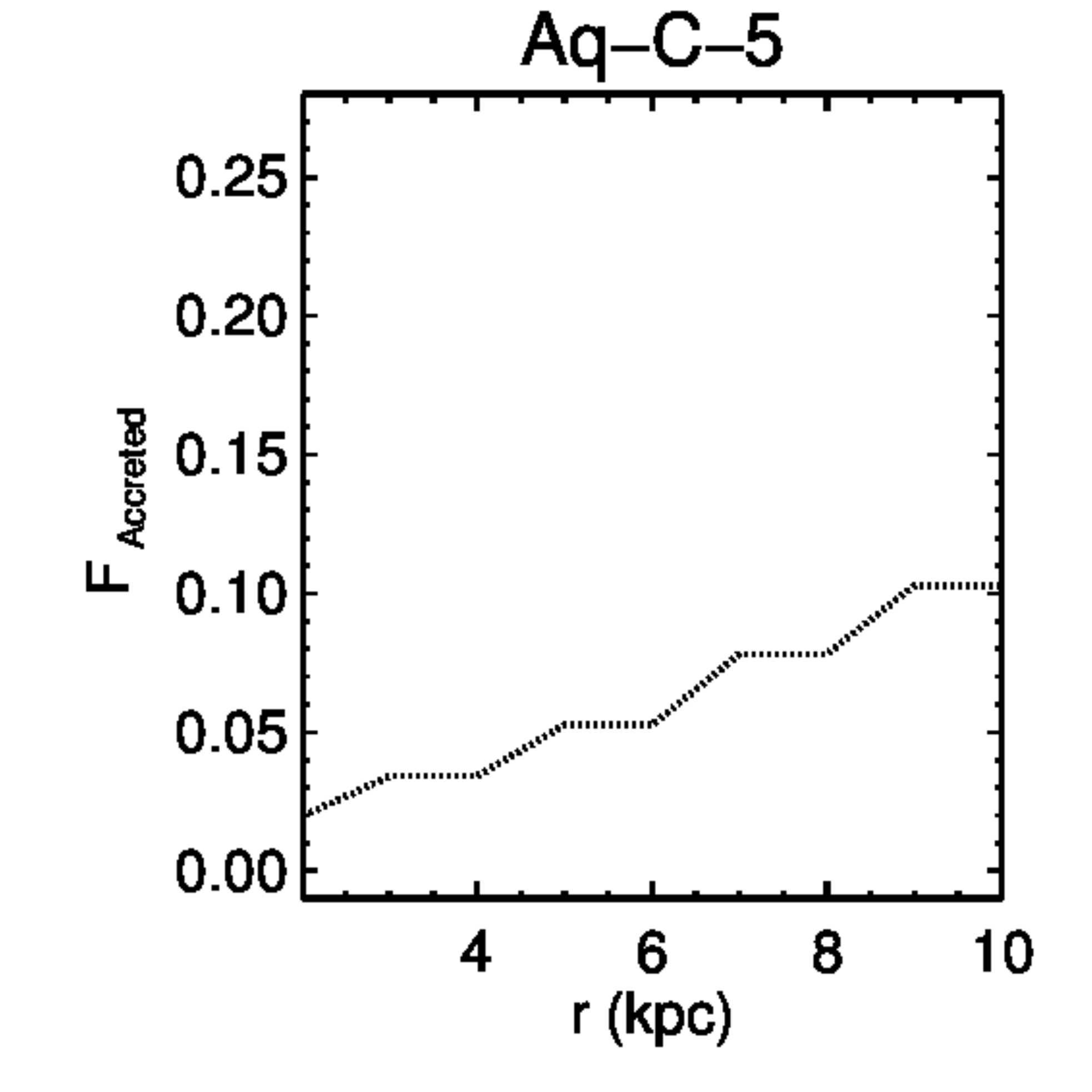}}
\resizebox{4.3cm}{!}{\includegraphics{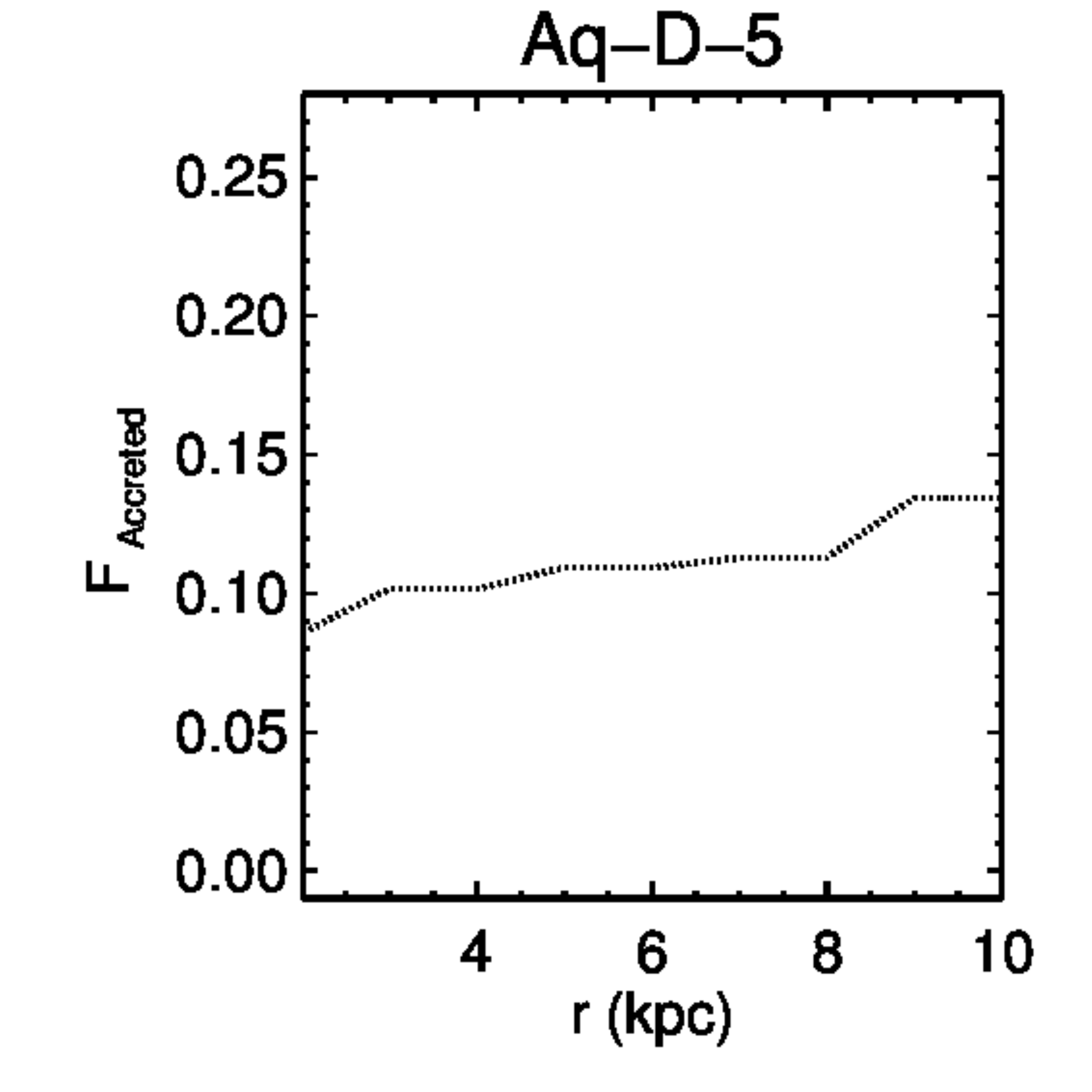}}
\caption{Upper panels: Mass fractions of all stars (i.e., accreted and in situ stars; black lines) 
and accreted stars alone (magenta lines) as a function of galactocentric distance in the four analysed haloes
(each subsample has been normalised to its total mass). Lower panels: The fraction of accreted stars with respect to the total mass of
the spheroidal component (i.e., bulge and stellar halo) as a function of
galactocentric distance. The frequency of accreted stars increases with
increasing galactocentric distance, as expected \citep{tissera2013,tissera2014}. } 
\label{mdfaccreted}
\end{figure*}

\subsection{The Simulated Galaxies}

We study in detail the main galaxies in virialized haloes identified by
using a Friends-of-Friend algorithm and {\small SUBFIND}
\citep{springel2001}. A main galaxy is defined as the more massive
system within a virial potential well at $z=0$. 
SUBFIND program \citep{springel2001}. The physical properties for main
galaxies are calculated within the galaxy radius, defined  to
enclose $\sim 80$ per cent of the baryonic mass of a galaxy. A
detailed discussion on the effects of numerical resolution on the
dynamics and chemical properties can be found in \citet{scan09} and
\citet{tissera2012}, respectively.

We use the dynamical decomposition performed by \citet{tissera2012}.
There are then three dynamical components for each system: bulge, disc
and stellar halo. For this purpose, we estimate the parameter $\epsilon$
of the star particles, defined as $\epsilon = j_{z} /j_{z,max}(E)$, where
$j_{z}$ is the angular momentum component in the direction of the total
angular momentum, and $j_{z,max}(E)$ is the maximum $J_{z}$ over all
particles of  a given binding energy, $E$. A particle on a perfect prograde
circular orbit in the disc plane has $\epsilon=1$. We consider particles
with $\epsilon$ greater than 0.65 to be part of a disc. 
The particles which do not satisfy this requirement are taken to belong
to the spheroidal component. By inspecting the $\epsilon - E$ plane, we
checked if the adopted limits are suitable to individualize the
rotational supported components. We also distinguish between the bulge
and the stellar halo components. The bulge  comprises the most
 gravitationally-bound particles, while the  less-bound
particles are considered to form the stellar halo.  For consistency
with our previous work,
we adopt the criteria defined by \citet[][]{tissera2012} and 
also used by \citet{tissera2013,
tissera2014,tissera2014b} to analyse the stellar haloes (see also
\citet{scan09} and \citet{scan10}). According
to these criteria, the bulges are defined by those particles that have
binding energy lower than the lowest found at the half stellar-mass
radius of a galaxy.  While the stellar haloes are comprises of particles with higher
binding energies, the disc particles are taken to have $\epsilon >0.65$
and located within the 1.5 times the galaxy radius. 

In this work we analyse all stars  supported by their velocity dispersion:
the bulge and stellar haloes within the inner 10 kpc region. This
region is defined by considering that the effective radius, $r_{\rm b}$, of
the analysed bulges are $\sim 3-4 $ kpc, as reported by \citet{scan10}.
In particular, we adopt those obtained from the dynamical
decomposition, since they are consistent with our analysis. The defined 10 kpc regions
represent about three times the $r_{\rm b}$, and ensure that we include the
bulge and the transition bulge-halo region. A single cut-off radius is
adopted for the sake of simplicity. These authors also detected bar
structures in some of the galaxies of the Aquarius run. In particular,
Aq-A and Aq-C have Bar/T ratios of $\sim 0.23$. 

The analysis carried out by \citet{tissera2012},  which classifies stellar
populations according to their origin as in situ or accreted stars, is
applied to the simulated galaxies. Accreted stars are those stars that
formed in separate galactic systems (i.e., outside the virial radius of
the main progenitor) and were accreted later on by the main progenitor
halo. The in situ stars are those formed within the virial radius of the
progenitor galaxies from gas which was directly accreted by the main
progenitor as gas inflows or was contributed by gas-rich mergers.  From
\citet{tissera2014}, we know that the stellar haloes  of the
  Aquarius haloes are dominated by
accreted stars, and that, in the inner stellar haloes, the fraction of
contributed in situ stars increases. These authors showed that the relative
contributions of both components depends strongly on the history of
formation of the halo and the main galaxy, while the metallicity of the
stellar haloes beyond $\sim 20$ kpc is determined by the characteristics
of the  accreted galaxy satellites.

\section{The inner 10 kpc  spheroid}

To analyse the stellar populations within 10 kpc supported by
velocity dispersion, the stellar populations are classified as in situ or
accreted subsamples according to the criteria described above. Figure
\ref{mapas} shows the projected density distributions of stars formed in
situ (upper panels) and in accreted satellites (lower panels). All
galaxies have been rotated so that their reference system is oriented
along the main axis of rotation with respect to the disc plane. As can
be seen from this figure, two systems exhibit clear bar structures.
\citet{scantun2012} analysed the bars and found that  they have
lengths and strengths, stellar density profiles, and velocity fields in
good agreement with observations \citep{gadotti2011} and dynamical
idealised simulations \citep{athana2002}. In a previous paper,
\citet{scan10} identified bars in Aq-A and Aq-C, as mentioned in Section
2. In the case of Aq-D there is also a very weak elongated distribution
\citep[see also][]{scan11},  although it was not classified as a bar
structure by \citet{scan10}. As can be seen from this figure, the bar
structures are clearly present in the in situ stellar populations.
However, the accreted stars exhibit more spherical distributions. In the
case of Aq-B, the distribution of in situ stars is more concentrated and
spherical, while the accreted stars show a more elongated distribution. 


In order to  provide insight on the distribution of stars in the central
regions of our simulated galaxies, the stellar mass fraction as a function of galactocentric
distances for  each of the four analysed haloes, and the corresponding
calculations considering the accreted stars, are shown in
Fig.~\ref{mdfaccreted}  (each subsample has been normalised to its
total stellar mass within the defined central region; upper panels). 
 The lower panels show the fraction of accreted
stars with respect to the total stellar mass at a given galactocentric
distance. In all analysed haloes, the frequency of accreted stars
increases with increasing galactrocentric distance, as expected
\citep{tissera2013}. For the analysed haloes, we found that $\sim 10-25$
per cent of the stellar mass  at $\sim 10$ kpc corresponds to
accreted stars (Table~\ref{tab1}).  These relative fractions are in
global agreement with other previous work, such as
\citet{pillepich2015}. We note that these authors included the disc
components in their estimations (their figure 4), and that their definitions of
bulge, inner, and outer haloes are based on spatial separations. Hence
the comparison can be made only on a generic basis. 

\begin{figure*}
\resizebox{4.3cm}{!}{\includegraphics{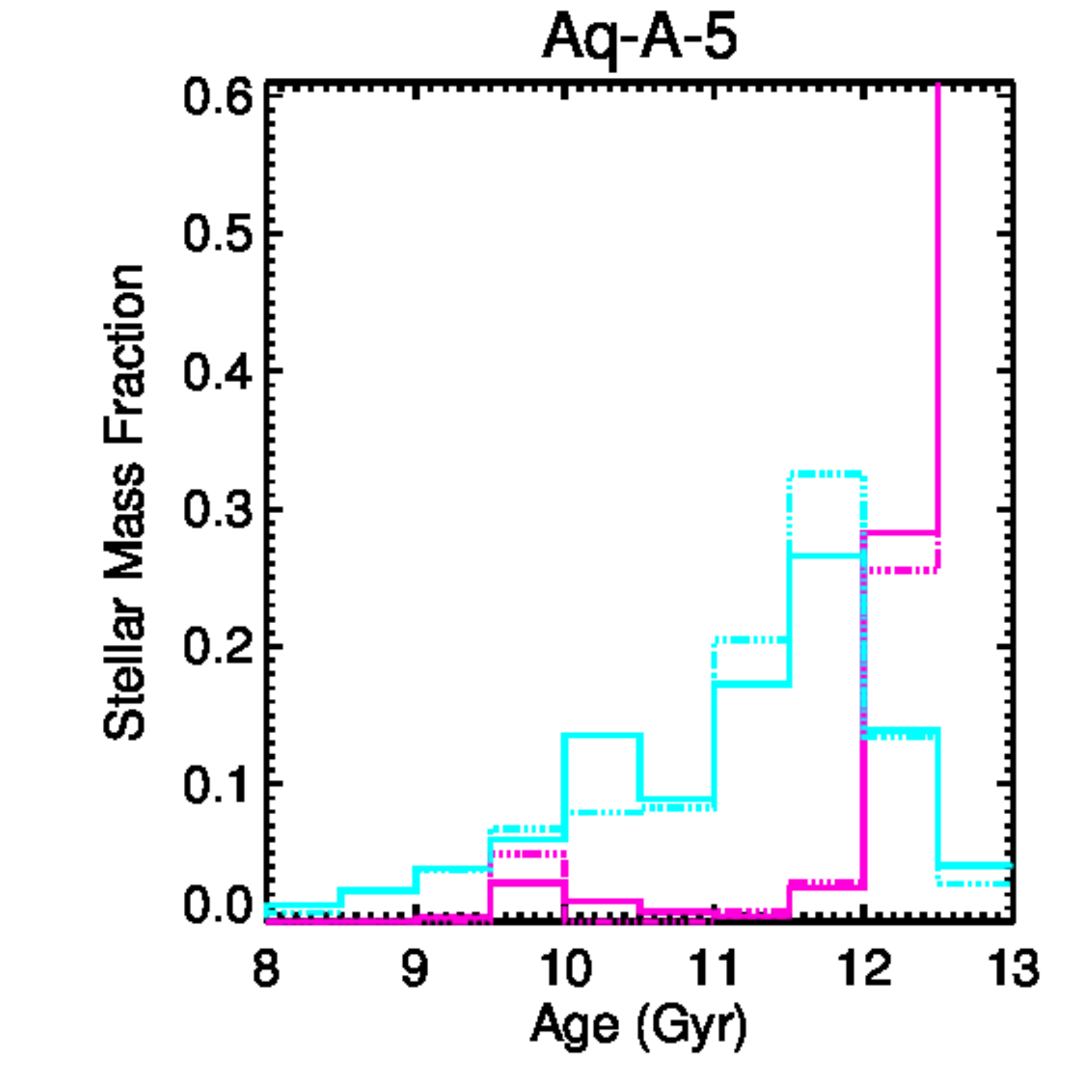}}
\resizebox{4.3cm}{!}{\includegraphics{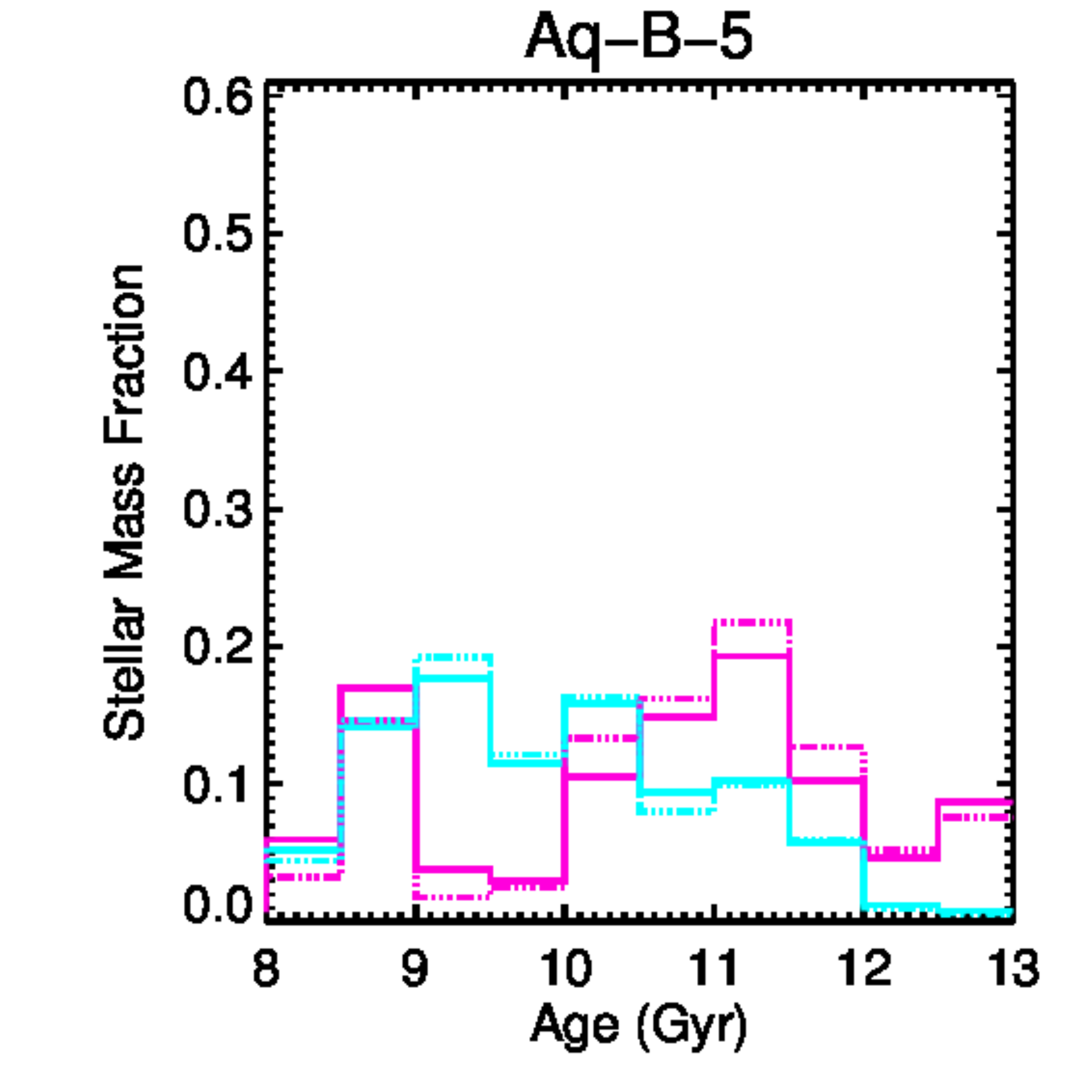}}
\resizebox{4.3cm}{!}{\includegraphics{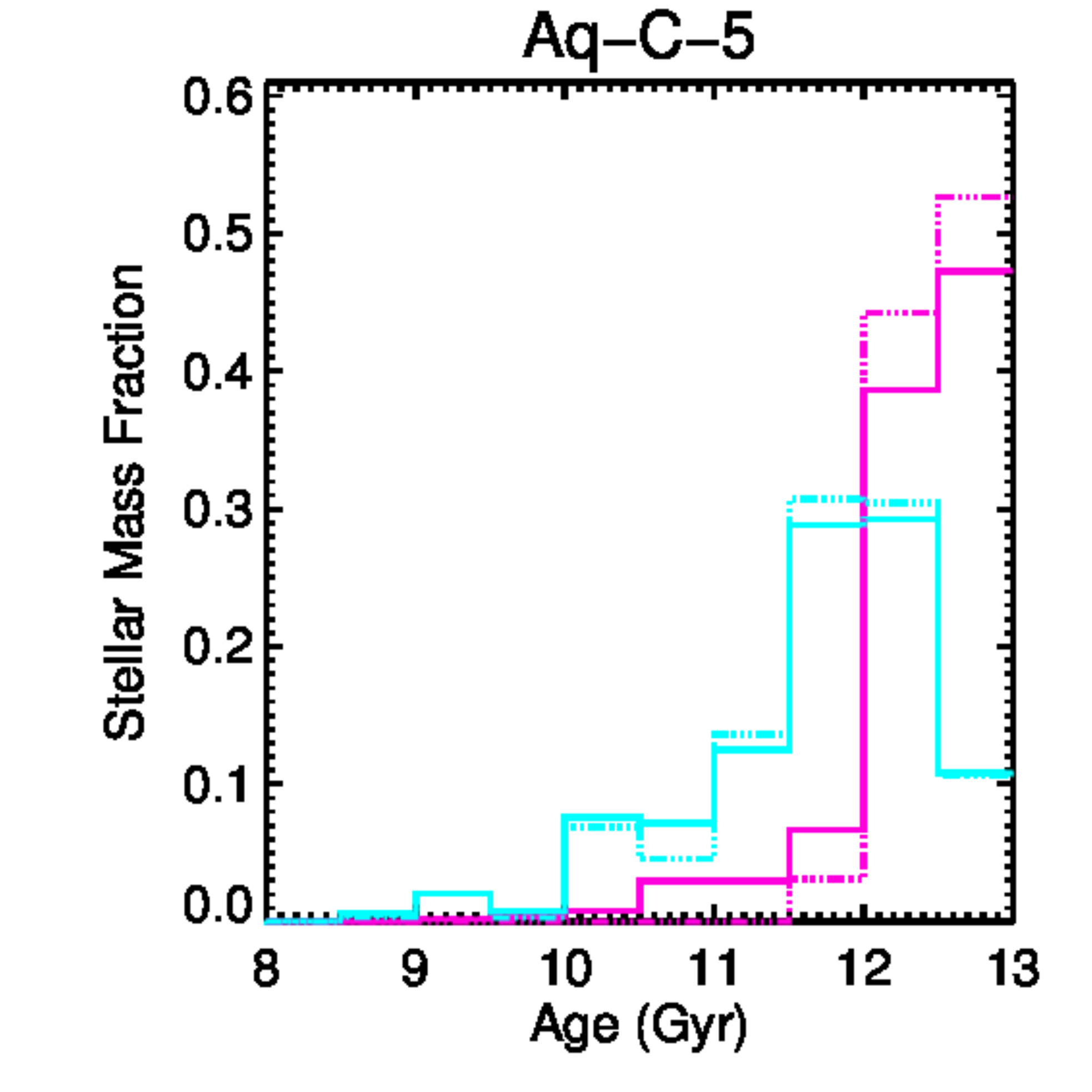}}
\resizebox{4.3cm}{!}{\includegraphics{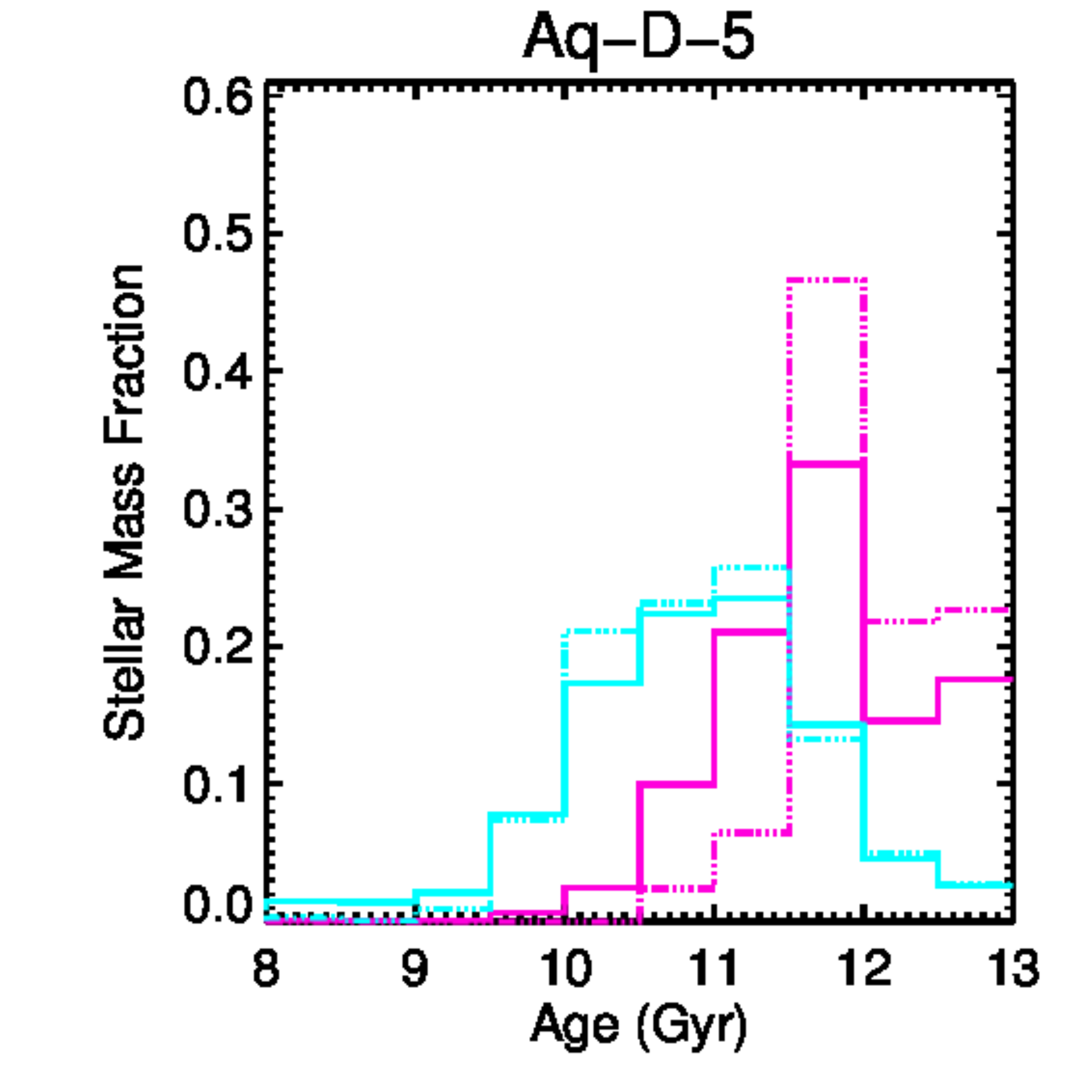}}
\caption{Stellar-mass weighted distributions of age for in situ (cyan
  solid lines) 
and accreted (magenta solid lines) stars for the analysed Aquarius
haloes within 10 kpc. We also show the stars classified as part of the
bulge components for comparison (dotted lines).
Accreted stars tend to be older than in situ stars by $\sim 1$ Gyr, on
average. Most of the in situ stars in Aq-A, Aq-C and Aq-D formed in a
main starburst. Aq-B has wider age distributions for both in situ and
accreted stars.} 
\label{mdf_age_All}
\end{figure*}

Figure~\ref{mdf_age_All} shows the distribution of ages for both in situ
and accreted stars  (each subsample has been normalised to its total
stellar mass to highlight the difference in their age distributions).
From this figure we can see that there is a slight difference between
the in situ and accreted stars. Aq-A, Aq-D and Aq-C show a clear trend for the
accreted stars to be older than the in situ stellar components by $\sim
1$ Gyr, on average. In situ and accreted stars in the central regions
formed mainly in starbursts,  occurring in the progenitor galaxy and
accreted satellites, respectively. The in situ stars have more extended
 starbursts. Conversely, Aq-B  exhibits a more uniform age distributions for both
types of stellar populations as the result of its assembly
  history involving an early massive merger which rejuvenate the
  central region.
 In Fig.~\ref{mdf_age_All} we also show the distributions for the
stars classified as belonging to the bulge components (dotted
lines). As can be seen the distributions are quite similar as expected
since this component dominates the central regions.

\begin{figure*}
\resizebox{4.3cm}{!}{\includegraphics{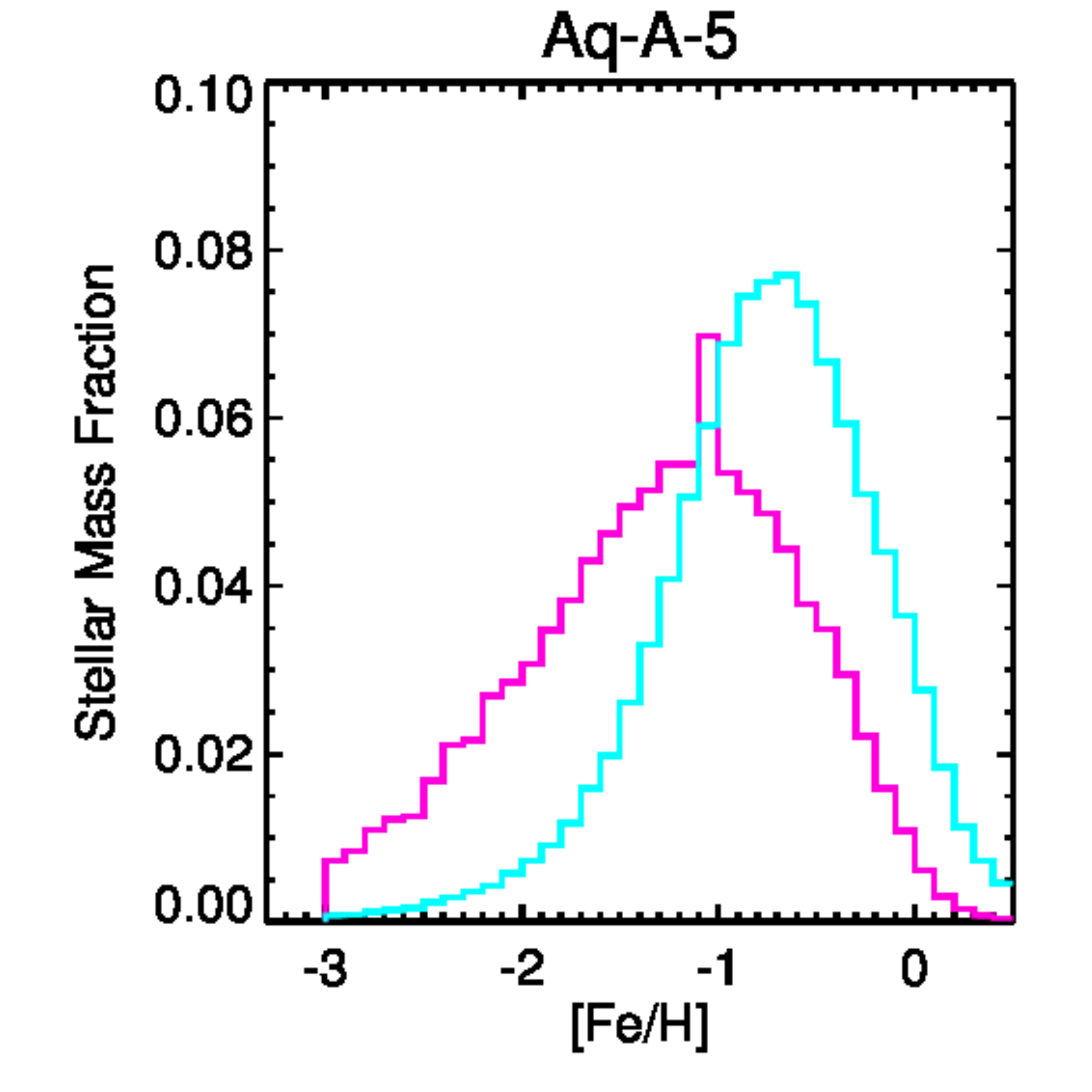}}
\resizebox{4.3cm}{!}{\includegraphics{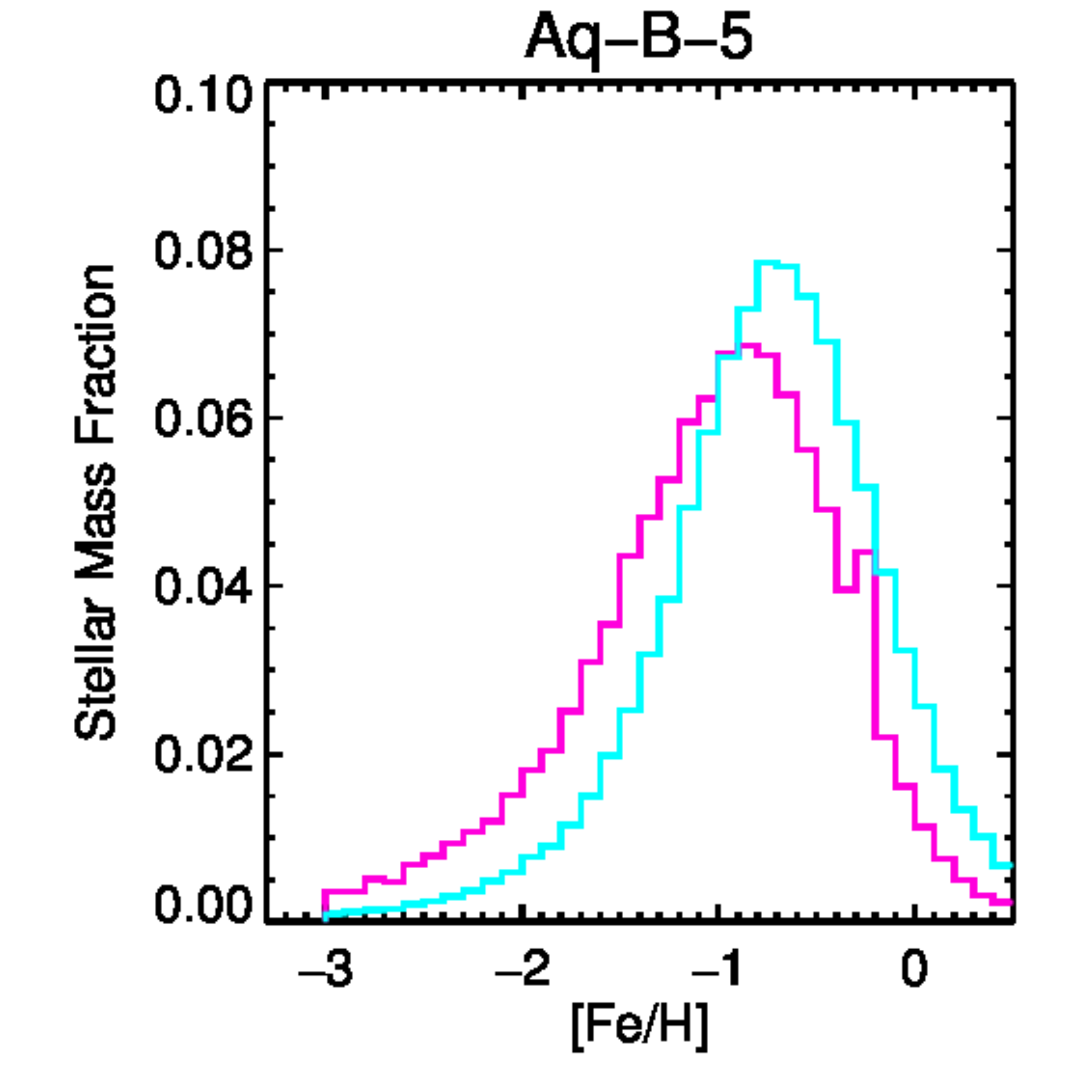}}
\resizebox{4.3cm}{!}{\includegraphics{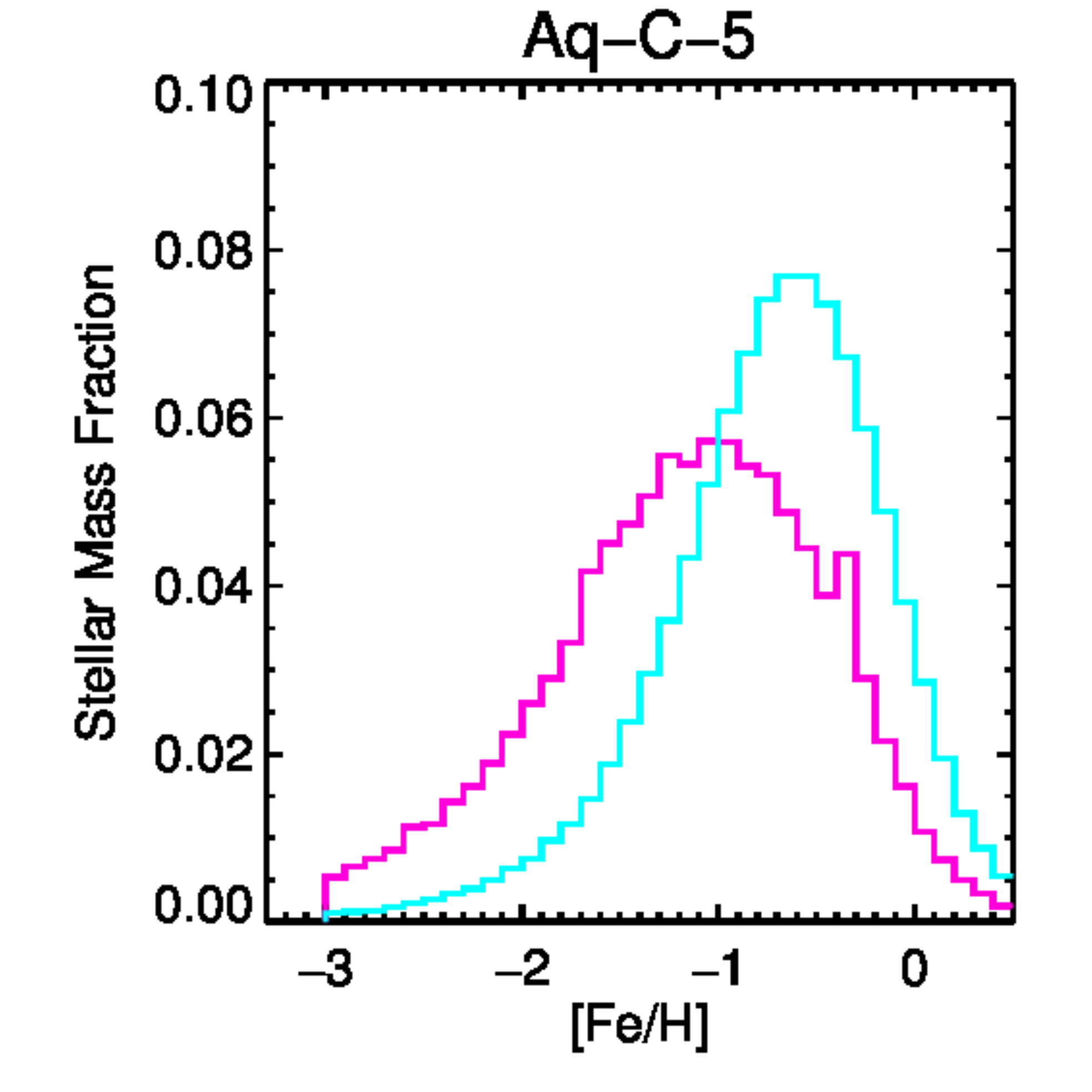}}
\resizebox{4.3cm}{!}{\includegraphics{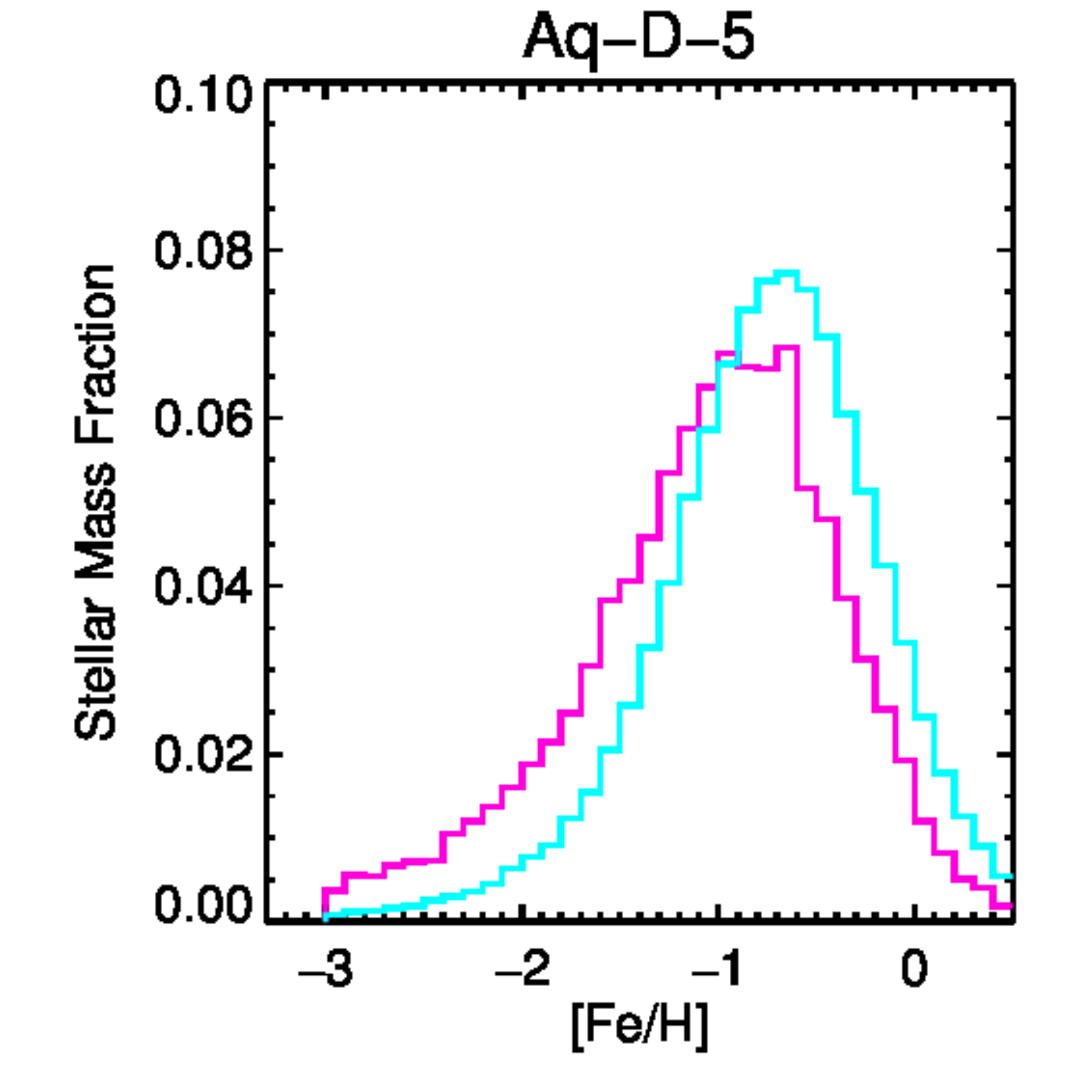}}
\caption{Stellar-mass weighted  distributions of [Fe/H] for in  situ
  (cyan lines) and accreted  (magenta lines) stars  for the analysed
  Aquarius haloes (normalised to the total stellar mass in each subsample). For all galaxies, the [Fe/H] distributions of accreted stars are shifted towards lower metallicities than
the distributions of  in situ  stars within 10 kpc. The shifts are larger for Aq-A and Aq-C.} 
\label{mdffeh}
\end{figure*}

Accreted stars tend to be less chemically enriched, as can be
appreciated from Fig.~\ref{mdffeh}. For all analysed systems, accreted
stars show a higher fraction of metal-poor stars with respect to the
[Fe/H] distribution of the in situ components. This is particularly
clear for Aq-A and Aq-C. The relative contribution of  low- and
high-metallicity stars in the accreted components is related to the masses of
the contributing satellites, as discussed in Section 3.2. These
metallicity characteristics are expected, considering the ages and the
trend for these stellar populations to be formed in starbursts. Due to
the  lifetime delay for SNIa,  short-duration starbursts do not allow SNIa
to enrich the remaining gas before star formation ceases in small
galaxies. The median [O/Fe] of these stellar populations are
super-solar, as expected, except for Aq-B, which exhibits lower mean [O/Fe] ratios
as a result of the contribution from a more massive satellite galaxy.

\begin{table*}
\begin{center}
\caption{Information on the inner 10 kpc regions of the analysed
  Aquarius haloes:  virial mass, 
stellar mass of the main galaxy, stellar mass within 10 kpc, mass of stars older than 10
Gyr within 10 kpc, mass of stars younger than 8 Gyr within 10 kpc, the mean age of the in
situ and accreted stars, the median [Fe/H] and
[O/Fe] abundances for the in situ and accreted stars, the fraction of stars older than 10 Gyr, the
fraction of the old stars which were accreted by galaxy satellites, the fraction of old, accreted stars which belong to the
bulge components, the fraction of stars younger than 8 Gyr, and the
fraction of young stars formed in situ (from the upper to lower raws). }
\label{tab1}
\begin{tabular}{lrrrr}\hline
  &                                 Aq-A-5 &Aq-B-5 & Aq-C-5 &  Aq-D-5 \\
\hline
  $M_{\rm 200} (10^{12}{\rm M_{\odot}h^{-1}}) $       & 1.10 &0.52&  1.18 & 1.09\\
  $M_{\rm gal}   (10^{10}{\rm M_{\odot}h^{-1}})$                   &  5.92 &2.53&5.93 &  4.41 \\
  $M_{\rm star,10 kpc}    10^{10}{\rm M_{\odot}h^{-1}})$ &5.21&1.94&4.76&3.39\\
  $M_{\rm old,10 kpc}   (10^{10}{\rm M_{\odot}h^{-1}})$  &4.30&0.90&4.56&2.89\\
$M_{\rm young,10 kpc}   (10^{9}{\rm M_{\odot}h^{-1}})$ &1.73&1.46&0.44&1.06\\
 <Age>$_{\rm in situ}$(Gyr)&11.44 &  10.82 & 11.85&11.12\\ 
 <Age>$_{\rm accreted}$(Gyr)&                                      12.62  &11.26&12.48   & 11.64\\
${\rm [Fe/H]_{\rm in situ}}$&                                      -1.02 &-1.06&-0.81&   -0.97\\
 ${\rm [Fe/H]_{\rm accreted}}$&                                  -1.44 & -1.13 &-1.24&-1.09\\
${\rm [O/Fe]_{\rm in situ}}$&                                      0.16 & 0.07&0.09& 0.09\\
 ${\rm [O/Fe]_{\rm accreted}}$&                                  0.35&0.07&0.33&0.17\\
${\rm F^{\rm old}}$&                                                 0.84&0.49&0.96&0.86\\
${\rm F^{\rm old}_{\rm accreted}}$&                                0.10  &0.35 &0.12 &0.21\\
${\rm F^{\rm old}_{\rm bulge,  accreted}}$&                                0.31  &0.72 &0.33 &0.36\\
${\rm F^{\rm young}}$&                                         0.04 &0.14 & 0.01& 0.03\\
${\rm F^{\rm young}_{\rm in situ}}$&                                             1 &0.93 & 1& 0.99\\
  
\end{tabular}
 \end{center}
\vspace{1mm}
\end{table*}

\subsection{Old Stars within the inner 10 kpc}
 
There is growing interest in understanding where and when the old
stellar populations formed, as they are expected to have formed in the
first galaxies. As mentioned above, in this study old stars are defined
by assuming a minimum age of 10 Gyr. In observational studies such old
stellar populations can be traced by RR Lyraes. As can be seen from
Fig.~\ref{mdf_age_All}, most of the stars in the inner spheroids are
older than 10 Gyr (except for Aq-B). Within the inner 10 kpc we find
that more than $\sim 85$ per cent are stars older than 10 Gyr (except
for Aq-B which exhibits a lower value, $\sim 50$ per cent of old
stars). Such a property is in global agreement with observational
results regarding the bulge \citep[][]{feltzing2000,zoccali2006} and
the inner region of the stellar halo \citep{santucci2015,carollo2016}
of the MW.

All simulated central regions have  a small fraction of stars younger
than $10^8$ yrs, which represents $\sim 1-3 $ per cent of the stellar
mass in the inner 10 kpc region. Interestingly, most of them are part of
the bulge component (i.e., they belong to the most gravitationally-bound
stellar populations). Recent discoveries of cepheids in the bulge of the
MW indicates the existence of young stars in the bulge
\citep{dekany2015}. A census of these young populations would be of
utmost importance to set  constraints on the formation of the MW and
large spirals in general. Our results show that, in the bulge components,
these young stars are  metal-rich in situ stars, with [Fe/H]$\sim
[-0.1, -0,4]$ and sub-solar [O/Fe] abundance ratios. They formed
from recycled material with a significant contribution from SNIa yields. 

\begin{figure*}
\resizebox{4.3cm}{!}{\includegraphics{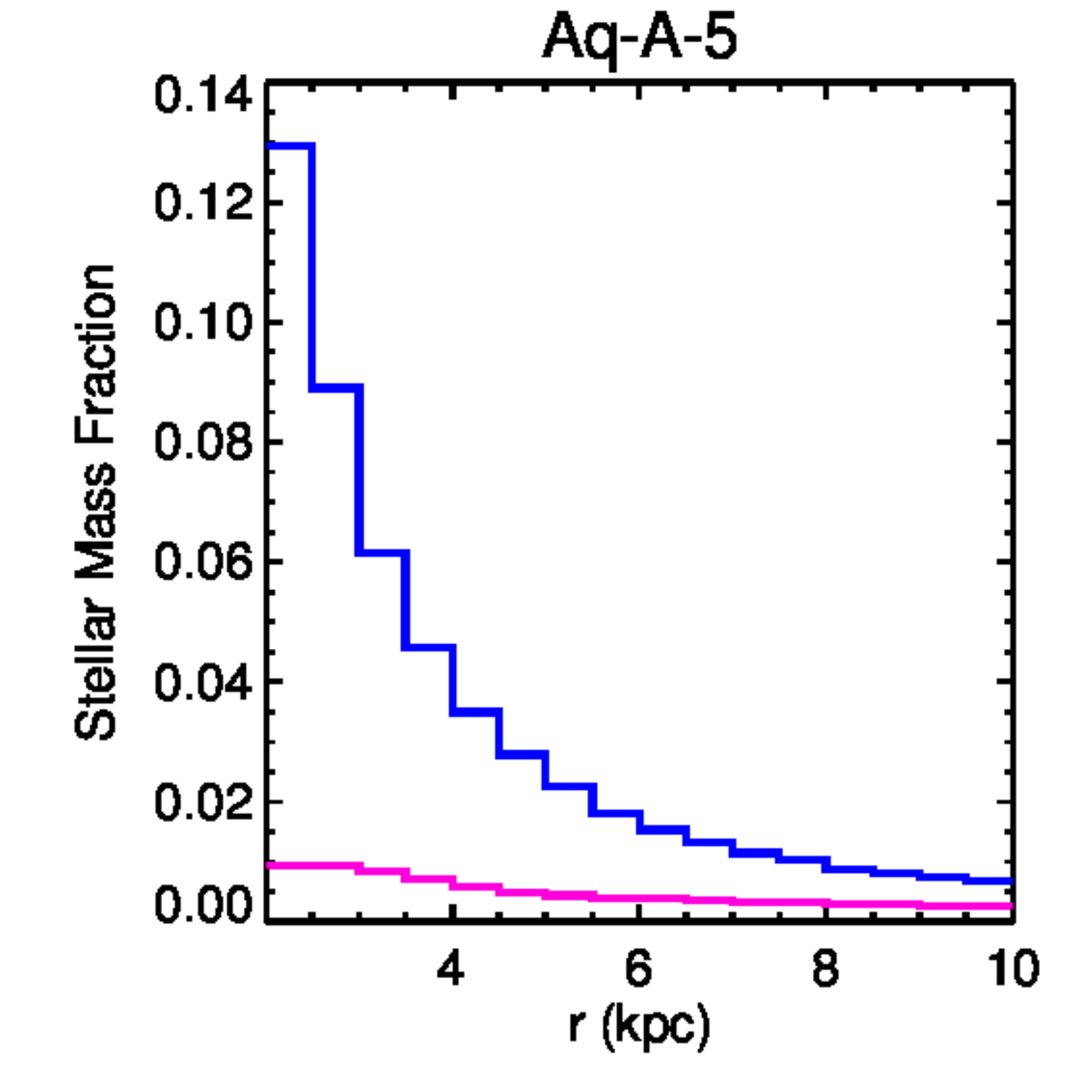}}
\resizebox{4.3cm}{!}{\includegraphics{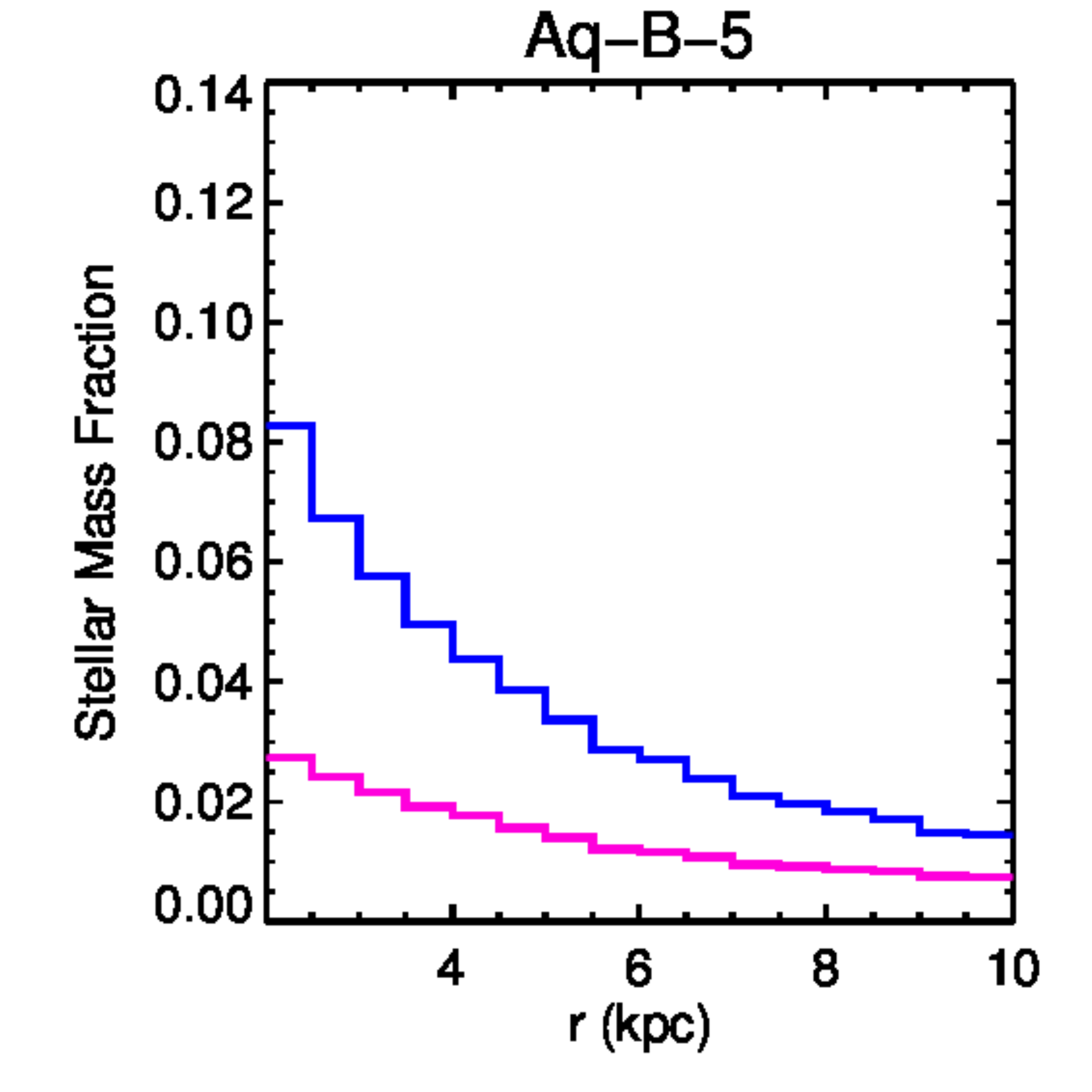}}
\resizebox{4.3cm}{!}{\includegraphics{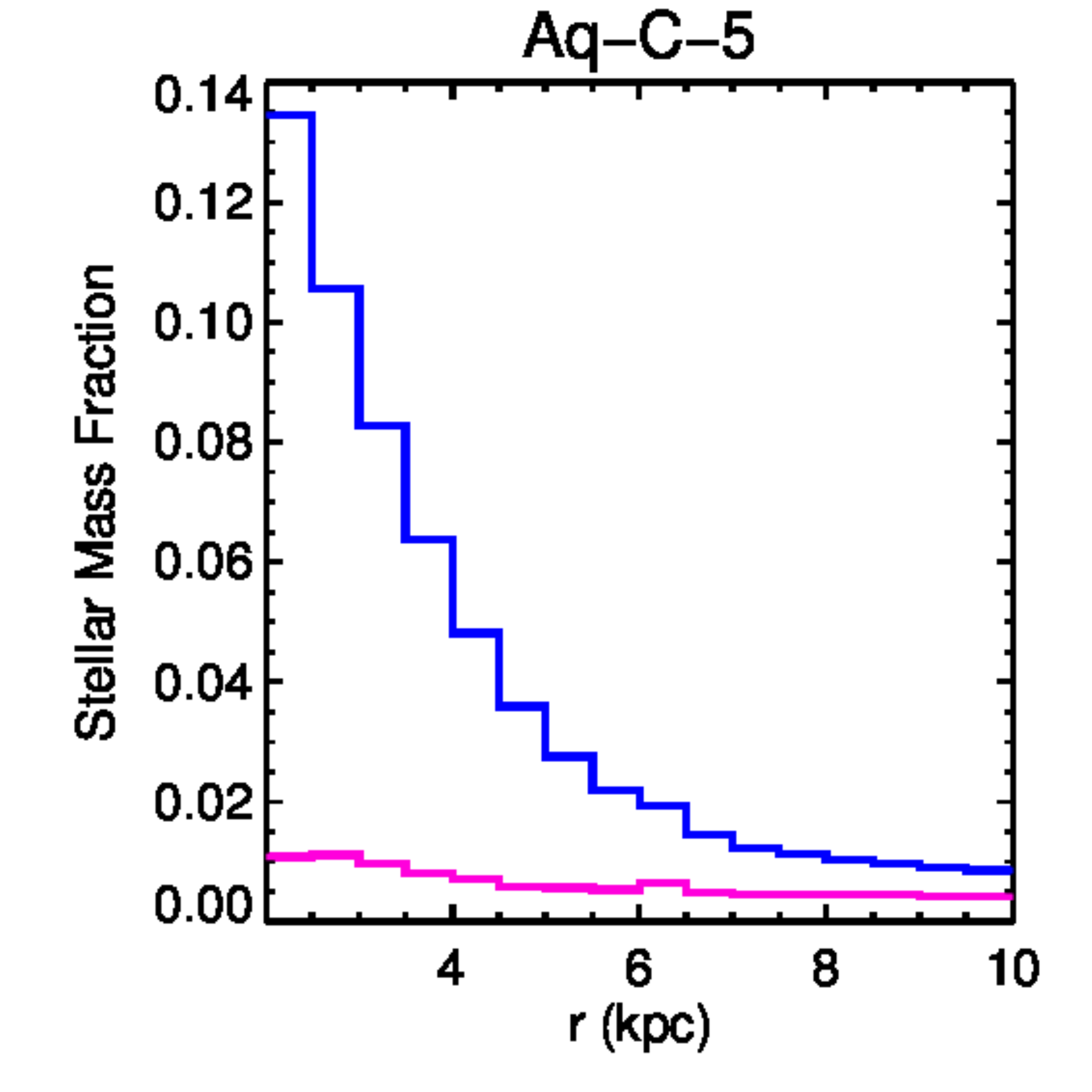}}
\resizebox{4.3cm}{!}{\includegraphics{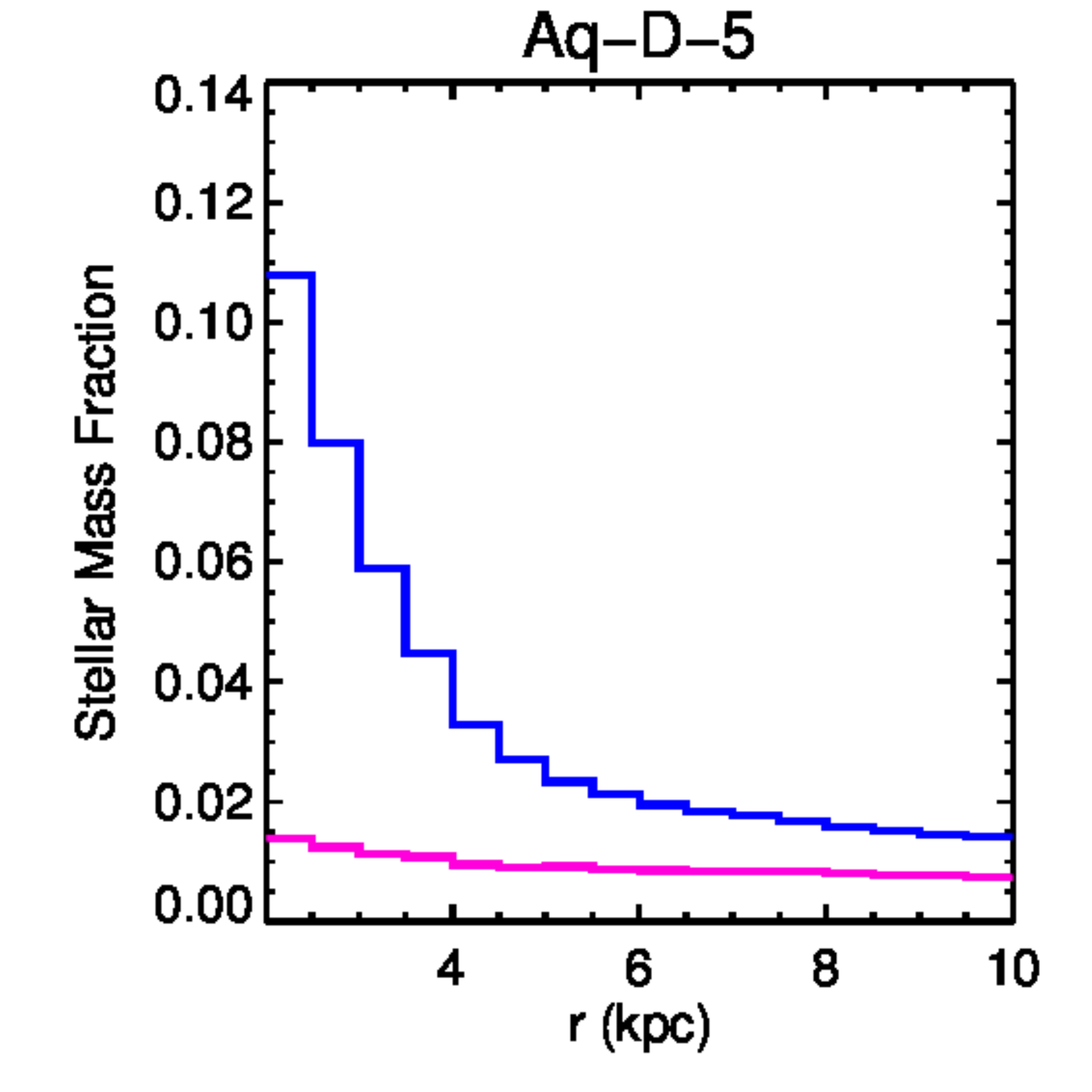}}\\
\resizebox{4.3cm}{!}{\includegraphics{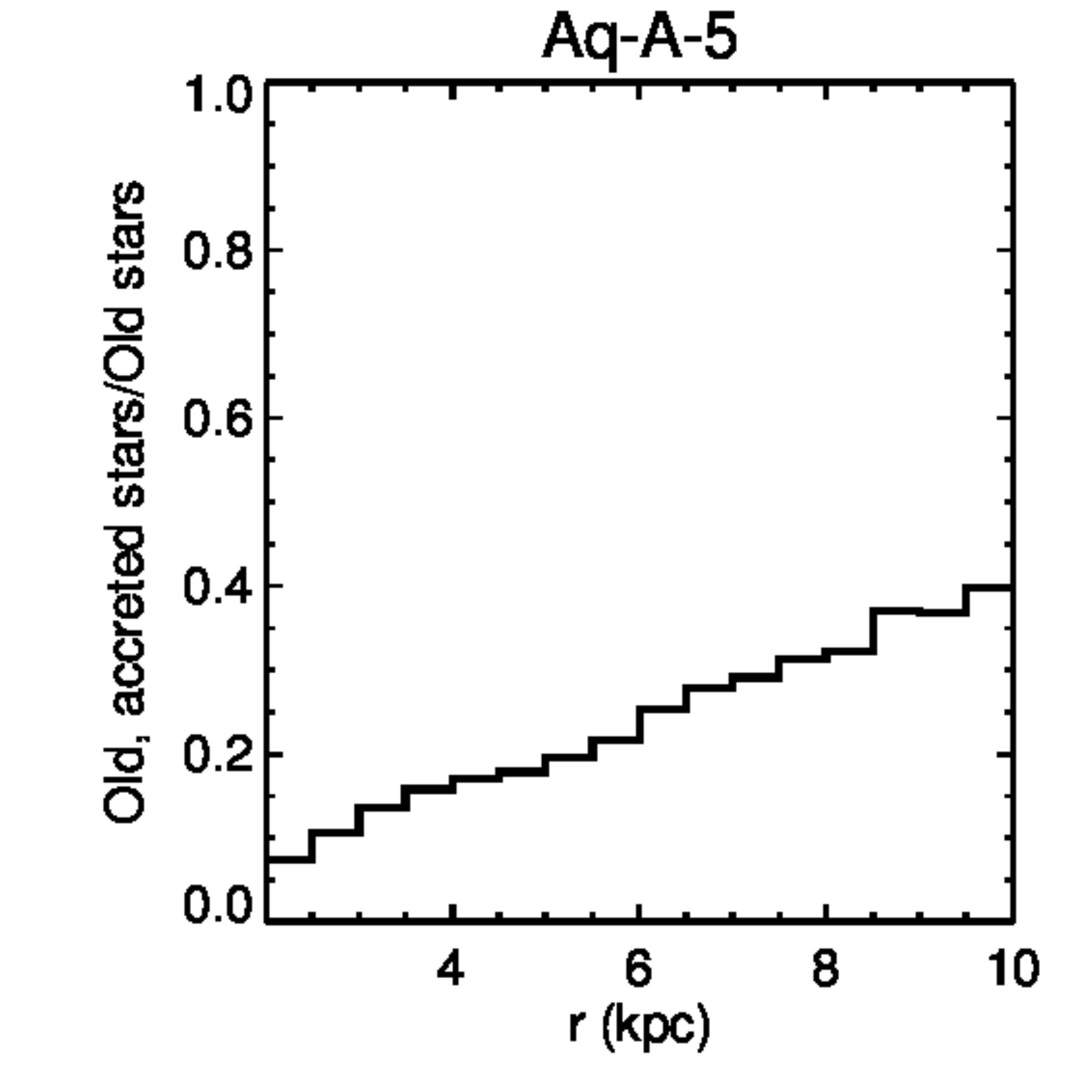}}
\resizebox{4.3cm}{!}{\includegraphics{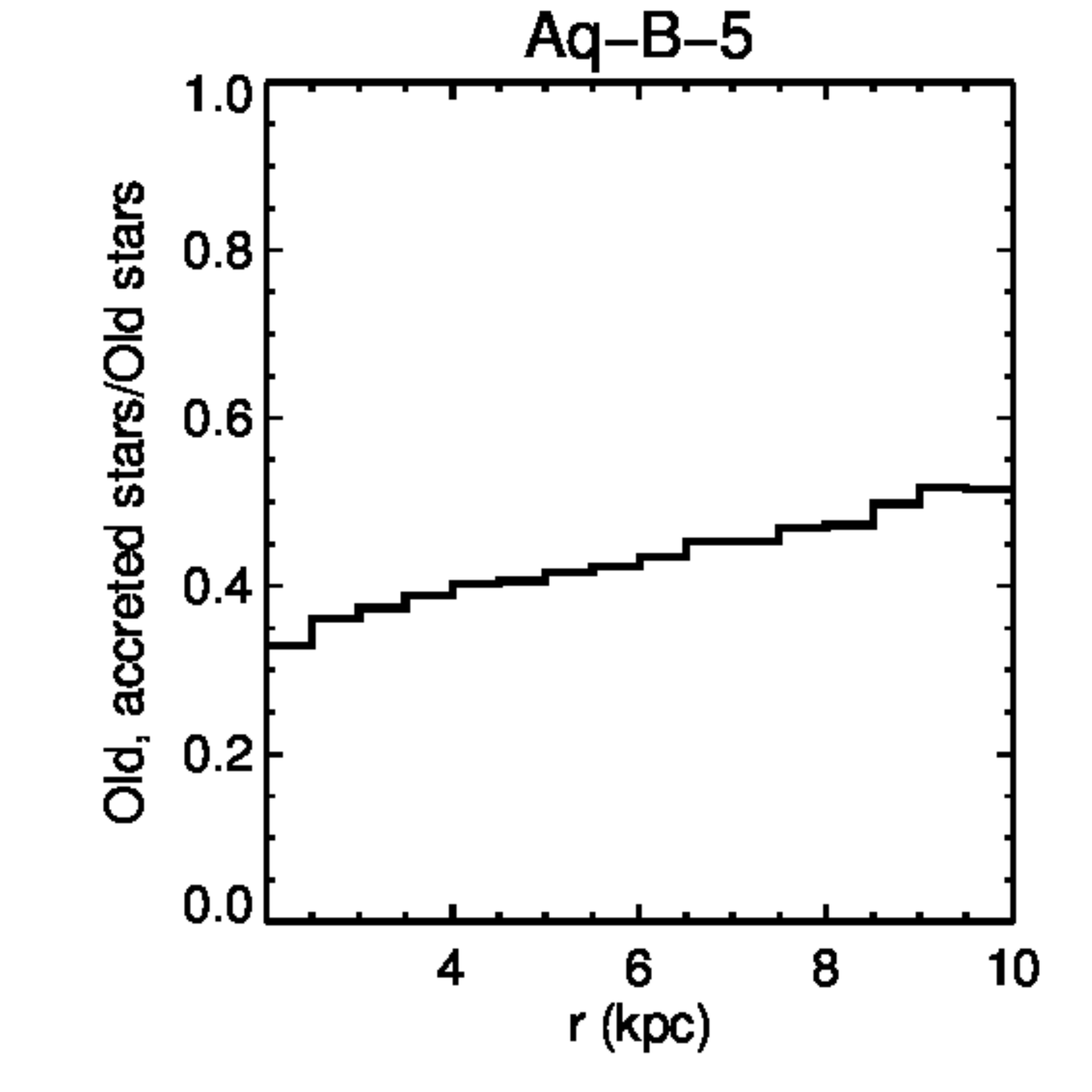}}
\resizebox{4.3cm}{!}{\includegraphics{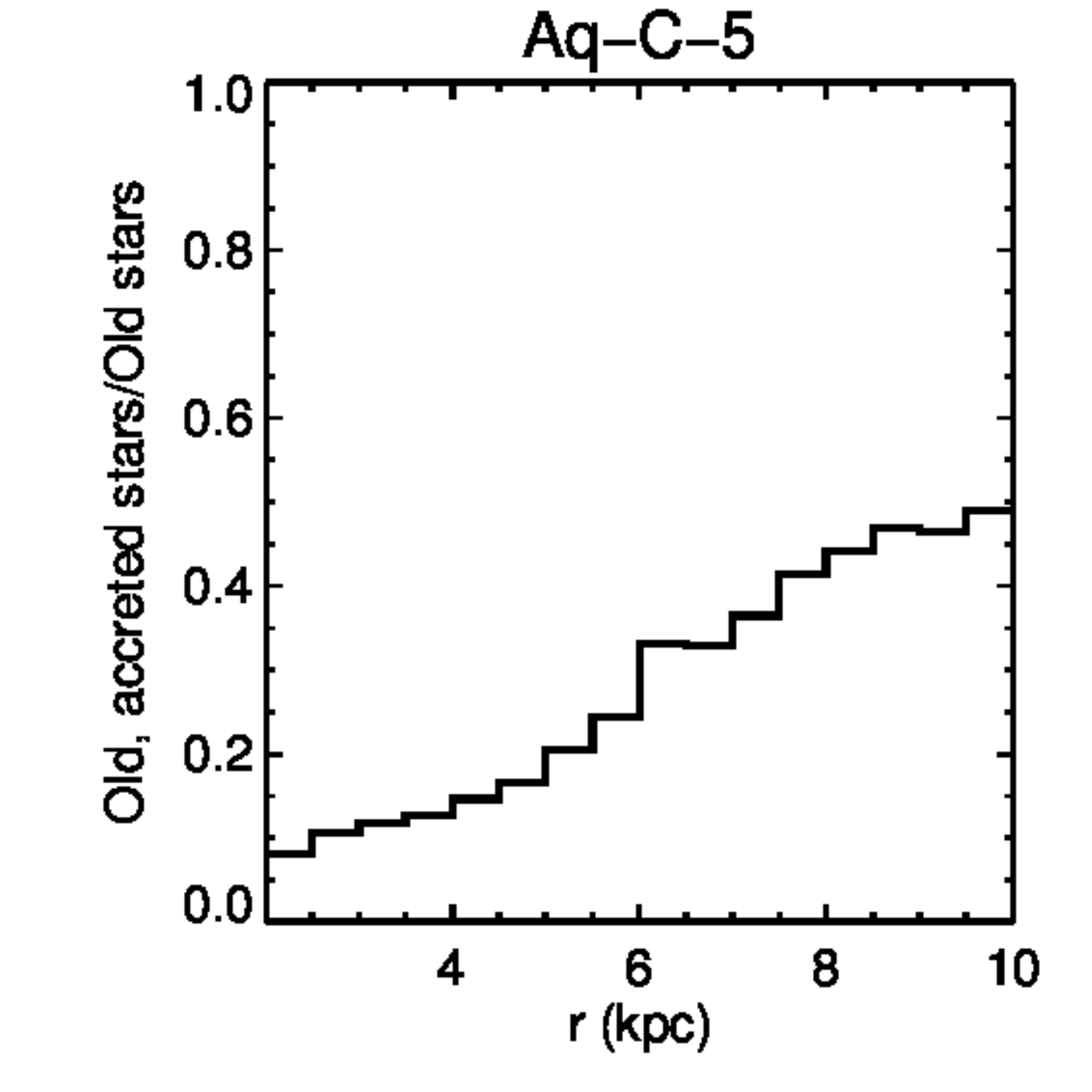}}
\resizebox{4.3cm}{!}{\includegraphics{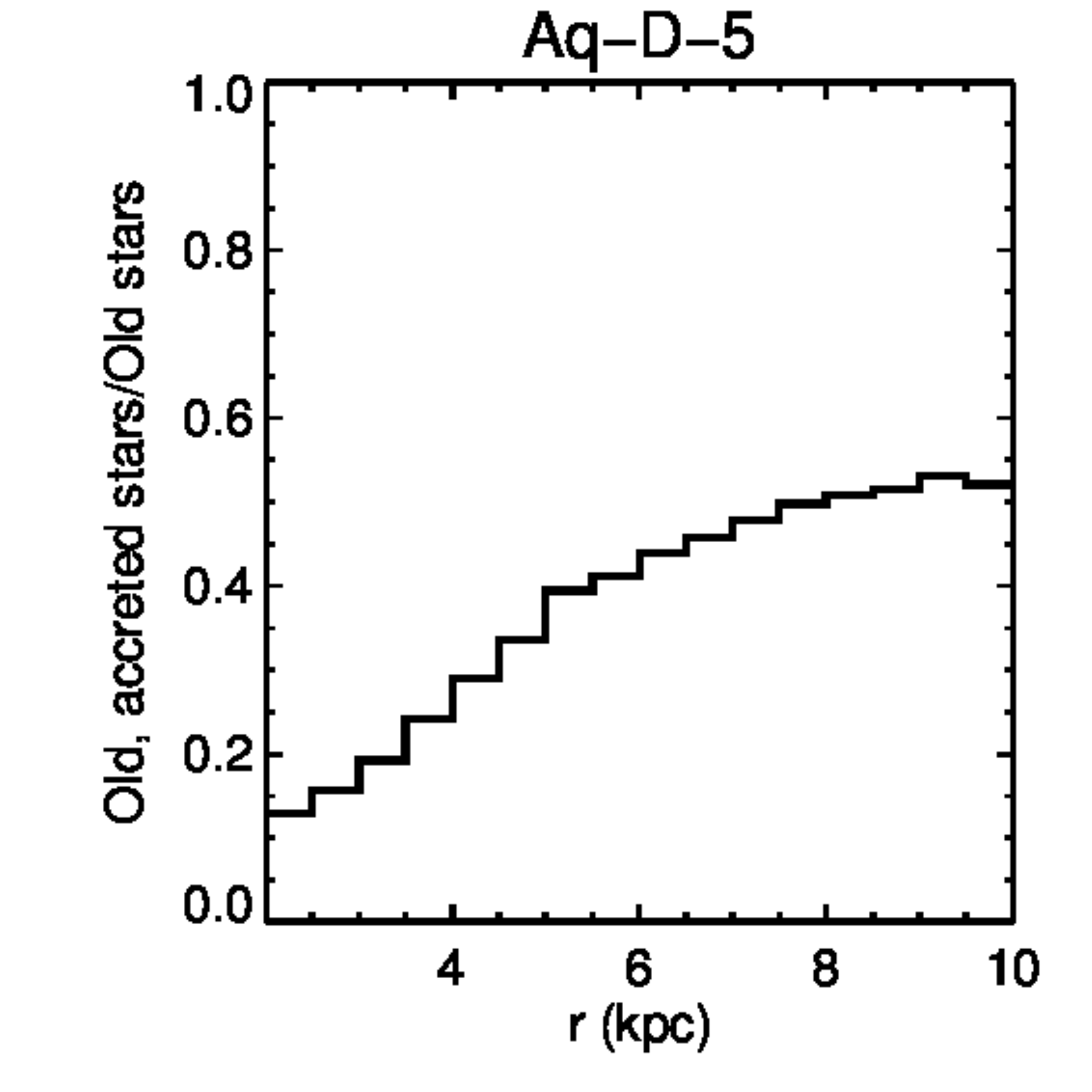}}
\caption{
 Distributions of stellar mass as a function of galactocentric distance for old  (blue lines) and old, accreted (magenta lines) stars for  the four analysed 
Aquarius haloes (upper panel) and the fraction of old, accreted stars with respect to  old stars as a function of galactocentric distance (lower panels). } 
\label{mdf}
\end{figure*}

Figure~\ref{mdf} (upper panels)  shows the distributions of old stars
(blue lines) and old, accreted stars (red lines) within the central 10
kpc. They tend to be concentrated in the central regions. However, as
one moves to larger  radii, the fraction of accreted stars increases
systematically, as shown by \citet{tissera2013} and \citet{tissera2014}.
This trend has been also reported by other work, such as
\citet{pillepich2015} and \citet{cooper2015}, where different codes and
subgrid physics are adopted. We note that different hydrodynamical
simulations do not agree on all the properties predicted for the stellar
haloes. But they do agree on the stellar haloes being formed mainly by the
aggregation of small galaxies and stars formed in situ. The
relative fractions of these populations are still under debate, as they
vary with the subgrid physics adopted (but not in a systematic fashion;
\citealt{cooper2015}).

The lower panel of Fig.~\ref{mdf} shows  the fraction of  old,
accreted stars with respect to old stars at a given radius. Globally,
the mass fraction of  old, accreted stars is larger at larger
galactocentric distances. A more detailed examination reveals  that
there are two kind of trends -- two of the analysed haloes show a monotonically
increasing fraction of old, accreted stars (Aq-A and Aq-C), while the
other two haloes exhibit  shallower increases, reaching plateaus at
about $\sim 50$ per cent contribution (Aq-B and Aq-D). The latter
suggests a significant contribution of old, accreted stars to the more
central regions. As we show in the next section, these two haloes have a
larger contribution from more massive dwarf galaxies contributing to the
inner regions. Globally, the mass fractions of  the old, accreted stars
represent $\sim 10-35$ per cent of the total old stars within the
central 10 kpc.

 Our simulations indicate that $\sim 40$ per cent of the old
stars would be accreted stars at $\sim 10$ kpc. This fraction is in
agreement with the observational results for the MW reported by
\citet{an2015} for the solar neighbourhood. We find that the
old, accreted stars are well-mixed within the central regions and less
gravitational bound than in situ stars.

An important aspect that can be appreciated from  inspection
of Fig.~\ref{mdf} is that the relative contributions of old, accreted
stars is different for different haloes, pointing to a strong dependence
on their assembly histories. The relative contributions of
accreted to in situ stars in the stellar haloes are important 
constraints for both improving the subgrid physics  modeling in numerical simulations
and  for understanding the assembly history of galaxies. There are
on-going surveys which will provide unprecedented information for the MW
\citep[e.g.,][]{helmi2016} to contrain the models and the subgrid
physics beyond what is currently possible.
 
\begin{figure*}
\resizebox{4.3cm}{!}{\includegraphics{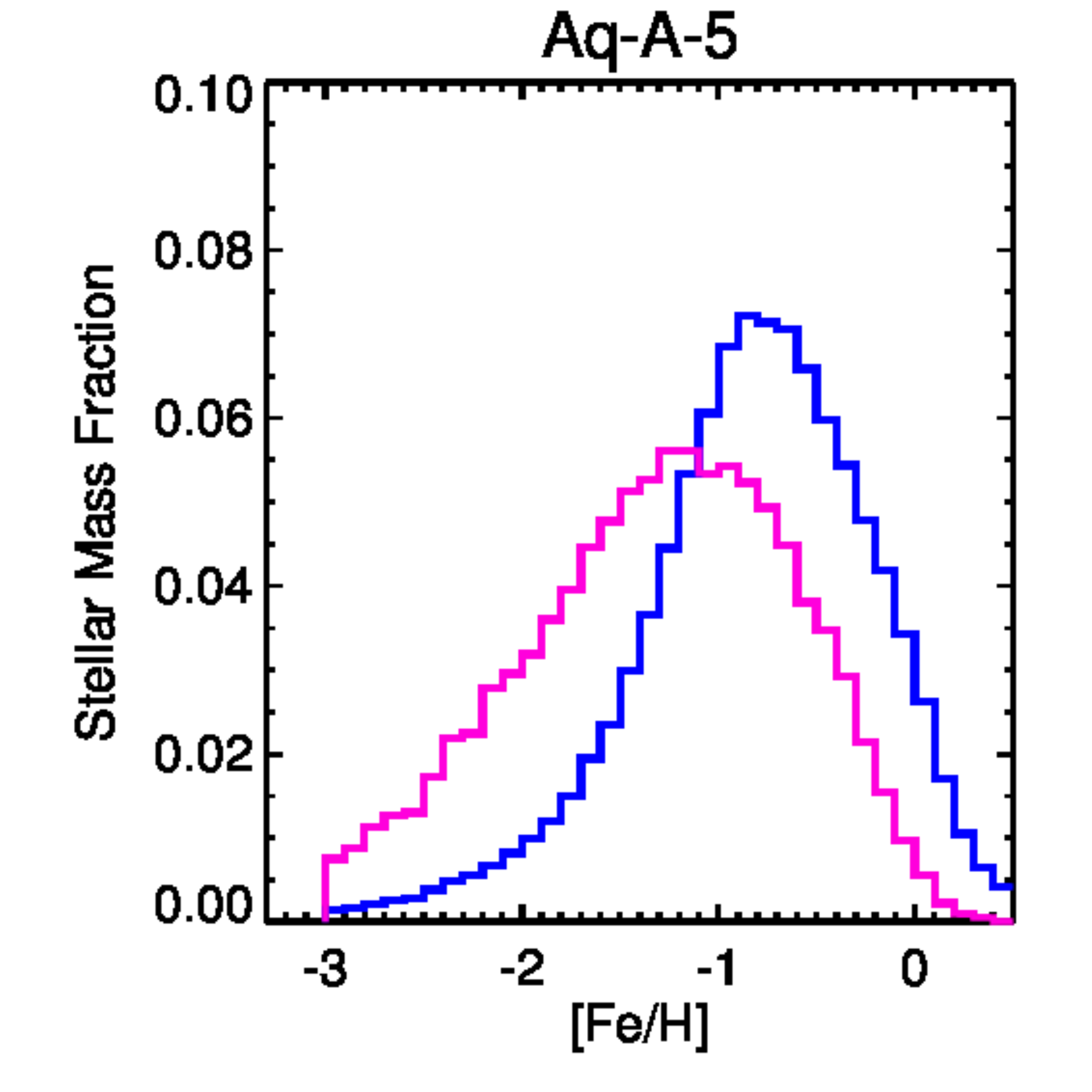}}
\resizebox{4.3cm}{!}{\includegraphics{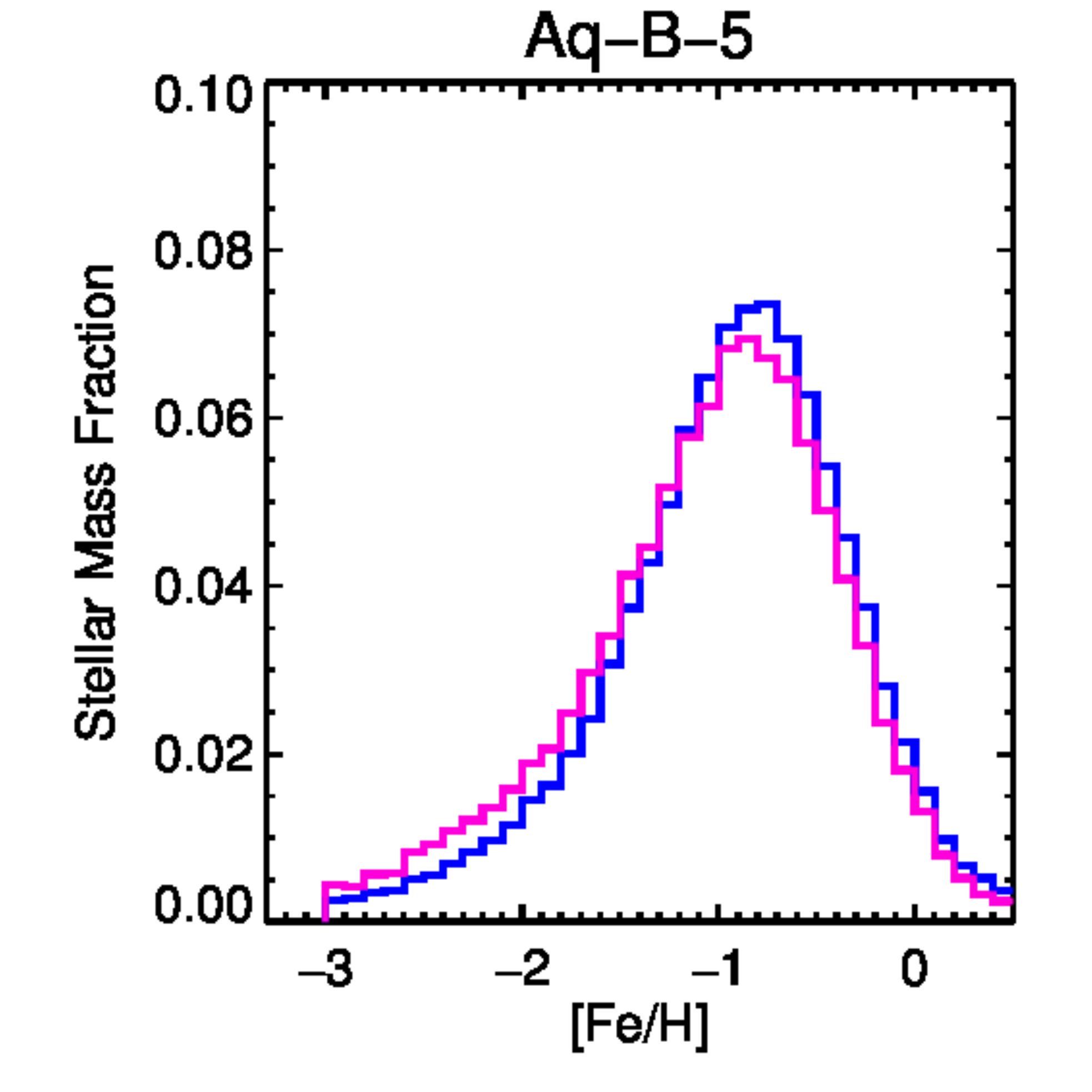}}
\resizebox{4.3cm}{!}{\includegraphics{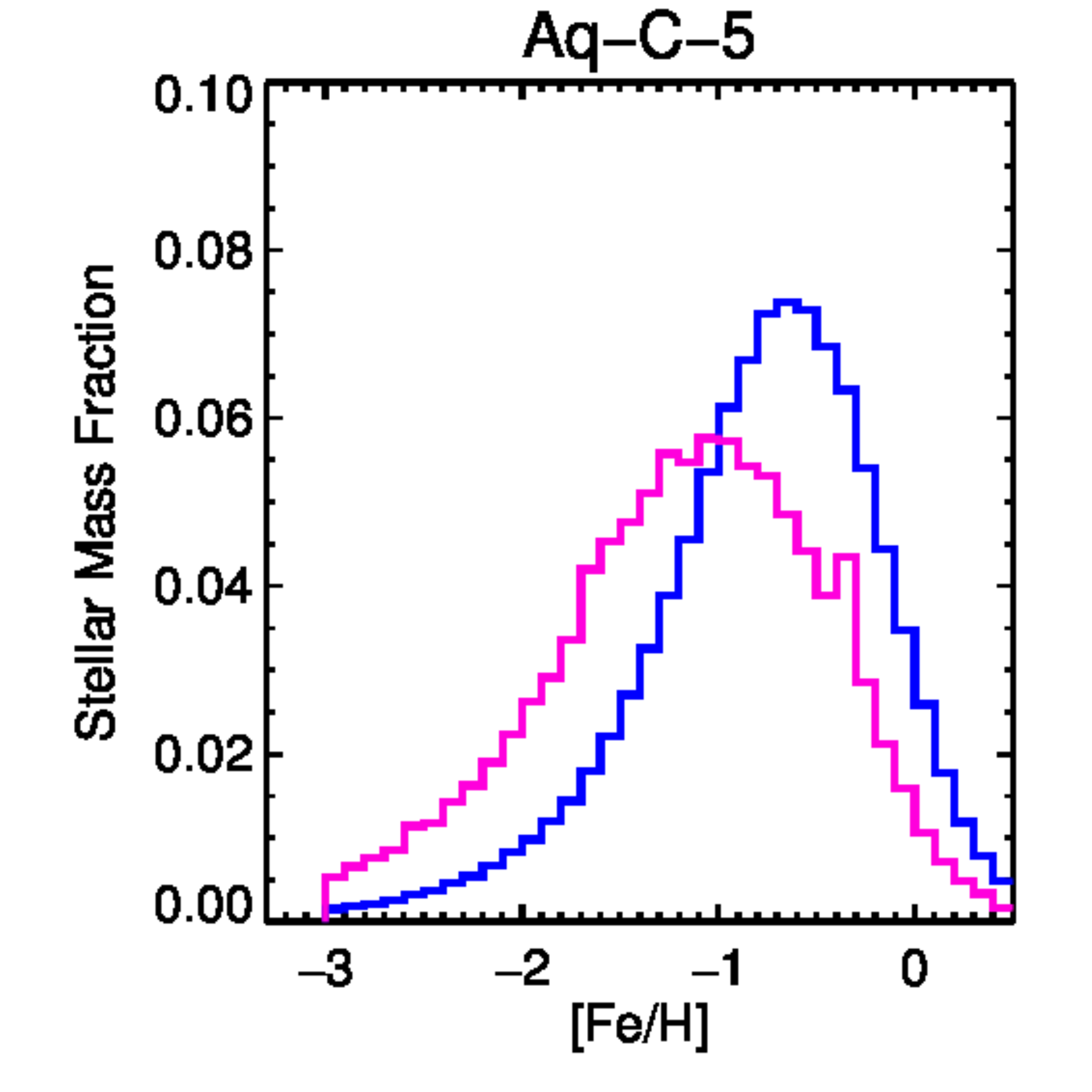}}
\resizebox{4.3cm}{!}{\includegraphics{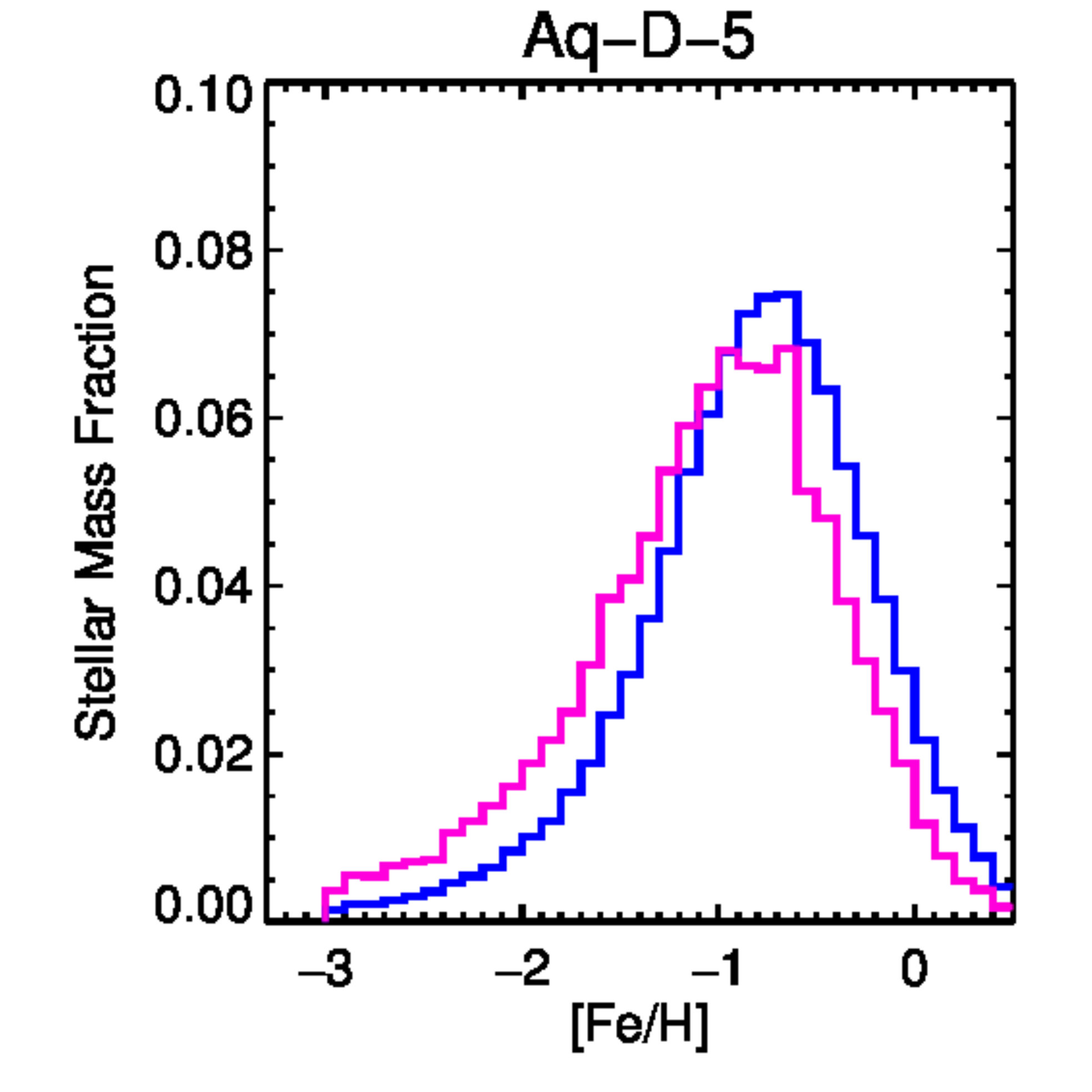}}
\caption{Mass-weighted distribution of [Fe/H] for old stars (blue lines) and old, accreted stars (magenta lines)  in the central region of the  four analysed Aquarius haloes (right panels).
Each subsample has been normalized to the its corresponding total mass.
This figure shows that old stars have metallicity distributions around  [Fe/H]$\sim -1 $ and that accreted old stars might be displaced towards lower metallicities [Fe/H]$\sim [-1.2,-1.5] $ ) depending on the history of assembly of the inner regions.} 
\label{fehdist}
\end{figure*}

In agreement with the global trends of Fig.~\ref{mdffeh}, we find that
old, accreted stars tend to be lower metallicity.  Fig.~\ref{fehdist}
displays the mass fraction of old stars (blue) and old, accreted stars (red
lines). Two of our haloes (Aq-B and Aq-D) exhibit a very weak shift towards
lower [Fe/H] abundances, while the other two (Aq-A and
Aq-C) have a larger contribution of low-metallicity,  accreted
stars.  Once again, this points to different histories of
assembly.
As can be seen, the distributions in Fig.~\ref{fehdist} are only slightly
different to those of the total stellar populations shown in
Fig.~\ref{mdffeh}. This is due to the fact that most of the stars in the
central regions are older than 10 Gyr, in agreement with recent results
reported for the MW by \citet{santucci2015} and \citet{carollo2016}.

For the set of analysed galaxies, we estimate that old, accreted stars
are older by $\sim 0.5-1.5$~Gyr compared to old, in situ stars. Old,
accreted stars tend to be less iron enriched, with median [Fe/H] lower by
$\sim 0.26$ dex, and more $\alpha$-enriched by $\sim 0.13$ dex than old,
in situ stars. 

These trends suggest that  the old, accreted stars formed in
smaller systems at early times that merged within the central bulk of
the forming galaxy. The quenching of star formation occurred at early
times, which explains both the lower metallicity content and high 
$\alpha$-element enrichment. Old, in situ stars formed during  more
extended 
starburst that occurred in the  progenitor(s).  
This led to higher metallicities and lower
$\alpha$-element enrichments (see Table \ref{tab1}).\\

\subsection{History of Assembly}

\begin{figure*}
\resizebox{4.3cm}{!}{\includegraphics{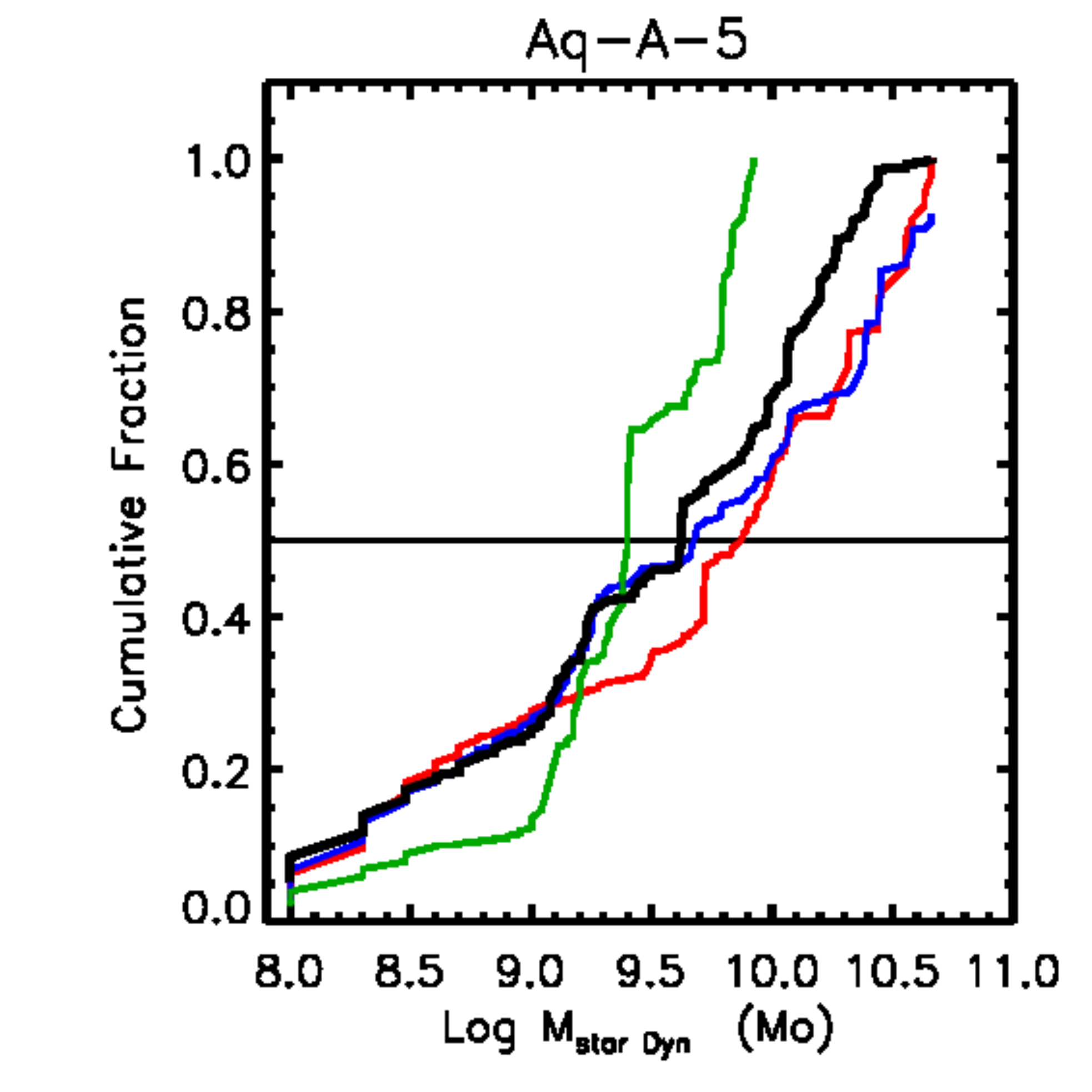}}
\resizebox{4.3cm}{!}{\includegraphics{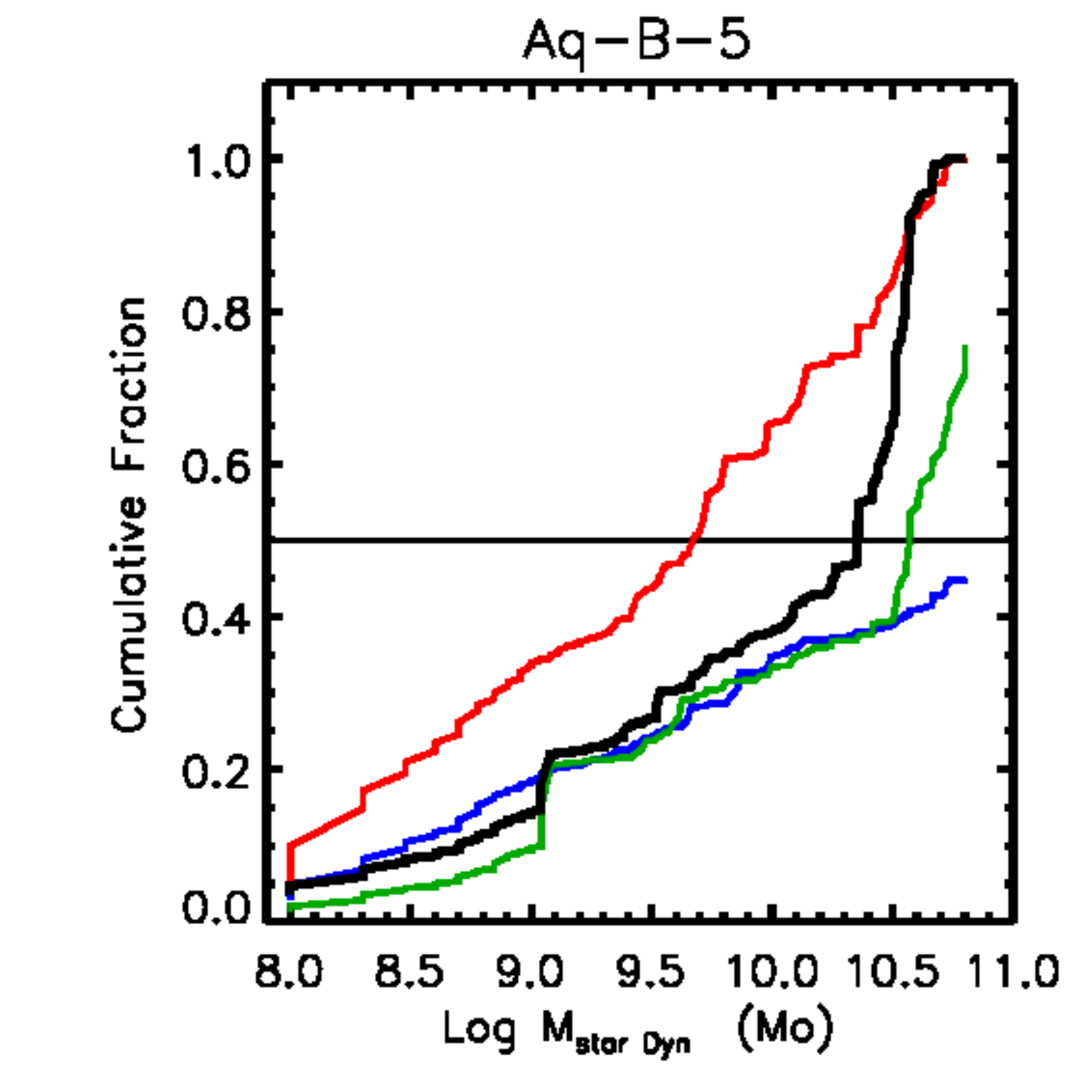}}
\resizebox{4.3cm}{!}{\includegraphics{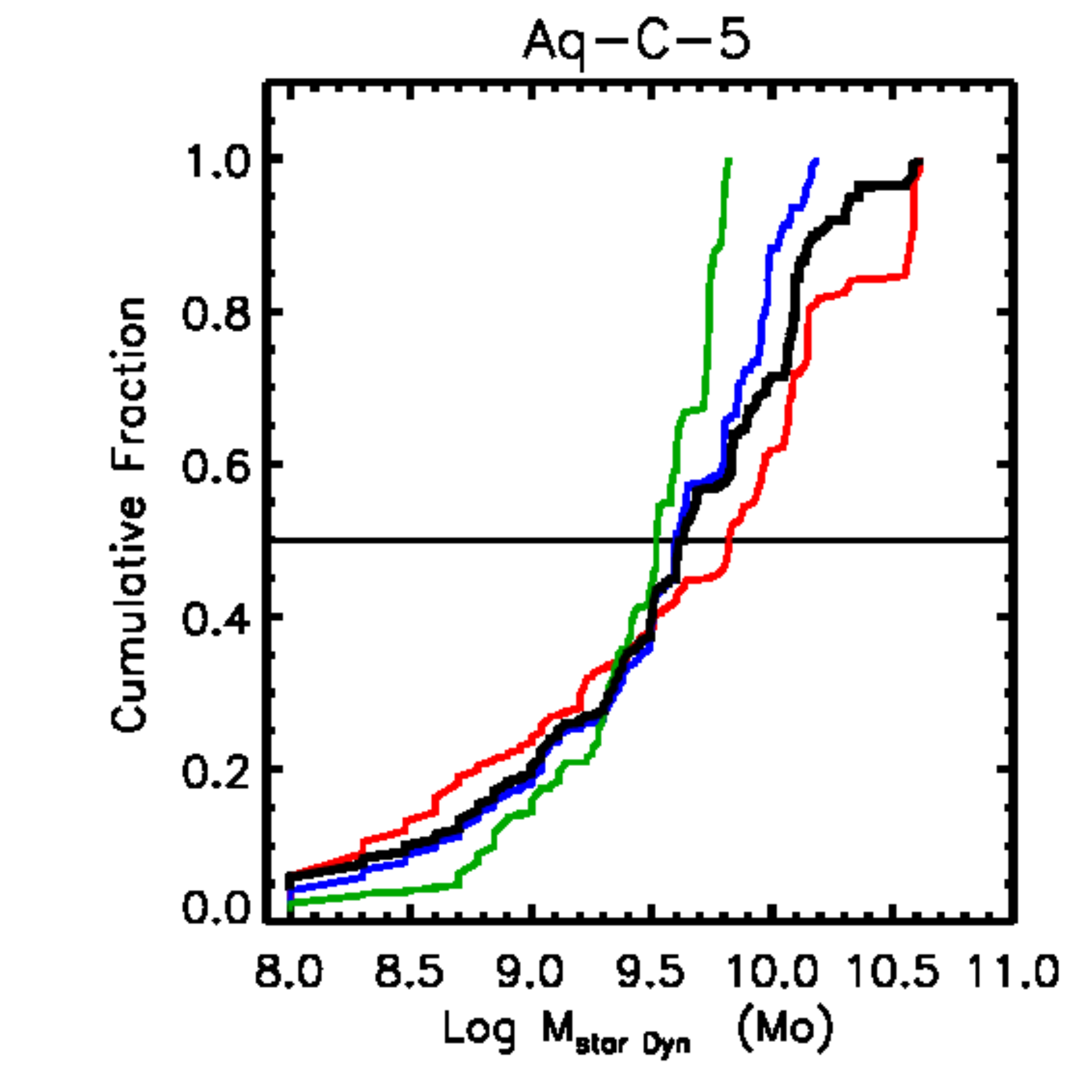}}
\resizebox{4.3cm}{!}{\includegraphics{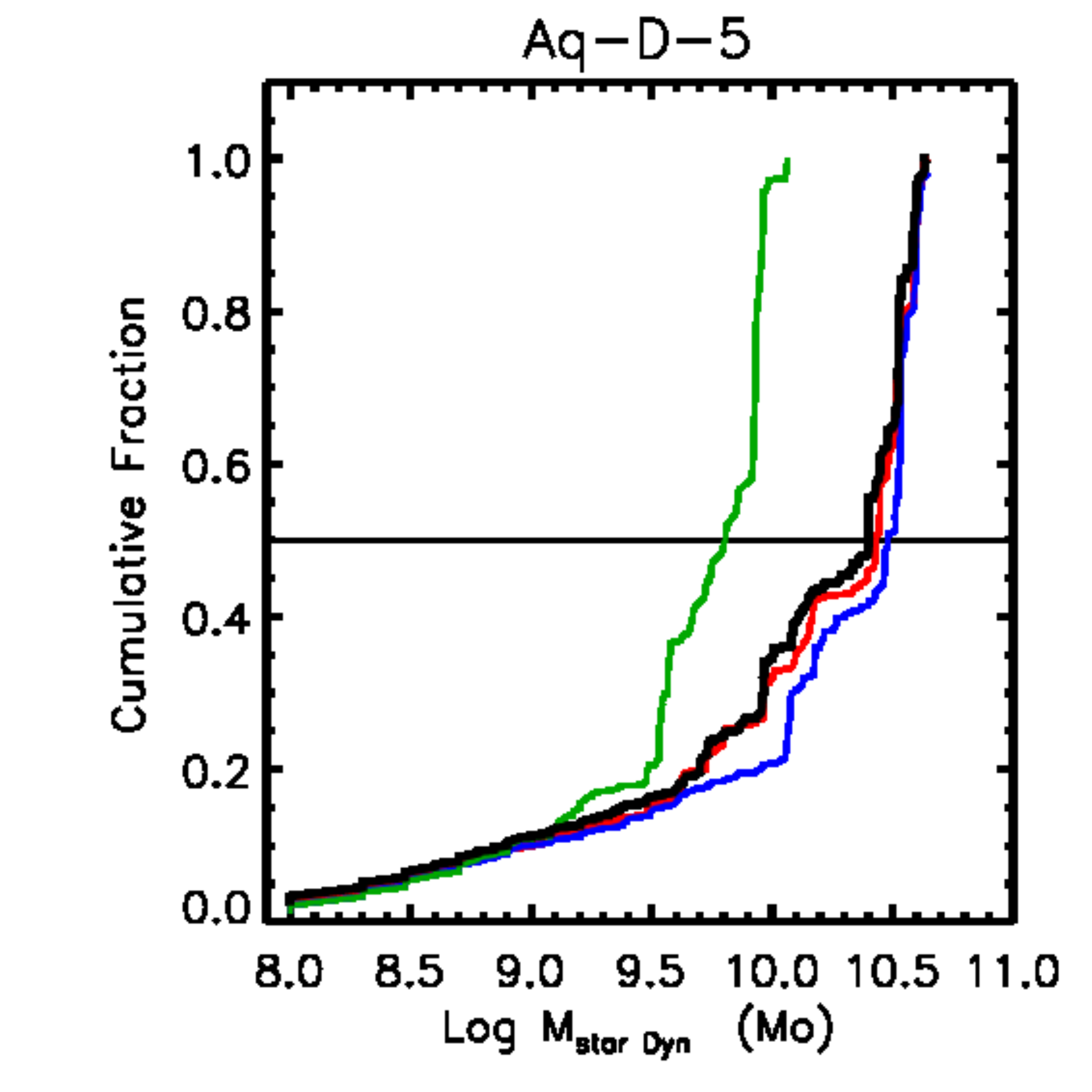}}
\caption{Cumulative  mass  fractions of old stars ($>10$ Gyr) within the inner 10 kpc
 regions, as a function of the dynamical mass of the accreted satellite where they
formed, at the time the satellite galaxies enter the virial radius of
the main galaxy (black lines). The distributions for accreted stars in
the bulges (green), inner haloes (blue), and outer haloes (red) are also
included. The central regions receive significant contributions of stars
from smaller satellites, while more than 50 percent arrives from
 a few more massive systems, $M_{\rm dyn} \> 10^{9.5}{\rm M_{\odot}}$. } 
\label{history}
\end{figure*}

To search for the satellites that contributed the old stars,
we followed back each of the stars with ages larger than 10 Gyr within
the central 10 kpc regions, identifying the galactic systems where they
originally formed.  Figure~\ref{history} shows the contributions of old
stars as a function of the dynamical mass of the accreted satellite
galaxies. The dynamical masses are taken at the time the satellite
galaxy enters the virial radius of the main galaxy. Dynamical friction
will modify the mass of the accreted satellites as they fall in,
depositing stars in the outer regions of the haloes before  they
reach into the central regions \citep{amorsico2017}. As can be seen from
Fig.~\ref{fehdist}, haloes with larger fractions of accreted,
low-metallicity stellar populations in the central regions have a larger
contribution from accreted satellites with $M_{\rm dyn} < 10^{10}
M_\odot$ (Aq-A and Aq-C).  When massive satellites contribute
significantly more stars, then their [Fe/H] distributions appear
more similar to the total abundance distributions  with a smaller
contribution of low-metallicity stars [Fe/H$<-1.1$. In this case, we found
that more than $\sim 50$ per cent of the accreted stellar mass in the
central regions arrives from satellites with $M_{\rm dyn} > 10^{10} M_\odot$ (Aq-B and Aq.D). 

 Figure ~\ref{history} also includes the stellar mass fractions of accreted stars 
identified as belonging to the bulge (green lines), inner halo (blue
lines), and outer halo (red lines),
as a function of the dynamical mass of the parent satellites. Recall
that this classification is based on a binding energy criterion, so that
stellar particles in the bulge components are those  that are more gravitationally
bound than those in the inner and outer haloes  (see Section 2). As
can be seen from this figure, the most bound accreted particles
(associated with the bulge component) are contributed by  smaller
satellites, with a significant impact from those around $M_{\rm dyn}
\sim 10^{9-9.5} M_\odot$. The exception is Aq-B, which received
contributions from more massive satellites in the inner regions as well
as in the most bound component. In the central regions the three dynamical
components co-exist, and each of them have different fractions of in
situ and accreted stars. As first shown by \citet{tissera2012}, the
bulges have the smallest contributions of accreted stars, followed by
the inner and the outer haloes, in increasing order. Considering  the
old, accreted stars within the inner 10 kpc, we found that for Aq-A, Aq-C and
Aq-D, most of them ($\sim 65$ per cent) belong to the inner region of
the stellar haloes, with $\sim 0.35$ per cent forming part of the
bulges. Aq-B shows a larger contribution to the bulge ($\sim 72$ per
cent), as a consequence of its particular history of formation. 


We calculated that, for Aq-A, Aq-C, and Aq-D, the accretion of satellites
into the inner 10 kpc regions occurred at $z >4$, and most of the
mergers involved small satellites ($M_{\rm dyn} < 10^{8} M_\odot$), with
only  a few ($\sim 2-3$) being in the range $M_{\rm dyn} \sim
10^{9.5}-10^{10} M_\odot$. These few more massive mergers contribute a
significant fraction of stars to the central regions, as can be seen
from Fig.\ref{history}. Aq-B experienced more massive accretion at a
lower redshift ($z \sim 3$), which determines their different properties
in comparison with the rest of the analysed haloes. 

\subsection{Kinematics of the Central Regions}

In the previous sections, we analysed the spatial, age, and metallicity
distributions of stars in the central regions, finding a difference
between accreted and in situ stars. Accreted stars were found  to
form spheroidal distributions, with lower metallicities and older
ages, on average. We showed that the differences can be ascribed to
their different assembly histories.

\begin{figure*}
\includegraphics[width=0.24\textwidth]{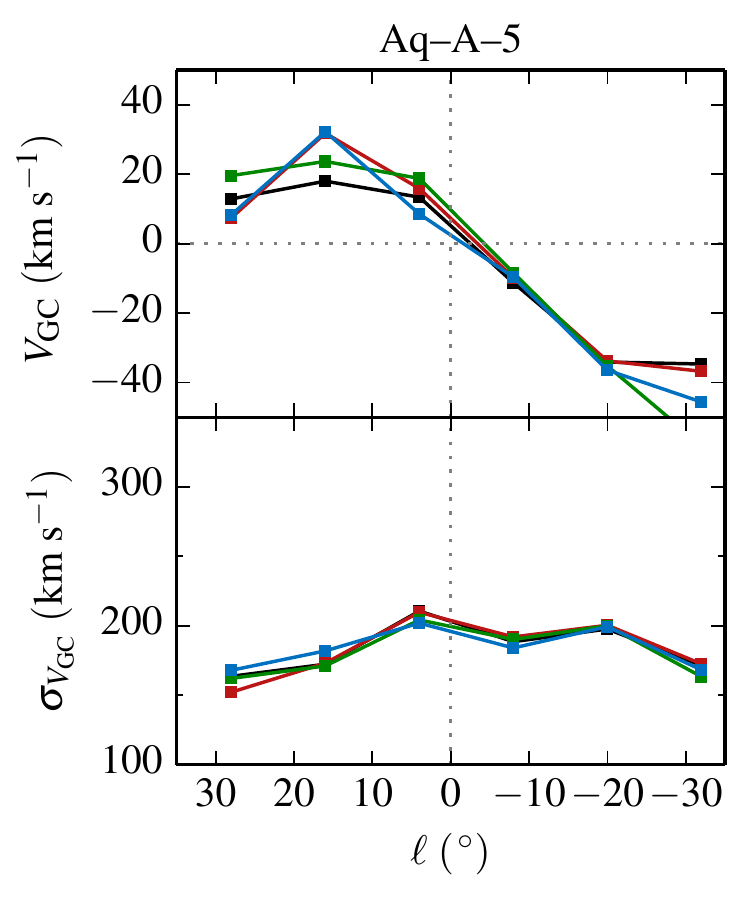}
\includegraphics[width=0.24\textwidth]{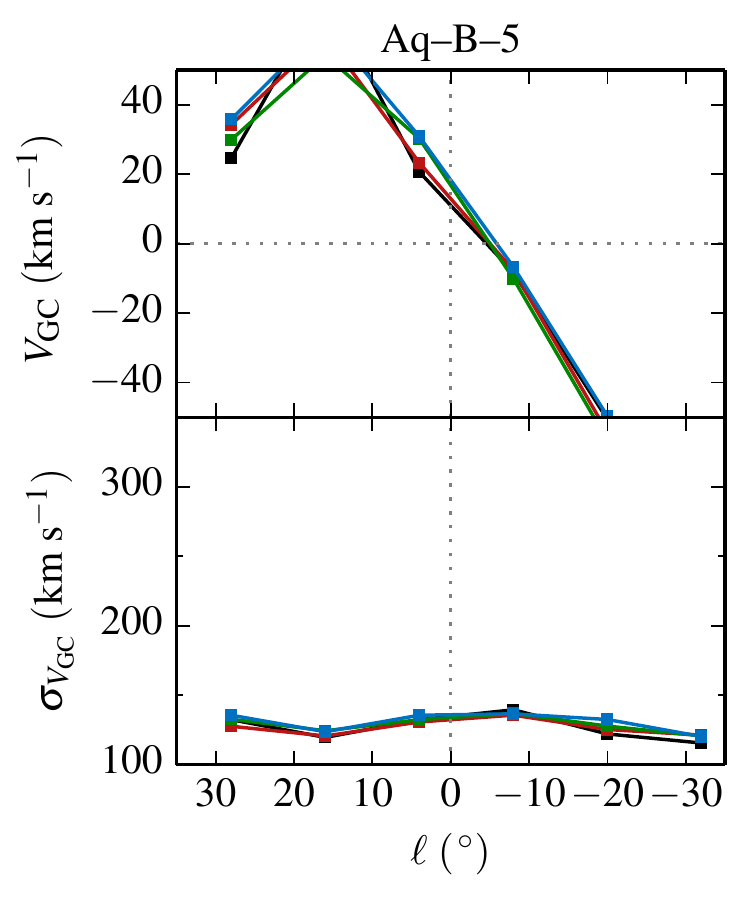}
\includegraphics[width=0.24\textwidth]{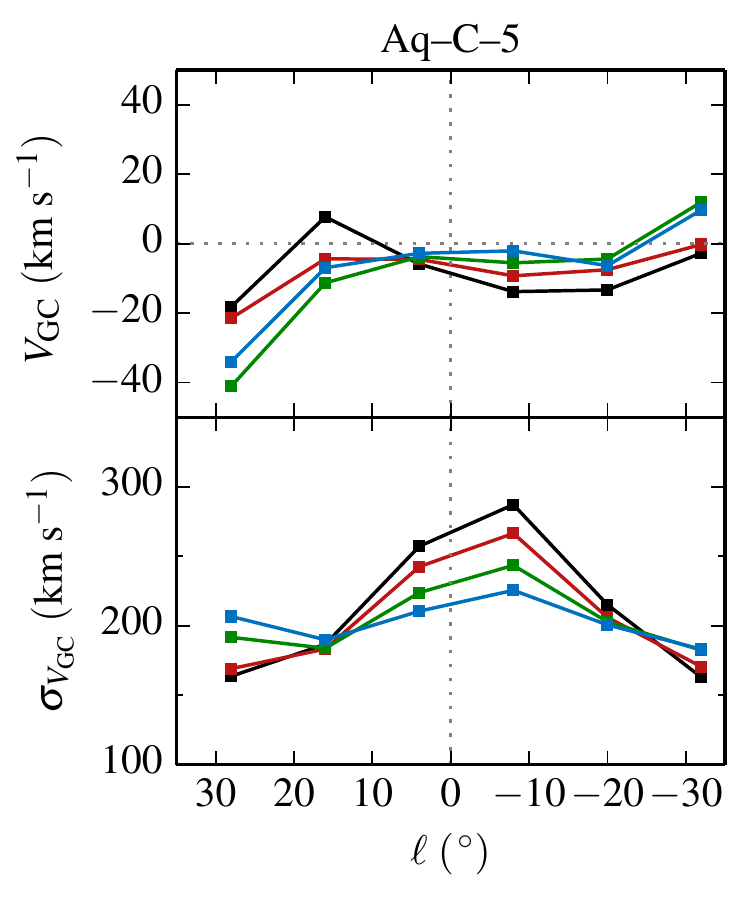}
\includegraphics[width=0.24\textwidth]{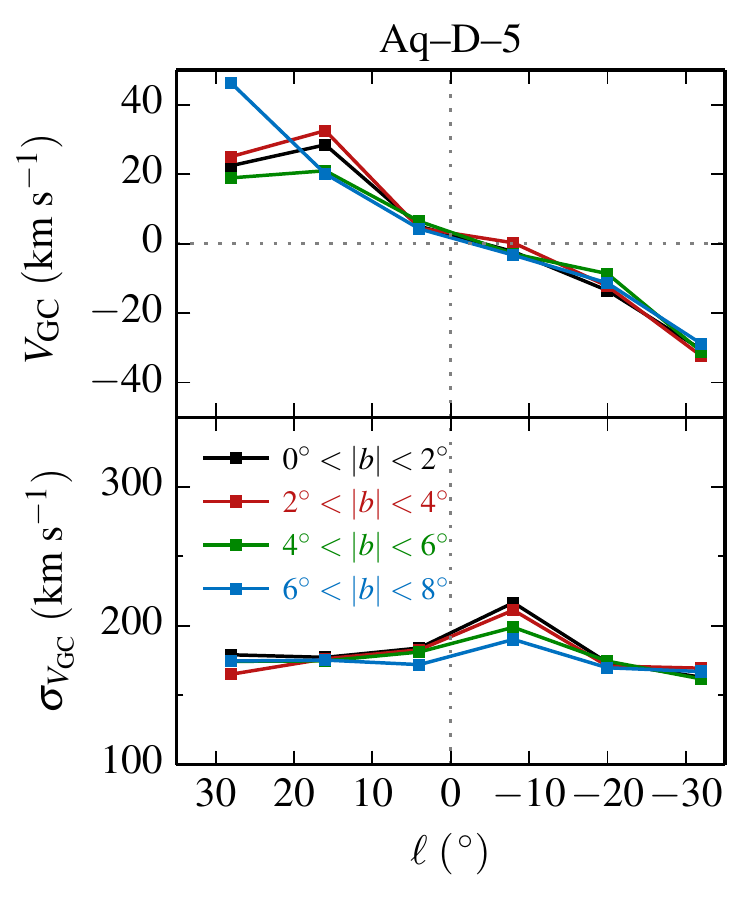}

\includegraphics[width=0.23\textwidth]{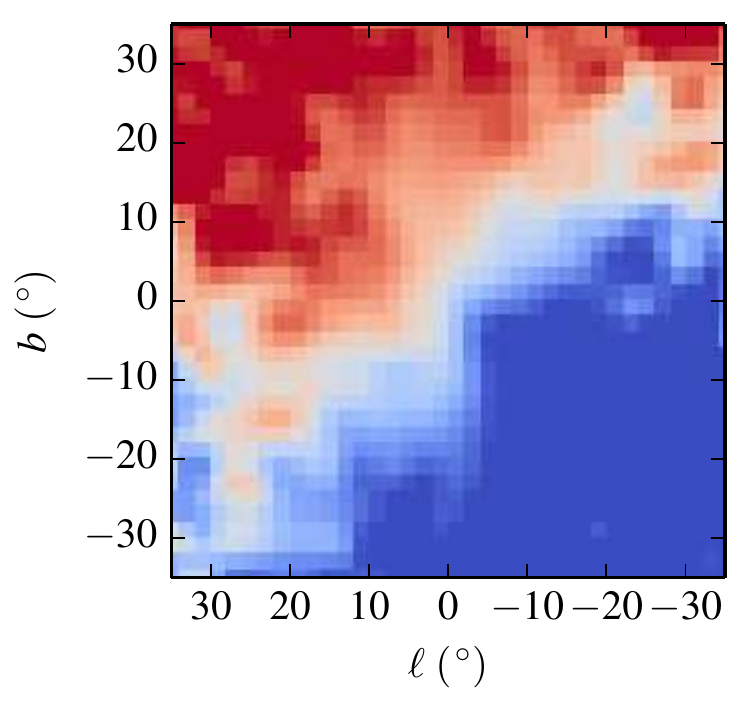}
\includegraphics[width=0.23\textwidth]{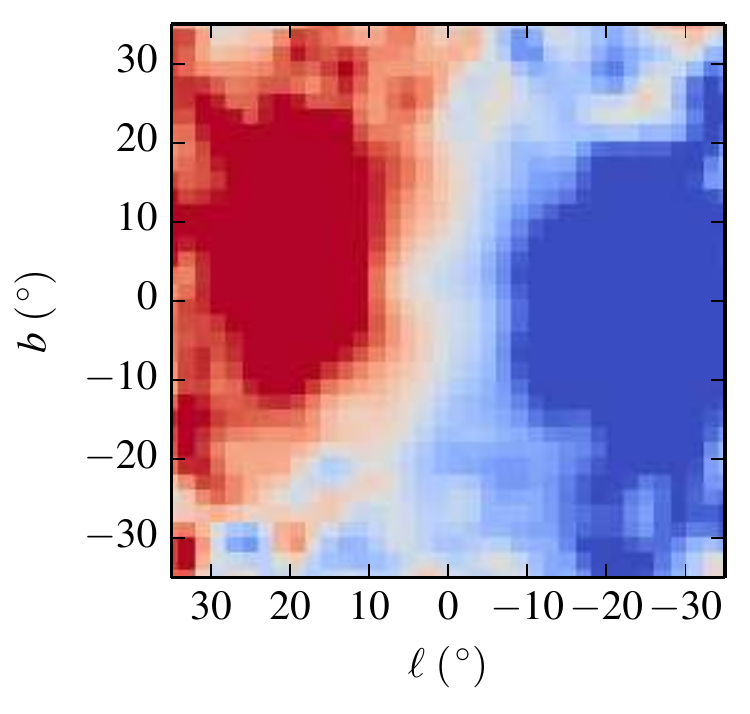}
\includegraphics[width=0.23\textwidth]{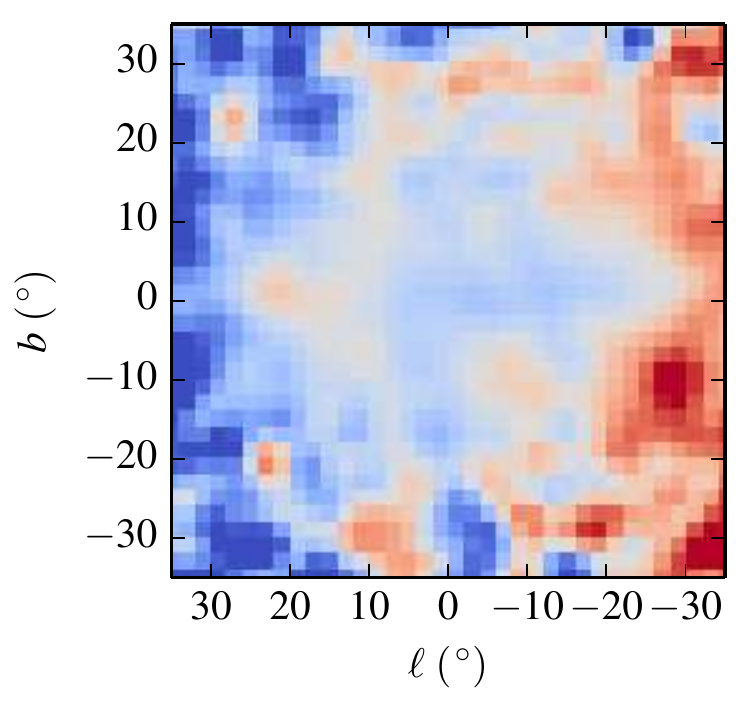}
\includegraphics[width=0.29\textwidth]{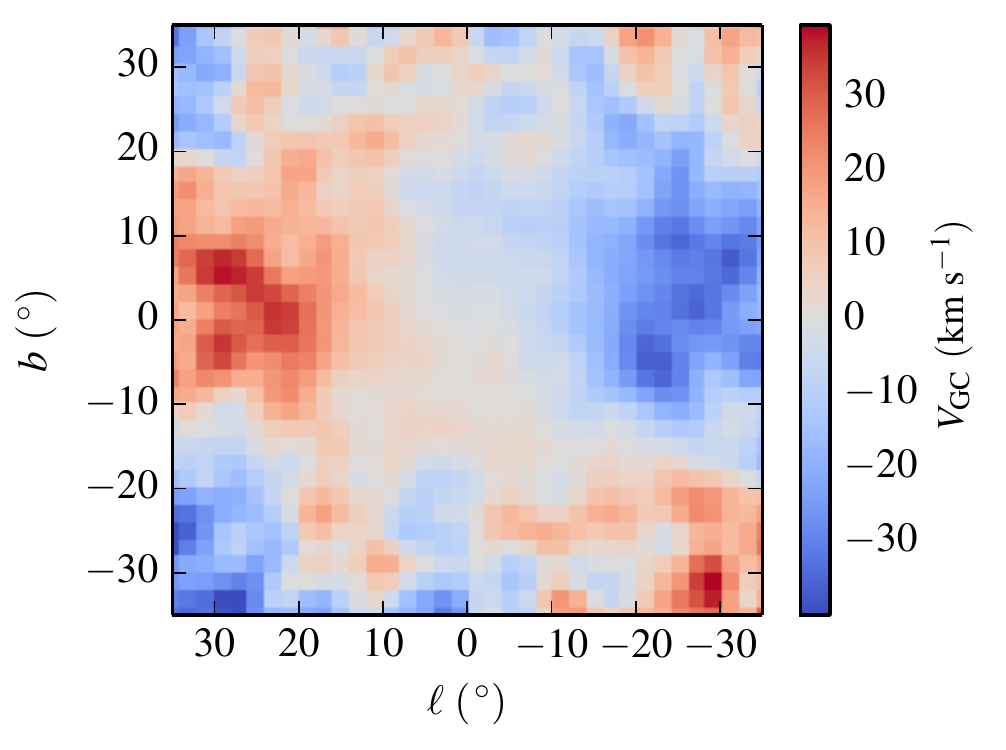}

\includegraphics[width=0.23\textwidth]{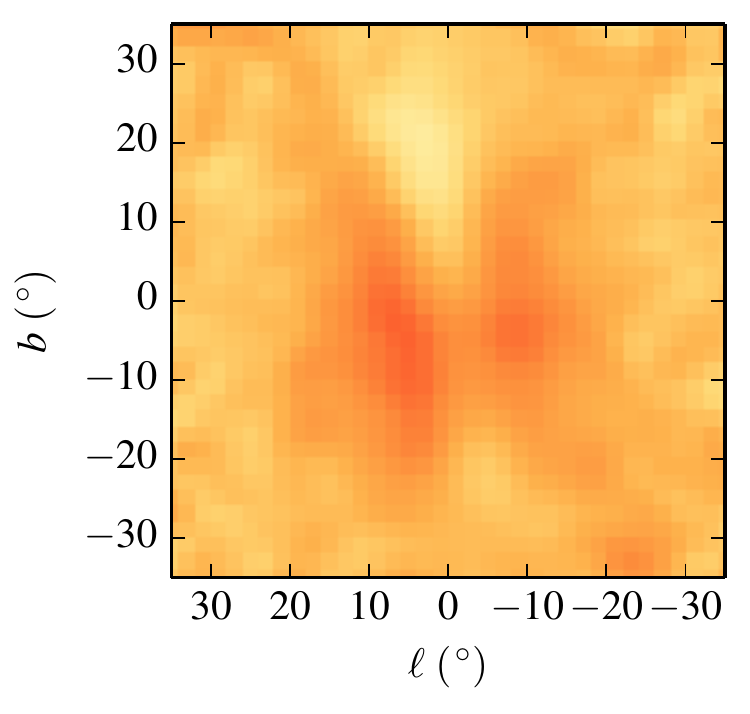}
\includegraphics[width=0.23\textwidth]{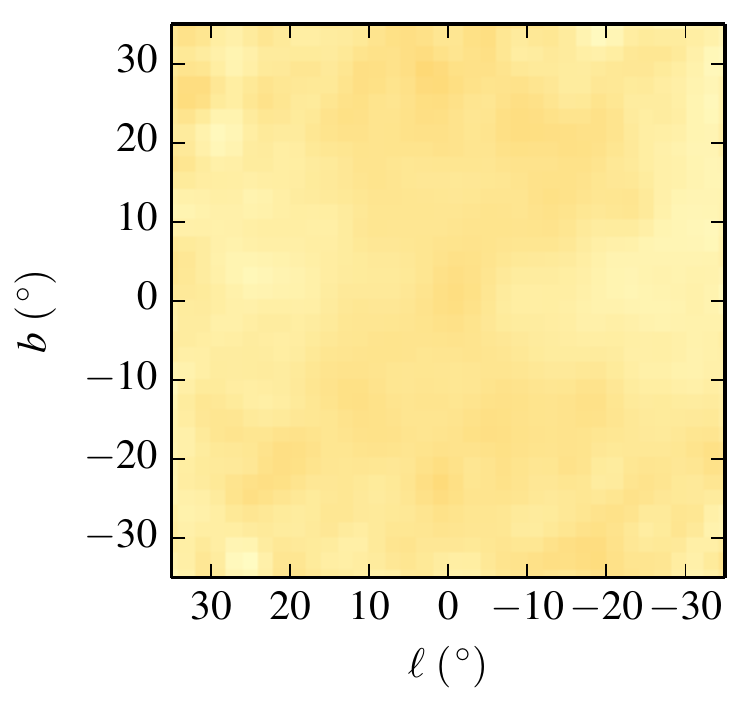}
\includegraphics[width=0.23\textwidth]{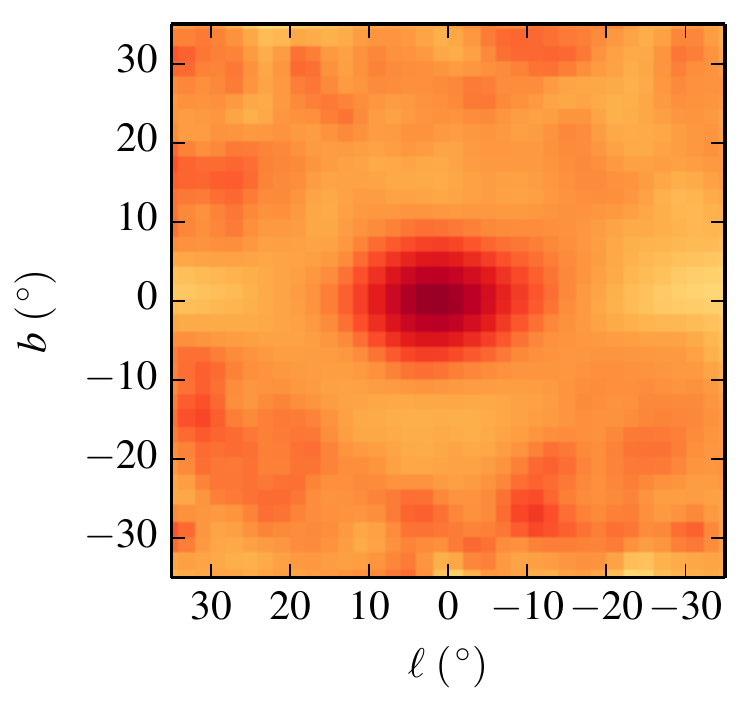}
\includegraphics[width=0.29\textwidth]{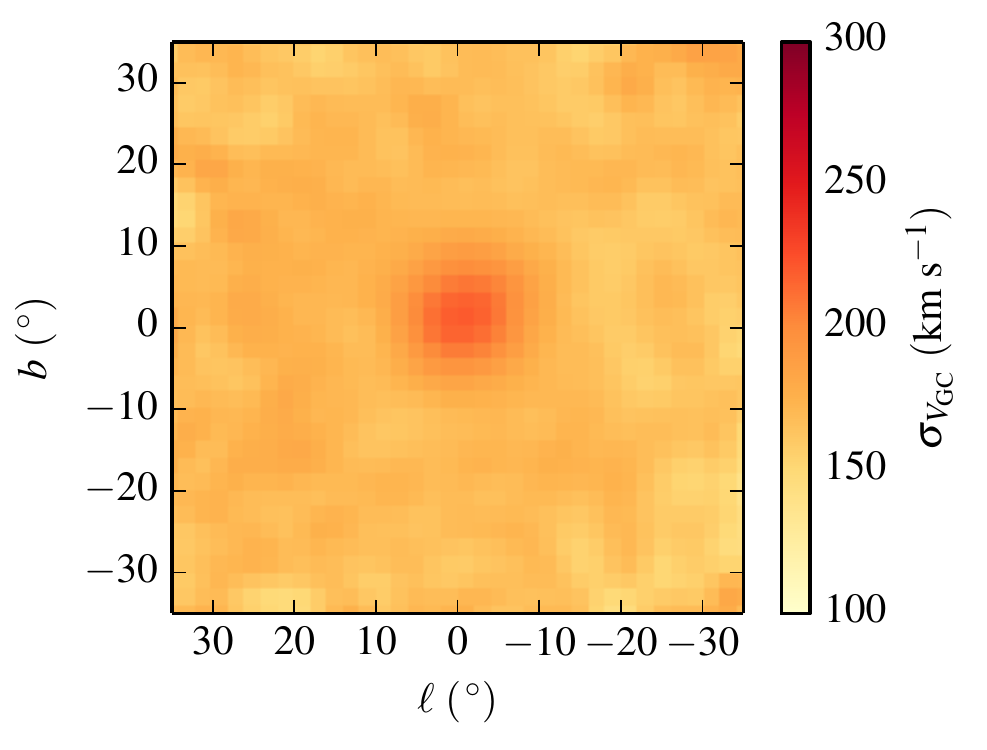}

\caption{Galactocentric rotational velocities ($V_{GC}$) and velocity dispersions 
($\sigma_{GC}$) for the {\it in situ} stellar populations in  the central
regions of the analysed Aquarius haloes. Upper panel: The projected mean
$V_{GC}$ and $\sigma_{GC}$ as a function of galactrocentric longitude
and latitude. Middle and lower panels: 2D projected maps of $V_{GC}$ and
$\sigma_{GC}$, respectively, within the inner 10 kpc  regions. In the
cases where there is a clear bar structure (Aq-C and Aq-D), the simulated velocity
distributions are in agreement with observations of the MW
\citep{zoccali2016}. }
\label{velocinsitu}
\end{figure*}

\begin{figure*}
\includegraphics[width=0.24\textwidth]{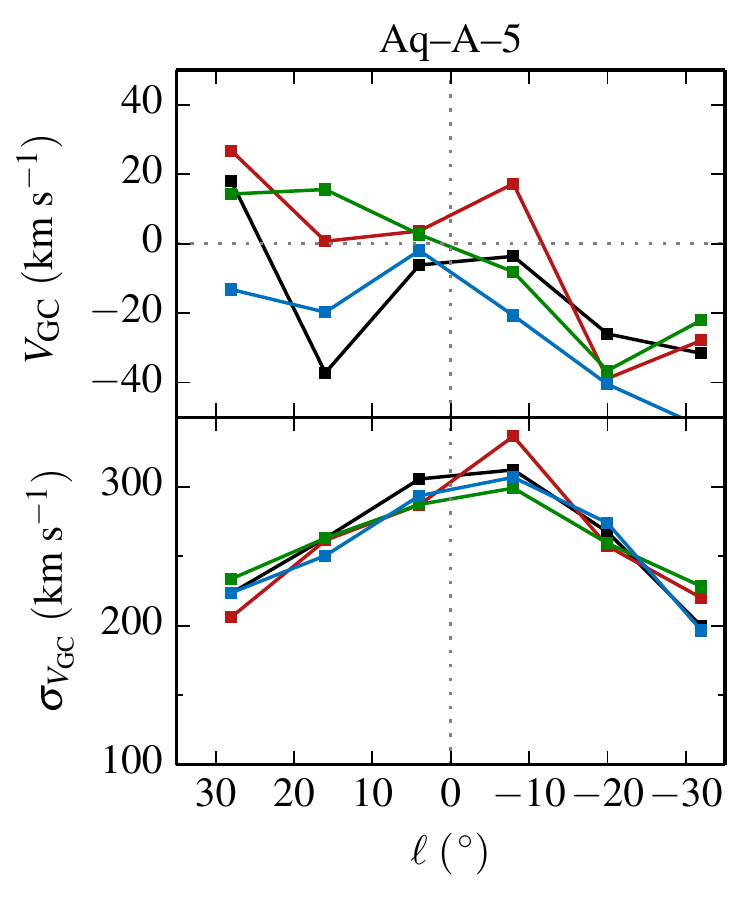}
\includegraphics[width=0.24\textwidth]{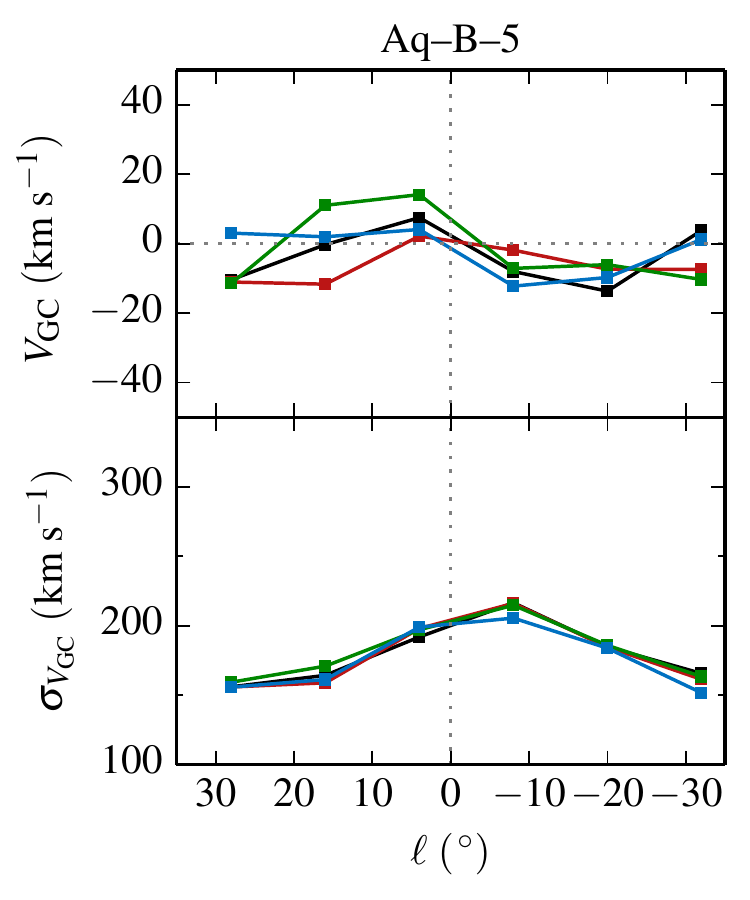}
\includegraphics[width=0.24\textwidth]{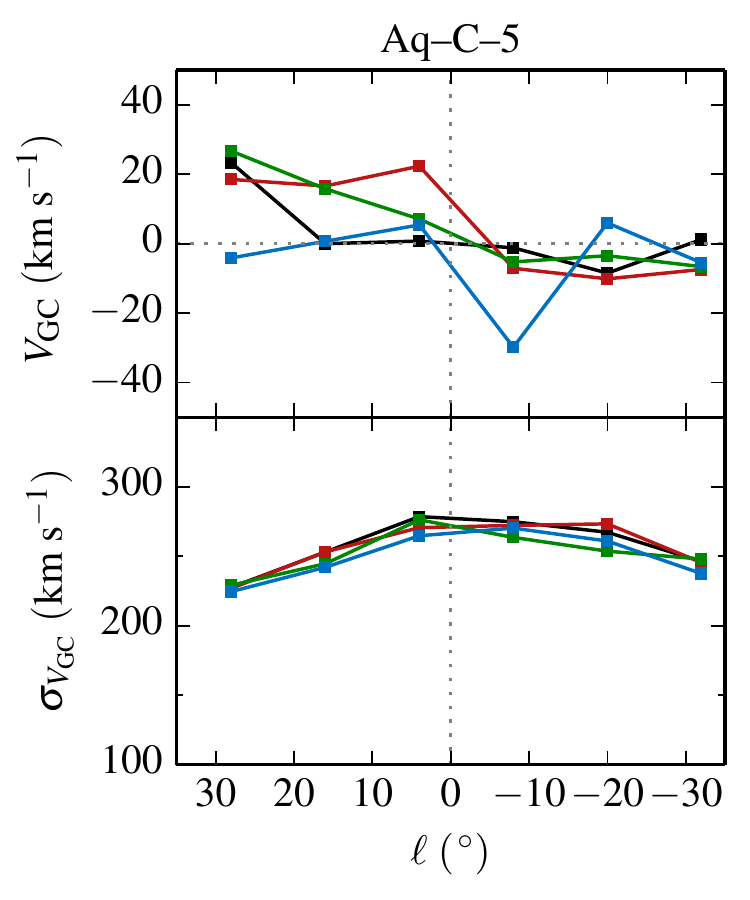}
\includegraphics[width=0.24\textwidth]{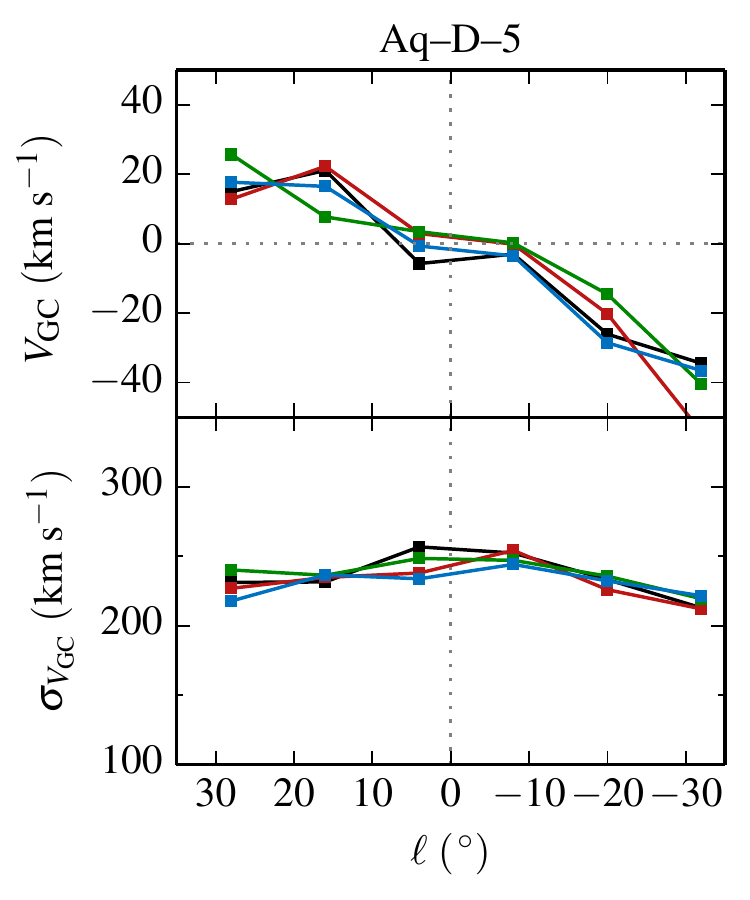}

\includegraphics[width=0.23\textwidth]{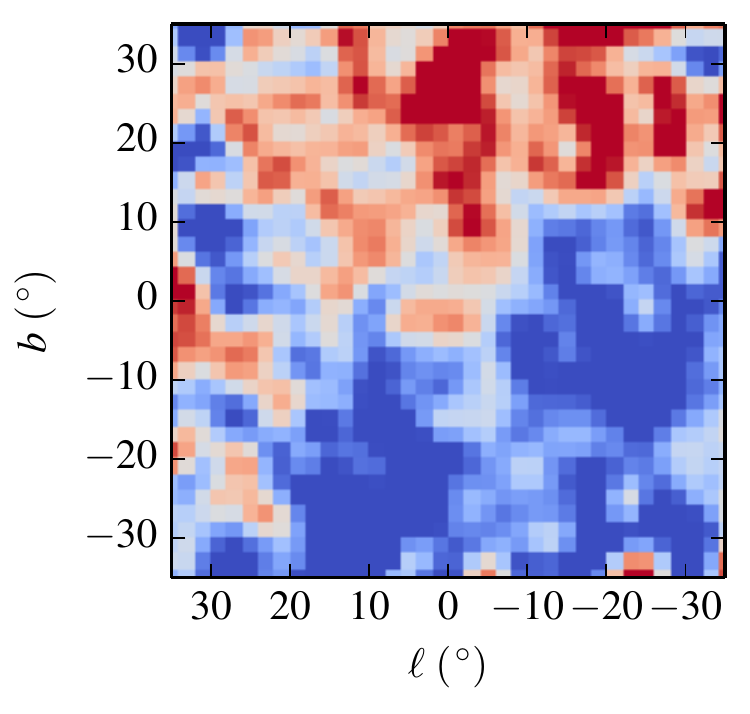}
\includegraphics[width=0.23\textwidth]{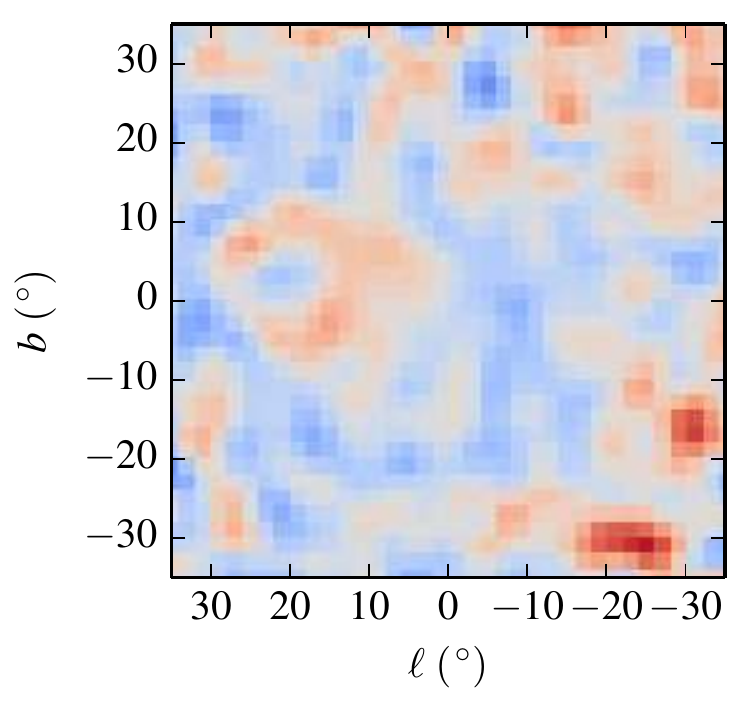}
\includegraphics[width=0.23\textwidth]{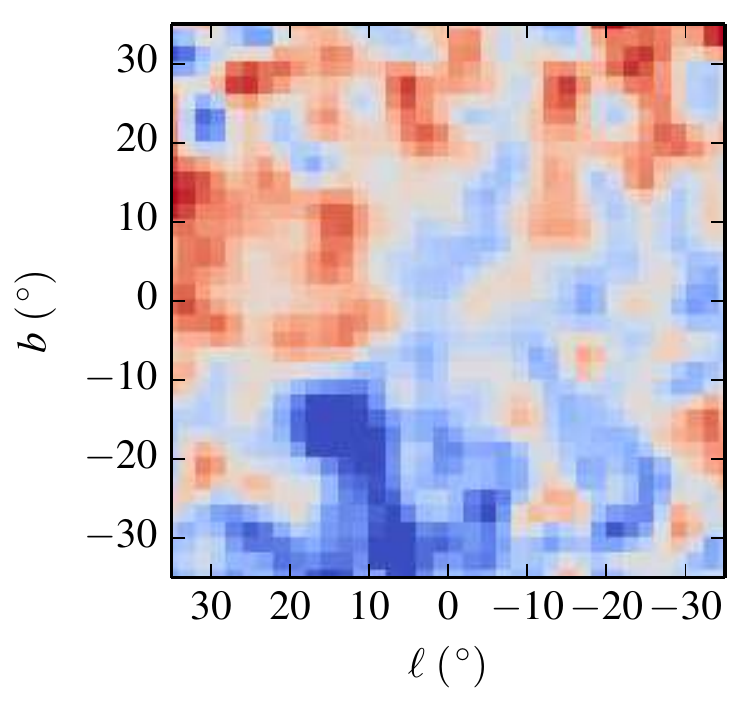}
\includegraphics[width=0.29\textwidth]{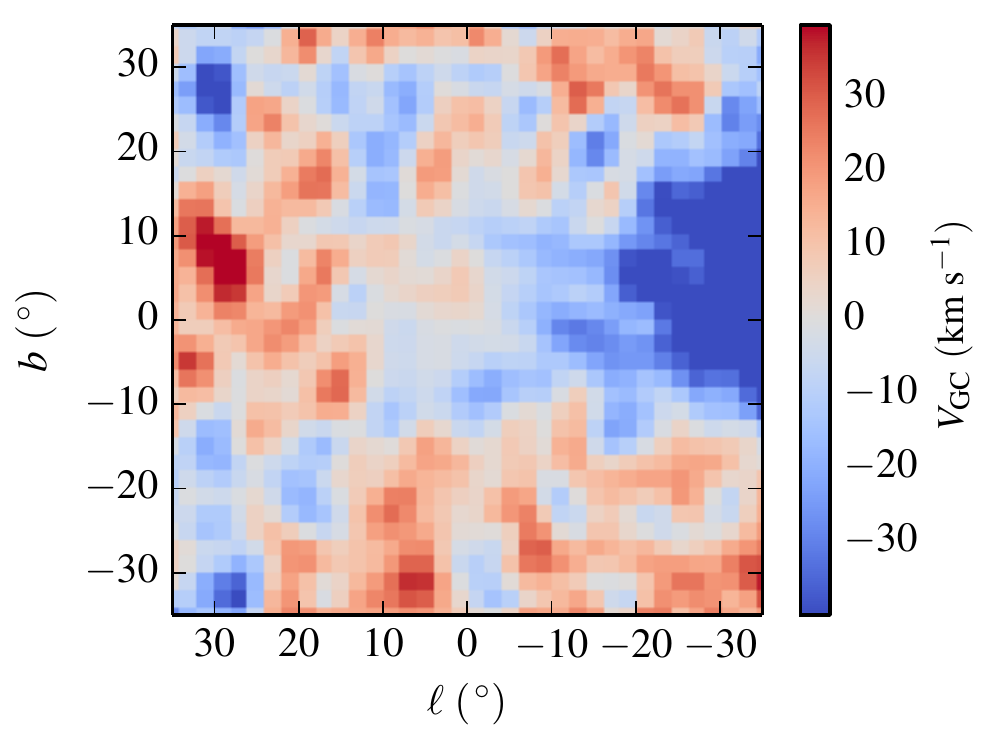}

\includegraphics[width=0.23\textwidth]{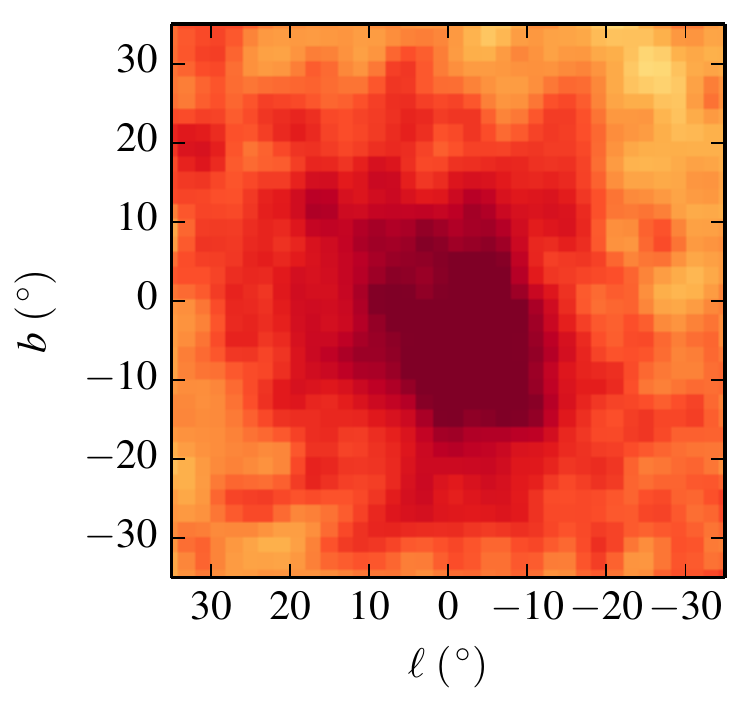}
\includegraphics[width=0.23\textwidth]{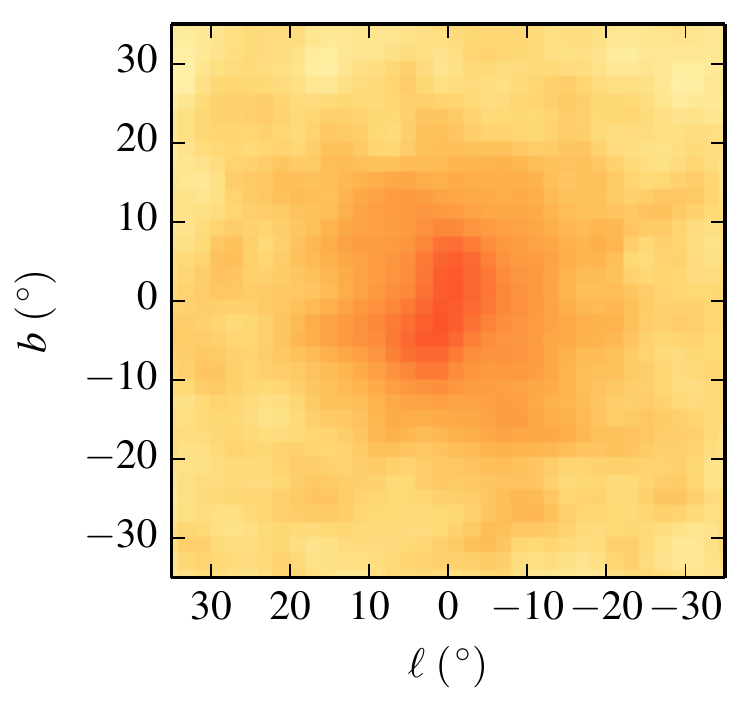}
\includegraphics[width=0.23\textwidth]{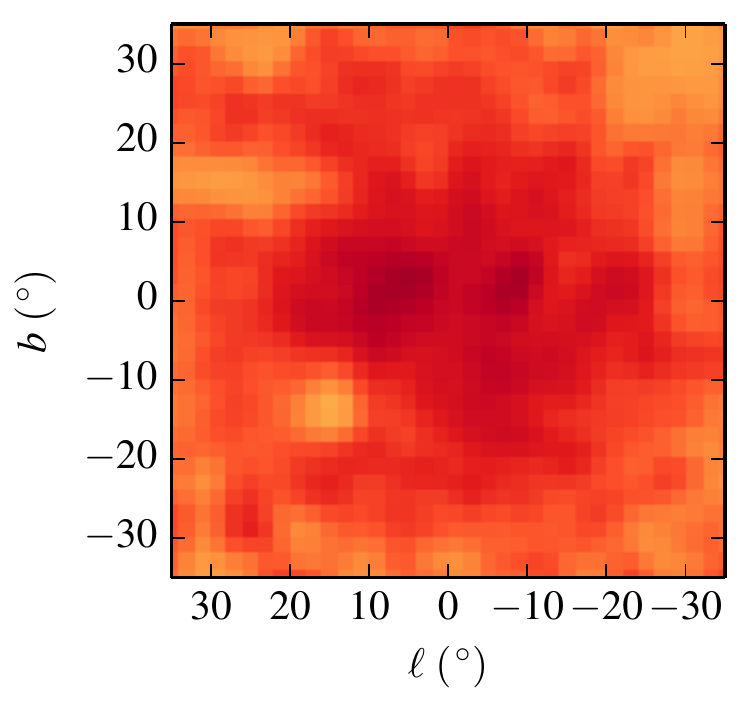}
\includegraphics[width=0.29\textwidth]{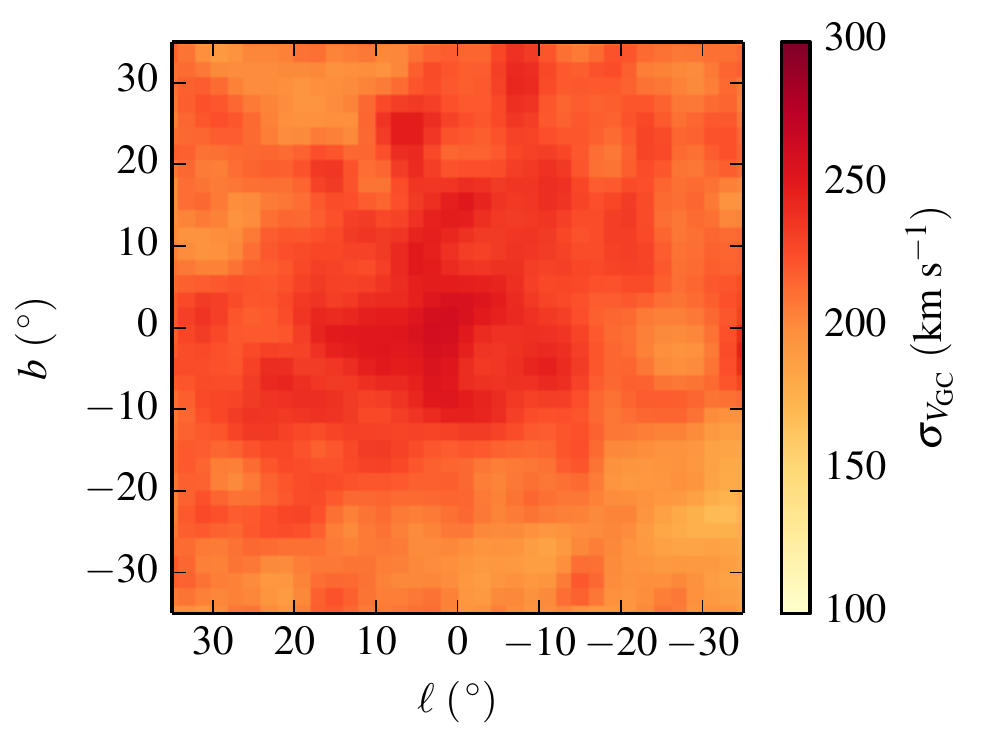}
\caption{ Galactocentric rotational velocities ($V_{GC}$) and velocity dispersions ($\sigma_{GC}$)
for the {\it accreted } stellar populations in  the central regions of the
analysed Aquarius haloes. Upper panel: The projected mean $V_{GC}$ and
$\sigma_{GC}$ as a function of galactrocentric longitude and latitude.
Middle and lower panels: 2D projected maps of $V_{GC}$ and
$\sigma_{GC}$, respectively, within the inner 10 kpc egions. Accreted
stars are dominated by velocity dispersion. } 
\label{velocaccreted}
\end{figure*}

Here we focus on the kinematical properties of stars in the central
regions. To make the analysis similar to Galactic observations \citep{zoccali2016},
Figs.~\ref{velocinsitu} and \ref{velocaccreted} show the
galactocentric rotational velocities, as seen from a location at  a
disc scale-length away from the galactic centre  as given in
\citet{scan10}), and the observers is located at a 20 degree angle
with respect to the bar. The mean rotational velocities and dispersions
are shown as a function of galactic longitude, $l$, for different ranges
of galactic latitude, $b$ (top rows). We also include their 2D projected
maps (middle rows and bottom rows), where each frame is approximately 10
kpc wide. 

For the in situ stellar populations, the central regions exhibit
different levels of rotation, while the velocity dispersions have a
maximum in the very central regions, as expected, since they are the
most dense and  concentrated (Fig.~\ref{velocinsitu}). However,
for Aq-B there is no clear central concentration, which is consistent
with the star formation history being more extended in time, as shown in
Fig~\ref{mdf_age_All}. Aq-C and Aq-D show the presence of the bar
structure in the velocity dispersion maps. The central regions of Aq-C
best resembles the results reported by
\citet{zoccali2016}.  However, we note that the velocity dispersions
are larger than those reported for the MW. This may be due to an
excess of stars in the central regions compared to observationally motivated
results \citep{scan09,moster2013}.  Nevertheless, it is very
encouraging that hydrodynamical cosmological simulations within the
current cosmological framework are able to naturally reproduce the
co-existence of a bar and spheroidal component.

Accreted stars are dominated by velocity dispersion, as can be
appreciated from Fig.~\ref{velocaccreted}. Aq-C and Aq-D exhibit more
spherical distributions, with weaker signs of rotation and an increase
of the central velocity dispersion. However, there is only a weak dependence on
galactic latitude, in contrast to the trends found for the in situ
components (Fig.~\ref{velocinsitu}).

\begin{figure}
\includegraphics[width=0.24\textwidth]{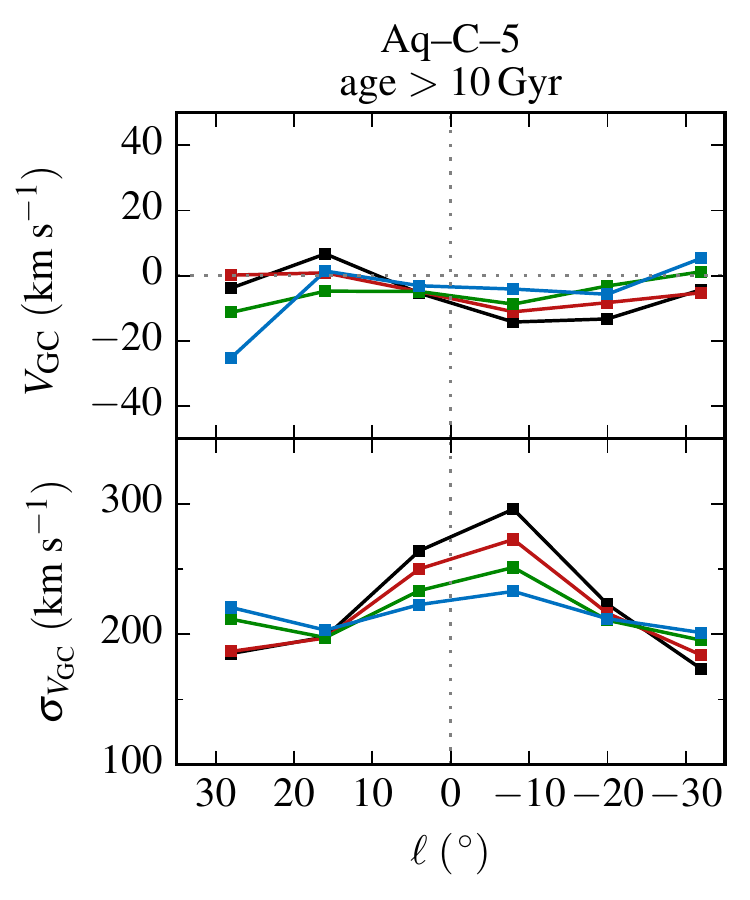}
\includegraphics[width=0.24\textwidth]{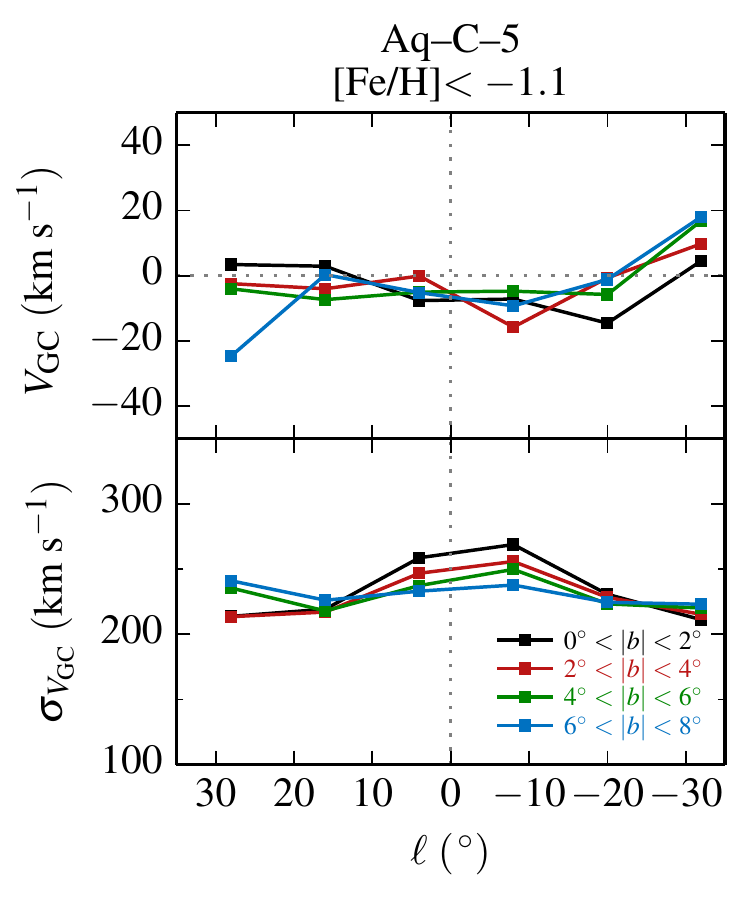}\\
\includegraphics[width=0.23\textwidth]{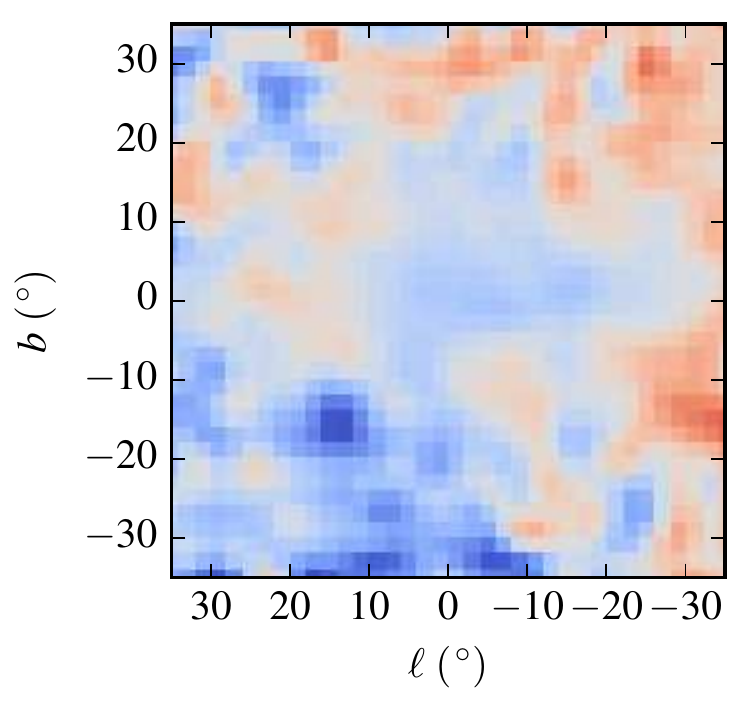}
\includegraphics[width=0.29\textwidth]{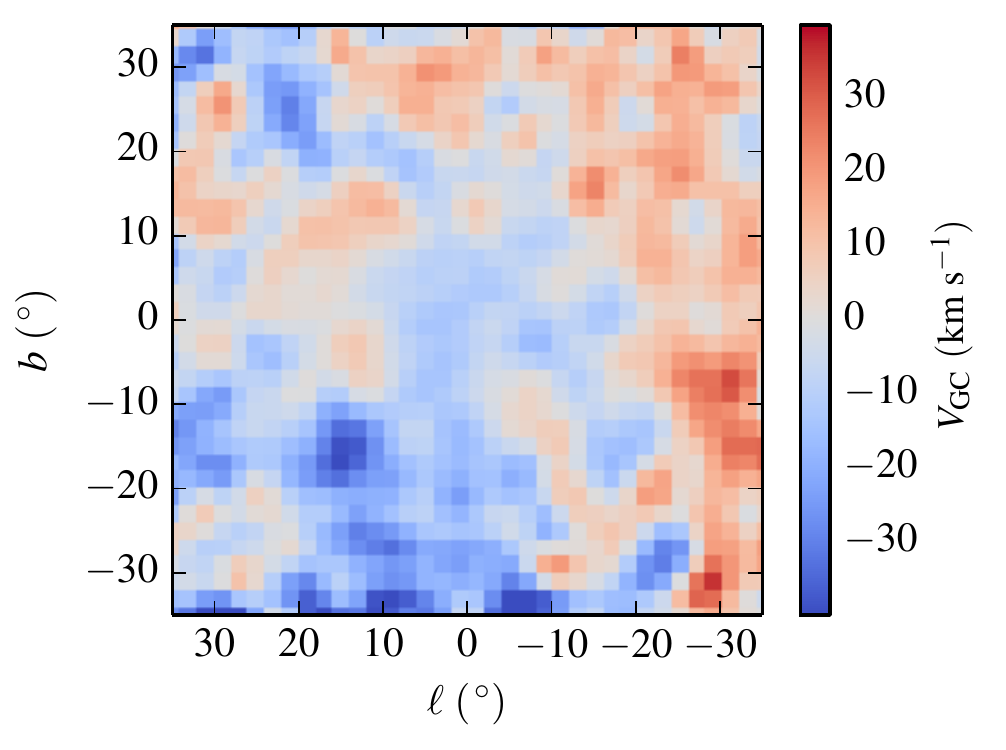}\\
\includegraphics[width=0.23\textwidth]{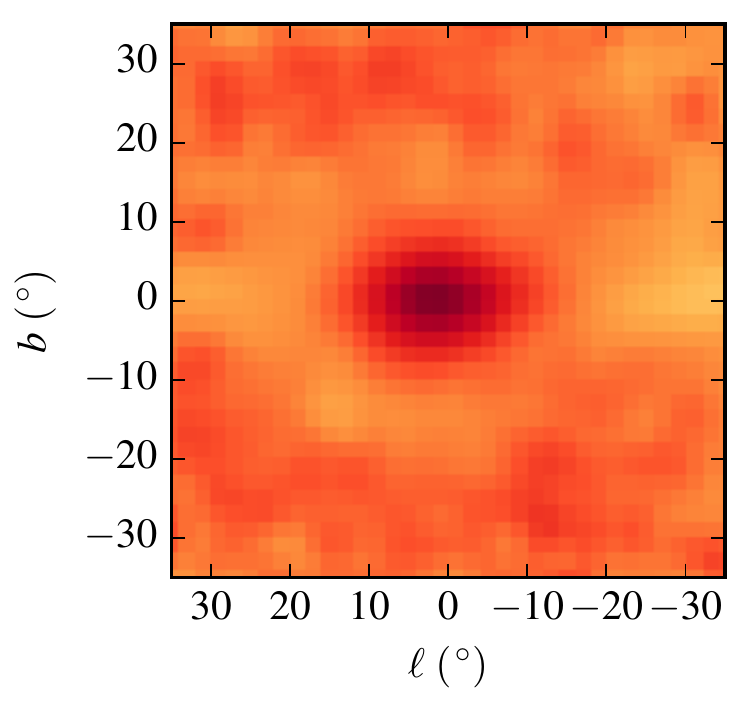}
\includegraphics[width=0.29\textwidth]{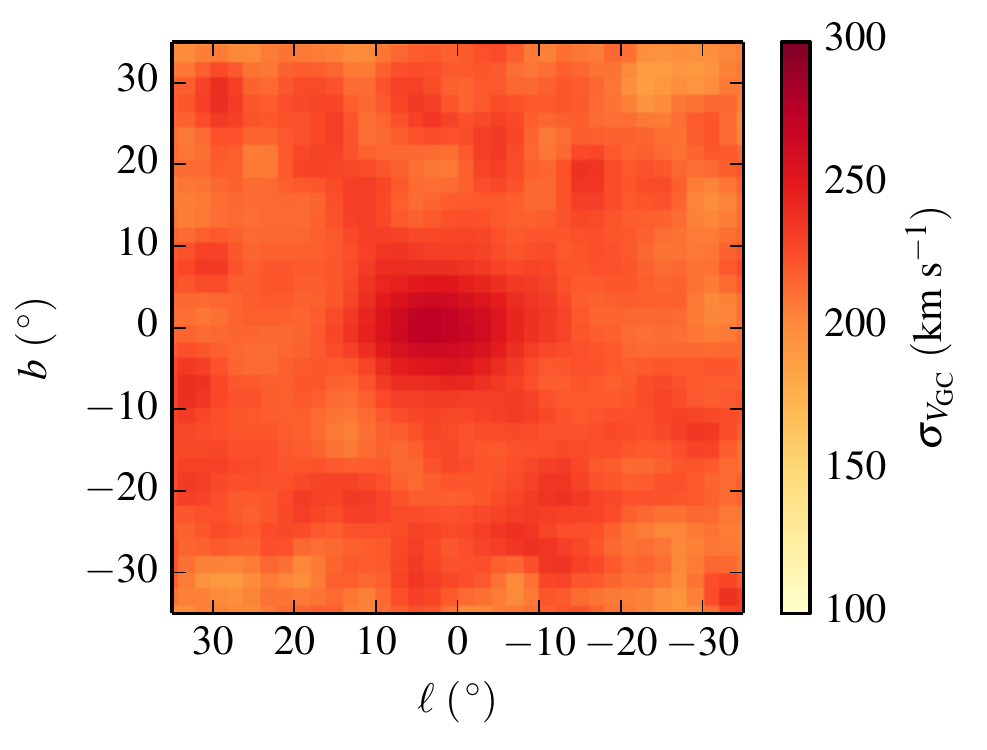}
\caption[]{Galactocentric rotational velocities ($V_{GC}$) and velocity dispersions
($\sigma_{GC}$) for the stellar populations in the central  spheroid
region of the analysed Aq-C-5  halo, selected according to [Fe/H]
$<-1.1$ (right column) and age > 10 Gyr  (left
column). Upper panel: The projected mean
$V_{GC}$ and $\sigma_{GC}$ as a function of  galactocentric
longitude and latitude. Middle and lower panels: 2D projected maps of
$V_{GC}$ and $\sigma_{GC}$, respectively, within the inner 10 kpc
region. 
}
\label{veloAqC}
\end{figure}

Observationally, it is not possible to easily distinguish between
accreted and in situ stars. Hence for the purpose of providing a more
suitable comparison with observations,  we estimate the  kinematic distributions
selecting stars according to [Fe/H] $<-1.1$ (similar results are
found by using Fe/H]$<-0.5$) and  stellar ages older than 10 Gyr,
separately, and without distinguishing between accreted and in situ components. Figure
~\ref{veloAqC} shows these distributions for Aq-C,
the halo that best reproduces the co-existence of a bar and a spheroidal
component. As
can be seen from this figure, old stars 
follow  more closely the distributions determined for the total in situ
component. In fact, this halo has a clear, old central knot.  Low-metallicity stars also show a central
concentration and velocity dependence in agreement with the presence
of a bar but the signal is weaker.

\section{Conclusions}

We  have analysed the inner 10 kpc spheroid of four haloes of MW mass-size
galaxies simulated within a cosmological context. We focussed our study
on the spatial  distributions, ages, metallicities, and velocity
distributions of  the in situ and accreted stars in these central regions.
Special attention was  paid to old stars ($> 10$ Gyr). These stars can
be traced by  RR Lyraes, and constitute a window into the first stages of
the evolution of galaxies. Our knowlege of the central region of the MW
has improved dramatically in the last few years, providing insights on the bulge
\citep[e.g.,][]{gran2016,zoccali2016} and the stellar halo
\citep[e.g.,][]{an2015, santucci2015,carollo2016,das2016}.

The analysis of the simulated stellar populations in the context of
these new observations, as well as forthcoming from  the Gaia
mission \citep[e.g.,][]{helmi2016} is of utmost importance to interpret
them within a cosmological framework. It is not only relevant to
understand the properties of the bulge and the stellar haloes
separately,  but also together. It is also  vital to confront
 models that set constraints on the subgrid physics used in the
simulations,  since present numerical codes show agreement {in their}
global behaviours, but important differences when examined in more
detail \citep[e.g.,][]{harmsen2017}.

For the central  spheroid region defined within the inner 10 kpc of the analysed
Aquarius subsample, we find the following results:

\begin{itemize}

\item The central regions of the simulated MW mass-sized galaxies are found
to be populated mainly by stars older than 10 Gyr in three of the
analysed haloes (Aq-B has $\sim 50 $ per cent  of its stars > 10
Gyr). Most of these stars formed in situ, with a fraction of old stars
in the range $\sim 0.10-0.35$ contributed by satellite accretion. The
spatial distributions of in situ and accreted stars are different --
accreted stars form centrally concentrated spheroidal distributions,
while the in situ components exhibit clear bar structures or elongated
distributions. The predicted kinematics are consistent with these
properties. In particular, the rotational velocity and velocity
dispersion distributions of the central region of Aq-C resembles those
reported by \citet{zoccali2016}.  

\item  Both in situ and accreted stars in the central regions formed in main starbursts
(in the progenitor and accreted satellites, respectively). However,
accreted stars formed early, so that there is mean age difference of
$\sim 0.5-1$ Gyr between the in situ and accreted stars. The age
distributions of in  situ stars exhibit a tail towards younger ages. Aq-B
has a more uniform star-formation history for both in situ and accreted
stars, since it is affected by a more recent merger event. The age
distributions are consistent with the central regions being dominated by
stars older than 10 Gyr. This finding is in agreement with the
observational results of \citet{santucci2015} and \citet{carollo2016}, who identified an ancient chronographic sphere
 in the MW, extending out to beyond the the Sun's location.

\item The metallicity distributions of the in situ and accreted stars are slightly different.
The metallicity distributions of  the accreted stars are shifted towards
lower abundances than those of the in situ stars, and the former have
higher $\alpha$-element enhancements. These trends  are consistent
with their formation history, since  the accreted stars are formed
in  satellite galaxies with dynamical masses smaller than
$10^{10}{\rm M_{\odot}}$ \citep{tissera2014}, while  the in situ
stars formed in the main progenitors. 

\item Regarding the old stars, they represent most of the stars in the central regions,
on average. In our simulations, old, accreted stars belong
preferentially to the stellar haloes within the inner 10 kpc. A mass
fraction of $\sim 35$ per cent are part of the bulge components. The
accretion histories show that Aq-A and Aq-C received larger
contributions from smaller satellites. On the other hand, the central
regions of Aq-B and A-D experienced a greater contribution from  a few
massive accreted dwarf galaxies. As a consequence, the former have more
metal-poor and $\alpha$-element enhanced stellar populations. At a
fiducial solar radius, we  estimate that $\sim 40$ per cent of the old
stars are accreted. This fraction is in reasonable agreement with
the recent results of \citet{an2015} for the solar neighbourhood. 

\item All the analysed central regions have a small fraction of stars
younger than  0.1 Gyr, $\sim 3$ per cent, on average. These stars are formed
in situ, and have high metallicity and low $\alpha$-element 
enhancements; they formed from heavily recycled material. It
would of great interest to measure the chemical abundances of the cepheids
stars  that have been detected in the bulge of the MW \citep{dekany2015}.

\item Among the analysed simulated galaxies, Aq-C is found to
exhibit kinematical distributions that are broadly similar to those
reported for the MW, albeit with a higher velocity dispersion. Despite this,
the Aq-C halo is a clear example of an inner region where a bar structure
and a spheroidal component co-exist. The former is dominated by in
situ stars while the latter hosts old stars formed in the first stages of
its assembly. The kinematic distribution shows a concentration of old
stars in the very central region of this halo. 

\end{itemize}



\section*{Acknowledgments}
PBT, DM, and AM acknowledge partial support from the Nucleo UNAB 2015
DI-677-15/N of Universidad Andres Bello. PBT acknowledges partial
support from Fondecyt Regular 1150334 and the Southern Astrophysics
Network (SAN) collaboration funded by Conicyt and PICT 2011-0959 and PIP
2012-0396 (Mincyt, Argentina). D.M. and M.Z. are supported by the BASAL
Center for Astrophysics and Associated Technologies (CATA) through grant
PFB-06, and the Ministry for the Economy, Development, and Tourism,
Programa Iniciativa Cientifica Milenio through grant IC120009, awarded
to the Millennium Institute of Astrophysics (MAS), and by FONDECYT
Regular grant No. 1130196.  DC and TCB acknowledge partial support for this
work from grant PHY 14-30152; Physics Frontier Center/JINA Center for
the Evolution of the Elements (JINA-CEE), awarded by the US National
Science Foundation. REGM acknowledges support from Ci\^encia sem Fronteiras (CNPq, Brazil).


\bibliography{Tissera_rv2}{}
\bibliographystyle{mnras}

\bsp  
\label{lastpage}
\end{document}